\documentclass[a4paper,11pt]{article}
\pdfoutput=1 
\usepackage{jcappub} 

\usepackage{epsfig}
\usepackage{graphics}
\usepackage{bm}
\usepackage{color}
\usepackage{dcolumn}   
\usepackage{bm}     
\usepackage{bbm}       
\usepackage{amssymb}  
\usepackage{amsmath}
\usepackage{latexsym}
\usepackage{float}
\usepackage{ifthen}
\usepackage{caption,subfig}
\usepackage{enumerate}
\usepackage{url}
\usepackage{caption,subfig}
\usepackage{amsopn}
\usepackage{hyperref}

\usepackage{amsfonts}
\usepackage{multirow}
\usepackage{array}
\usepackage{booktabs}
\usepackage{rotating}

\usepackage{ulem}
\def\clap#1{\hbox to 0pt{\hss#1\hss}}

\def\bea{\begin{eqnarray}}
\def\eea{\end{eqnarray}}
\def\be{\begin{equation}}
\def\ee{\end{equation}}
\def\mpl{M_{\rm Pl}}
\def\d{\mathrm{d}}
\newcommand{\bpm}{\begin{pmatrix}}
\newcommand{\epm}{\end{pmatrix}}

\newcommand{\lp}{\left(}
\newcommand{\rp}{\right)}
\newcommand{\lb}{\left[}
\newcommand{\rb}{\right]}

\newcommand{\F}{{\mathcal F}}

\newcommand{\Xt}{\tilde{X}}
\newcommand{\Yt}{\tilde{Y}}

\newcommand{\lambdab}{\bar{\lambda}}

\newcommand{\vPhi}{\vec{\Phi}}

\renewcommand{\geq}{\geqslant}

\newcommand{\mS}{\mathcal{S}}

\newcommand{\lag}{\mathcal{L}}
\newcommand{\mH}{\mathcal{H}}

\newcommand{\tr}[1]{{\mathrm{Tr}\lb  #1\rb}}
\newcommand{\nM}{\hat{\nu}}
\newcommand{\cM}{\hat{C}}
\newcommand{\pM}{\hat{N}}
\newcommand{\mM}{\hat{M}}
\newcommand{\eM}{\hat{E}}
\newcommand{\tM}{\hat{\theta}}
\newcommand{\gM}{\hat{G}}
\newcommand{\iM}{\hat{I}}
\newcommand{\eps}{\epsilon}
\newcommand{\Awkb}{\hat{A}_{\text{wkb}}}
\newcommand{\bA}{\bar{A}}
\newcommand{\Bwkb}{\hat{B}_{\text{wkb}}}
\newcommand{\Ak}{\hat{A}_{\text{large-k}}}
\newcommand{\Bk}{\hat{B}_{\text{large-k}}}

\newcommand{\cgw}{c_{\rm GW}}
\newcommand{\agw}{\alpha_{\rm GW}}
\newcommand{\fGW}{f_\text{GW}}
\newcommand{\dLgw}{d^\text{GW}_{L}}
\newcommand{\dLem}{d^\text{EM}_{L}}

\newcommand{\dc}{\Delta c}
\newcommand{\dm}{\Delta m}
\newcommand{\dnu}{\Delta \nu}
\newcommand{\dmu}{\Delta \mu}
\newcommand{\bdnu}{\bar{\Delta \nu}}
\newcommand{\onu}{\omega_\nu}
\newcommand{\bonu}{\bar{\omega}_\nu}
\newcommand{\omu}{\omega_\mu}
\newcommand{\bomu}{\bar{\omega}_\mu}
\newcommand{\mtot}{\mu_{\rm tot}}
\newcommand{\bdmu}{\bar{\Delta \mu}}
\newcommand{\bmtot}{\bar{\mu}_{\rm tot}}

\newcommand{\Od}{\mathcal{O}}

\definecolor{orange}{rgb}{1,0.5,0}

\author[a]{Jose Beltr\'an Jim\'enez,}
\affiliation[a]{Departamento de F\'isica Fundamental and IUFFyM, Universidad de Salamanca, E-37008 Salamanca, Spain.}
\emailAdd{jose.beltran@usal.es}
\author[b]{Jose Mar\'ia Ezquiaga}
\affiliation[b]{NASA Einstein Fellow, Kavli Institute for Cosmological Physics and Enrico Fermi Institute, The University of Chicago, Chicago, IL 60637, USA.}
\emailAdd{ezquiaga@uchicago.edu}
\author[c]{and Lavinia Heisenberg}
\affiliation[c]{Institute for Theoretical Physics, 
ETH Zurich, Wolfgang-Pauli-Strasse 27, 8093, Zurich, Switzerland.}
 \emailAdd{lavinia.heisenberg@phys.ethz.ch}

\title{Probing cosmological fields with gravitational wave oscillations}
\abstract{
Gravitational wave (GW) oscillations occur whenever there are additional tensor modes interacting with the perturbations of the metric coupled to matter. These extra modes can arise from new spin-2 fields (as in e.g. bigravity theories) or from non-trivial realizations of the cosmological principle induced by background vector fields with internal symmetries (e.g. Yang-Mills, gaugids or multi-Proca).
We develop a general cosmological framework to study such novel features due to oscillations. 
The evolution of the two tensor modes is described by a linear system of coupled second order differential equations exhibiting friction, velocity, chirality and mass mixing. We follow appropriate schemes to obtain approximate solutions for the evolution of both modes and show the corresponding phenomenology for different mixings. 
Observational signatures include modulations of the wave-form, oscillations of the GW luminosity distance, anomalous GW speed and chirality. 
We discuss the prospects of observing these effects with present and future GW observatories such as LIGO/VIRGO and LISA.
}
\date{\today}

\keywords{gravitational waves, dark energy, modified gravity.}

\begin{document}

\maketitle
\flushbottom

\section{Introduction}

After more than 100 years of scrutiny, General Relativity (GR) still stands out as the best contender to explain gravitational phenomena in a broad range of scales. The core of its experimental confirmation is conformed by the three classical tests (the perihelion shift of Mercury, the deflection of light and the gravitational redshift), although there are nowadays many other probes at different regimes \cite{Will:2014kxa}. 

The vast majority of these tests however probed the non-radiative sector of the theory, while gravitational waves (GW) have remained more elusive as a consequence of the weakness of the gravitational interaction. Nevertheless, the celebrated quadrupole formula, calculated as early as 1916 by Einstein, was confirmed very precisely by measurements of the period variation of the Hulse-Taylor binary pulsar \cite{Wex:2014nva}. This is considered to be a first, albeit indirect, proof for the existence of GWs. Additionally, these measurements permitted to confirm  the predominantly quadrupolar nature of the gravitational radiation as it corresponds to a spin-2 field, and to constrain its propagation speed to deviate from the speed of light at most by a factor $10^{-2}-10^{-3}$. This already puts some of the prominent effective field theories of gravity into a corner \cite{Jimenez:2015bwa}.

We had to wait a century since the inception of GR for the major breakthrough achieved by the LIGO team \cite{TheLIGOScientific:2014jea} with the direct detection of the first GW ever observed \cite{Abbott:2016blz}. This observation allowed to test gravity in the dynamical, strong-field regime \cite{TheLIGOScientific:2016src}. 
After the VIRGO team \cite{TheVirgo:2014hva} joined the LIGO network, the sensitivity to the polarization of the GWs was increased and improved information about the source position became available \cite{Abbott:2017oio}. More recently, the first detection of the signal from the merger of two neutron stars inaugurated the era of multimessenger astronomy as it was possible to obtain the signal of the event in GWs \cite{TheLIGOScientific:2017qsa} as well as its electromagnetic counterpart \cite{GBM:2017lvd}. Among many other revolutionary discoveries, this observation posed a direct constraint on the difference in the propagation speeds of GWs and photons, which can only differ by one part in $10^{-15}$ \cite{Monitor:2017mdv}, with strong implications for theories featuring an anomalous propagation speed for GWs \cite{Lombriser:2016yzn,Ezquiaga:2017ekz,Creminelli:2017sry,Sakstein:2017xjx,Baker:2017hug} (see \cite{Heisenberg:2018vsk,Ezquiaga:2018btd,Kase:2018aps} for recent reviews). This new stringent constraint is many orders of magnitude tighter than the ones available from binary pulsars and remarkably is at the same order as the ones derived from the absence of Cherenkov radiation \cite{Moore:2001bv}. 
 
 Since the era of GW astronomy has commenced we can utilize it to extract the properties of our universe with data that complements the valuable information we already have from observing the electromagnetic spectrum. The present work partially covers this task by investigating the effects on the GW signals induced by the presence of a helicity-2 partner in our universe. The origin of this helicity-2 companion can be diverse and, depending on its underlying nature, its interplay with the GWs can lead to a rich phenomenology with potentially discriminating signatures on GW observations. Throughout this work we will have in mind mainly two theoretical scenarios where an additional helicity-2 mode could arise. The first one is to simply consider a second spin-2 field so that the extra helicity-2 mode directly follows from this spin-2 field. The paradigmatic framework for this scenario is provided by massive bi-gravity theories \cite{deRham:2010kj,Hassan:2011zd}\footnote{We could also mention higher order curvature theories that generally propagate an additional massive spin-2 field. However, one of the spin-2 fields is necessarily a ghost so that we will not consider those scenarios.}. The second scenario we have in mind gives rise to a helicity-2 companion in a somewhat less evident manner, based on a non-trivial realization of the cosmological principle. This mechanism occurs when the universe contains some fields whose vacuum expectation values break homogeneity and/or isotropy, so their background configuration does not comply a priori with the cosmological principle. It is possible however to have a homogeneous and isotropic universe if these fields contain some internal symmetries that allow to {\it hide} the apparent violation of the cosmological principle. The helicity-2 partner then originates from this non-trivial background configuration that enables some perturbations to arrange themselves into a helicity-2 mode, even if the original theory does not have any spin-2 field. A paradigmatic example for this mechanism, that we will explain at some length in the core of the manuscript, corresponds to models with vectors fields (see e.g. \cite{Maleknejad:2011jw,Adshead:2012kp,Maleknejad:2012fw,BeltranJimenez:2018ymu,Piazza:2017bsd}).\footnote{Solid inflation \cite{Endlich:2012pz} is another example where the apparent violation of the cosmological principle occurs and it is restored by some internal symmetries. In that case however there are no additional helicity-2 modes.}
 
Although one could expect to have many different sources of GWs from different cosmological scenarios, in this work we focus on individual detections and will not consider stochastic backgrounds. Furthermore, we will be interested in studying the oscillations between GWs and the helicity-2 partner as the main effect, although there could be some other effects induced by the background. GW oscillations have been previously considered in the context of bigravity \cite{DeFelice:2013nba,Narikawa:2014fua,Max:2017flc,Max:2017kdc,Belgacem:2019pkk} and gauge field dark energy \cite{Caldwell:2016sut,Caldwell:2018feo}. The oscillations of GWs into photons mediated by magnetic fields have also been explored \cite{Cembranos:2018jxo}. We extend those analysis and set up our general framework to study GW oscillations in Section \ref{sec:genFrame}. There, we parameterize all distinctive features realized in different theories and establish our working assumptions. The contributions from friction, velocity, chirality and mass will be discussed in detail in the corresponding subsections. Special classes of gravity theories containing a second tensor mode, such as massive bi-gravity \cite{deRham:2010kj,Hassan:2011zd}, Yang-Mills \cite{Cervero:1978db,Galtsov:1991un,Darian:1996mb}, Abelian multi gauge fields in a gaugid configuration \cite{Piazza:2017bsd} and multi Proca fields \cite{ArmendarizPicon:2004pm,Hull:2014bga,Allys:2016kbq,Jimenez:2016upj} interactions will be then introduced in Section \ref{sec:theorLandscape}, where they will represent one or a combination of the distinctive features in terms of friction, velocity, chirality and mass. Whereas bigravity leads to a mass mixing, the cosmological gauge fields gives rise to a richer phenomenology, including friction and chiral mixing. Some of the vector-tensor theories can even induce a velocity mixing. The phenomenological implications of these mixings on GW observations will be investigated in Section \ref{sec:pheno}. This includes modulations of the wave-form, oscillations of the GW luminosity distance, anomalous GW speed and chirality. We will conclude and give future prospects in Section \ref{sec:conclu}.

\section{General framework for gravitational waves oscillations}\label{sec:genFrame}

\begin{figure}[t]
\centering 
\includegraphics[width=\textwidth]{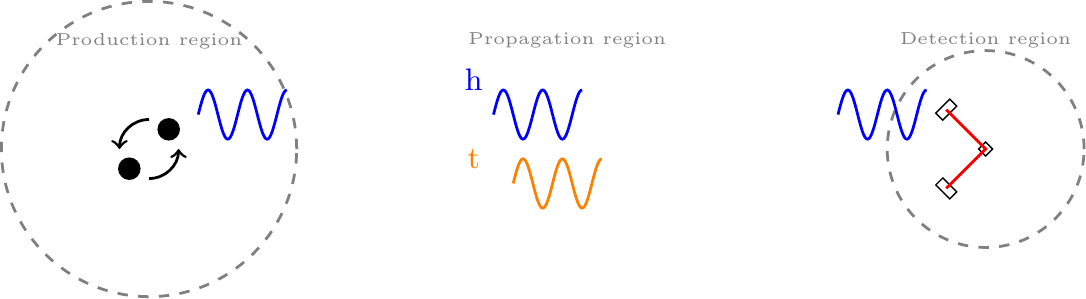}
 \caption{Schematic representation of the different regions between the source and the detector: production, propagation, detection. We assume that \emph{(i)} GWs are generated as in GR, \emph{(ii)} the second tensor $t$ is only excited in the propagation region and \emph{(iii)} only $h$ couples to matter and, thus, the detector.}
 \label{fig:propagation_region}
\end{figure}

In this section we will introduce the general framework to study the cosmological propagation of GWs in models with an additional tensor mode. Theoretical scenarios featuring these extra tensor degrees of freedom will be explored later, in Sec. \ref{sec:theorLandscape}.   
We will thus develop here an effective parameterization for the cosmological propagation of a system with two helicity-2 modes. These tensor modes will be assumed to propagate on a homogeneous and isotropic background described by the Friedmann-Lema\^itre-Robertson-Walker (FLRW) metric, 
\be
\d s^2=a^2(\eta)\Big(-\d\eta^2+\delta_{ij}\d x^i\d x^j\Big)
\label{eq:FLRW}
\ee 
in conformal time $\d\eta=\d t/a$. On top of these homogeneous and isotropic backgrounds we will consider small perturbations. They can be decomposed into scalar, vector, and tensor perturbations according to the irreducible representations of the background $SO(3)$ symmetry. The decomposition theorem allows to study each sector independently at linear order so we will only consider the tensor modes. The tensor perturbations of the metric then read
\be
g_{ij}=\bar{g}_{ij}+h_{ij}(\eta,\vec{x})\,,
\ee
where $\bar{g}_{ij}$ is the FLRW metric given in \eqref{eq:FLRW} and the tensor perturbations are transverse and traceless, i.e. $\partial^i h_{ij}=h^i{}_i=0$. As placeholders for the two respective polarizations we will use $h_{+,\times}$ or $h_{L,R}$ in the usual circular and helicity basis respectively. As mentioned above, the tensor perturbations can only couple to other tensor sectors but are completely decoupled from the scalar and vector perturbations at linear order. The cosmological scenarios that we are interested in contain a second tensor mode that we will denote as $t_{ij}$. Precisely the presence of this tensor companion, that in turn mixes with the metric perturbation $h_{ij}$, enables the possibility of GW oscillations. The complete phenomenology for these cosmological scenarios from the emission of GWs by astrophysical systems (mainly the merger of binary black holes) to the detection of the signal in GW interferometers can be overwhelmingly cumbersome. Thus, in order to connect with observations, we will consider the following assumptions throughout the work:
\begin{enumerate}[\itshape(i)]
\item only the metric perturbation $h_{ij}$ interacts with matter, while the helicity-2 partner $t_{ij}$ lives in a decoupled sector, thus guaranteeing that the interferometers are only directly sensitive to $h_{ij}$;

\item the production of GWs follows that of GR, as observationally supported by the decay of the orbit of binary pulsars \cite{Wex:2014nva}, at least in the region in which the post-Newtonian expansion holds\footnote{This assumption will break near the merger, where the GW scattering is large (disregarding other possible effects such as absorption, dispersion and diffraction). The modifications in the production of GWs go beyond the scope of this work and generally needs highly involved numerical analysis.};

\item the production and detection regions are small compared to the propagation zone (see Fig. \ref{fig:propagation_region}) so that we can consider the GW propagating over the cosmological background from emission to detection;
\item there are not significant deviations of the cosmological background in the propagation zone.
\end{enumerate}
These requirements allow us to write down an effective quadratic action to describe the dynamics of the tensor sector that we can parametrize as follows:
\begin{eqnarray}\label{ActionTensorModes}
\mathcal{S}=\frac12\int \d\eta \d^3x&\Big[&\mathcal{K}^{ab}H'_{a\,ij}H_b'^{\,ij}+\nu^{ab}H_{a\,ij}H_b'^{\,ij}+\mathcal{C}^{ab}\partial_kH_{a\,ij}\partial^kH_b^{\,ij}+\mathcal{M}^{ab}H_{a\,ij}H_b^{\,ij}\nonumber\\
&&+\epsilon^{ijk}\Big(\mathcal{N}^{ab}H_{a\,im}\partial_j H_{bk}{}^m+\tilde{\mathcal{N}}^{ab}H'_{a\,im}\partial_j H_{bk}{}^m\Big)\Big]\,,
\end{eqnarray}
where $a,b$ stand for a {\it flavor} index so that $H_{a\,ij}=(h_{ij},t_{ij})$. We have also introduced the time-dependent matrices in flavor space $\mathcal{K}^{ab}$, $\nu^{ab}$, $\mathcal{C}^{ab}$, $\mathcal{M}^{ab}$, $\mathcal{N}^{ab}$ and $\tilde{\mathcal{N}}^{ab}$, that will be determined by background quantities for each specific model and we have extensively exploited the rotational symmetry of the background that only permits to use the two $SO(3)$-invariant tensors $\delta_{ij}$ and $\epsilon_{ijk}$ to contract spatial indices. This background symmetry further allows to simplify the quadratic action by going to some specific helicity basis, being the circular $(+,\times)$ and the chiral $(L,R)$ polarization basis the most convenient ones. Due to their properties under parity, the first line in \eqref{ActionTensorModes} will be the same for both helicity modes in either basis, while the second line will differentiate between the two polarizations. In the circular basis there will be a mixing of both polarizations $H_+$ and $H_\times$, but this mixing disappears by going to the chiral basis with $H_{L,R}$, in which case a relative minus sign appears for both polarizations. Thus, we can write
\begin{eqnarray}\label{ActionTensorModes2}
\mathcal{S}=\frac12\sum_\lambda\int \d\eta \d^3x&\Big[&\mathcal{K}^{ab}H'_{a\lambda}H'_{b\lambda}+\mathcal{B}^{ab}H_{a\lambda}H_{b\lambda}'+\mathcal{C}^{ab}\partial_kH_{a\lambda}\partial^kH_{b\lambda}+\mathcal{M}^{ab}H_{a\lambda}H_{b\lambda}\nonumber\\
&&+\sum_{\lambdab}g^{\lambda\lambdab}\epsilon^{ijk}\Big(\mathcal{N}^{ab}H_{a\lambda}\partial_j H_{b\lambdab}+\tilde{\mathcal{N}}^{ab}H'_{a\lambda}\partial_j H_{b\lambdab}\Big)\Big] \,,
\end{eqnarray}
where $g^{\lambda\lambdab}= \text{diag}(1,-1)$ in the chiral basis, while in the circular basis we have $g^{++}=g^{\times\times}=0$ and $g^{+\times}=g^{\times +}=1$. We then see explicitly the aforementioned effect of the parity breaking terms in the second line of \eqref{ActionTensorModes} for the tensor polarizations in the different basis. It is important to emphasize that the parity breaking terms appearing in the quadratic action do not necessarily originate from parity violating operators in the original theory, but they can also arise due to the specific field configuration even in perfectly parity-preserving theories. We will give explicit examples below.

One important remark about \eqref{ActionTensorModes2}  is that not all the components of the matrices are independent since they can be related in different ways. One obvious procedure to establish relations is via integrations by parts. For instance,  a term $\mathcal{B}^{hh}h_\lambda h'_\lambda$ can be transformed into $-\frac12\partial_\eta \mathcal{B}^{hh}h_\lambda^2$ which then contributes to $\mathcal{M}^{hh}$. One could proceed similarly to absorb $\mathcal{B}^{tt}t_\lambda t'_\lambda$ into $\mathcal{M}^{tt}$. Thus, without loss of generality we could assume that $\mathcal{B}^{ab}$ only contains off-diagonal terms. Notice however that the presence of off-diagonal terms in $\nu^{ab}$ will present an obstruction to fully eliminate it in favor of other terms in the action via integration by parts if $\mathcal{B}^{ht}\neq\mathcal{B}^{th}$. One could also be tempted to use a flavor basis that diagonalizes the kinetic matrix $\mathcal{K}^{ab}$, but that could potentially introduce couplings of $t_{ij}$ to the matter fields. However, in view of our assumptions, we want to prevent such couplings in favor of a more direct connection with the phenomenological effects and their observable signatures with GW interferometers. 

In the effective quadratic action \eqref{ActionTensorModes} we have neglected terms containing higher spatial derivatives\footnote{Needless to say that we only consider local operators, so the expansion in derivatives will always come in with positive powers. Let us also emphasise that we do not consider higher time-derivatives because they would introduce new (pathological) degrees of freedom, but this does not happen with higher spatial derivatives. An example where these higher spatial derivatives could be relevant is provided by ghost condensate-like models \cite{ArkaniHamed:2003uy} where the dispersion relation is in fact governed by quadratic terms with four spatial derivatives, thus leading to a quartic dispersion relation $\omega^2\propto k^4$. It would be interesting to explore this possibility, but we will not pursue it in the present work.} of the schematic form $(\partial^{2n}H)^2$ for $n\geq1$ because these contributions are expected to be suppressed by a factor $(\partial^2/\Lambda^2)^n$ with $\Lambda$ some scale that depends on the UV physics of the model. Since we assume the EFT theory to be valid at LIGO/VIRGO frequencies, this scale should be $\Lambda\gg f_{\rm LIGO}$ so that these operators are negligible. 

Bearing in mind all the above considerations, it is clear that the main observable of interest for us will be the transfer function of the amplitude $T(\eta,k)$ and the phase of the wave $\theta(\eta,k)$ defined by means of
\be
h_{+,\times}(\eta,k)=h_{+,\times}^{\text{GR}}(\eta,k)\cdot\vert T_{+,\times}(\eta,k) \vert e^{i\int\theta_{+,\times}(\eta,k)\d\eta}\,,
\ee
where $h^\text{GR}$ denotes the GR signal. One should notice that, although in this analysis we are neglecting any modification in the emission or in the cosmological background, those effects could be incorporated by complicating the schema of zones of Fig. \ref{fig:propagation_region}. For instance, if the emission is modified, one only needs to take the appropriate function as the initial condition of the propagation region. On the other hand, if there is a region in which the background is not FLRW, one would need to add an additional transfer function in this new zone. 

At this point it is worthwhile to mention that in the following we are going to solve the evolution assuming a stationary phase approximation. In reality, compact binary mergers produce wave packets of a given duration. In the case in which there is a modified dispersion relation and the frequency of the wave changes rapidly (near the merger for instance), there could be interference within the wave packet \cite{Max:2017kdc}. This could lead to new observational effects. In the inspiral part, however, we expect these corrections to be small, specially for long signals in the detector. 
Finally, let us emphasize that we are limiting our analysis to only one additional tensor mode. Nevertheless, the formalism we are going to present can be straightforwardly extended to cosmologies with multiple extra tensor modes.

\subsection{Solving the evolution}
Without specifying the underlying covariant theory, we can write down the most general coupled equations of motion on top of a cosmological background and work in terms of these parameterized quantities.
For the evolution of the linear tensor perturbations $h_{ij}$ and $t_{ij}$ we will assume a system of coupled, second order differential equations. For the subsequent discussion, it will be convenient to present it in matrix notation and omit the verbose subscripts. The general equations of motion that will govern the evolution of the mixed flavor tensor modes can be compactly written as
\be \label{eq:generalequation}
\lb\frac{\d^2}{\d\eta^2} + \nM\frac{\d}{\d\eta}+\cM k^2 + \pM k +\mM\rb\bpm h \\ t \epm=0\,,
\ee
where we have defined the friction matrix $\nM$, the velocity matrix $\cM$\footnote{We refer to $\cM$ as the velocity matrix because it determines the propagation speeds of high frequency modes. Strictly speaking, $\nM$ and $\mM$ also contribute to the actual propagation speeds of the modes, but their effect will become increasingly negligible as we consider higher frequency.}, the chirality matrix $\pM$, the mass matrix $\mM$, and we are evolving in conformal time $\eta$. These equations characterize a more general situation than the system described by the quadratic action \eqref{ActionTensorModes2} and, for the sake of generality, our subsequent analysis will be based on the equations \eqref{eq:generalequation} rather than on the quadratic action. In principle, the friction matrix $\nM$ could be accompanied by a term linear in $k$ that would correspond to $\tilde{\mathcal{N}}^{ab}$ in \eqref{ActionTensorModes2}. Such a term could lead to interesting chiral effects, but for the sake of simplicity we will not consider it in detail in this work, where we will only study chiral effects originating from $\pM$. One should note that we have dropped the indices because the equations are the same for the two transverse, traceless polarizations. Whenever there are not parity violating terms $\pM$, we will be implicitly working in the usual $+$ and $\times$ basis (although the equation will not change for $L$ and $R$ polarizations). On the contrary, when $\pM\neq0$, we will refer to the circular polarizations left $L$ and right $R$ because in this case the polarization basis matters and this term will have a relative sign for the two helicity modes. 

Before proceeding, some comments are in order. Firstly, if the equations \eqref{eq:generalequation} were derived from an action, there would be some relations amongst the elements of the matrices that can be straightforwardly obtained. For instance, one could impose the equality of cross functional derivatives, i.e., $\delta \mathcal{E}_h/\delta t=\delta \mathcal{E}_t/\delta h$, where $\mathcal{E}_{h,t}$ are the equations of each tensor mode, in order to attain the required relations. Secondly, by working at the field equations level we avoid the ambiguities associated to total derivates in the action or, in other words, to the relations between the different mixing matrices via integration by parts. Finally, we could have written a general matrix for the principal symbol of the equation \eqref{eq:generalequation} that would correspond to $\mathcal{K}^{ab}$ in the action. However, since this matrix must be non-degenerate (otherwise one of the tensor modes would be non-dynamical), we can always multiply the equation by its inverse to remove it. Notice that this is different from diagonalizing and performing a canonical normalization in the action. In this case however the presence of off-diagonal terms will give rise to a coupling of the matter fields to the second tensor mode, in conflict with one of our assumptions. This can be easily understood if we remember that the source for the equations will be along the $h$ direction in flavor space, in accordance with our assumption that the interaction with matter only occurs via the usual coupling to the energy-momentum tensor as $h_{\mu\nu} T^{\mu\nu}$. However, multiplying this source term by a non-diagonal matrix will generically make it acquire some component along the $t-$direction, thus sourcing also the $t-$flavor. Moreover, these non-diagonal components are prone to generate an anomalous propagation speed for GWs and, consequently, they will be tightly constrained. 

 On the other hand, we could impose additional conditions to reduce the number of independent components for the matrix elements of $\nM$, $\cM$, $\pM$ and $\mM$. A particularly interesting set of constraints could be obtained by imposing diffeomorphisms gauge symmetry (i.e., that the system derives from a covariant theory), Lorentz invariance or some additional internal symmetries for the sector that gives rise to the helicity-2 partner $t_{ij}$. We will not delve into these interesting theoretical analyses, since our main goal is to give a comprehensive account of the phenomenological consequences, and we will simply take \eqref{eq:generalequation} as our starting point. In Sec. \ref{sec:theorLandscape} we will give some examples for which we can explicitly give the matrix elements and that will help clarifying some of the above points.

In general, any of the matrices appearing in \eqref{eq:generalequation} can be non-diagonal and thus trigger a mixing of the two modes. Moreover, the entries of these matrices are also generically time-dependent. As a consequence, there will be no exact, analytic solutions. For that reason, in order to understand the physics of the problem, we present different schemes to obtain approximate solutions: one based on a Wentzel-Kramers-Brillouin (WKB) expansion in Section \ref{sec:WKB}, and another on a large wavenumber $k$ expansion in Section \ref{sec:largek}. As a warm-up for the unfamiliar reader, we present a summary of the equivalent one-dimensional problem in appendix \ref{app:1D}.

Concretizing down to a model would mean fixing the coefficients of these matrices in a specific way in terms of the background evolution. In some cases, some of the entries will be even associated with each other. The aim will be to break the degeneracies between the parameters using the full-fledged observational information. In most cases, this will require the combination of various observational channels. We will discuss the rich phenomenology of GW oscillations in section \ref{sec:pheno}. 

\subsubsection{WKB expansion}
\label{sec:WKB}

In this work, we are interested in modified scenarios at cosmological scales. Therefore, the typical time variation of the parameters will be of order of the inverse Hubble constant $H_0$. On the other hand, the frequency of a GW from a compact binary merger scales as
\be
\fGW\sim 1\text{kHz}\lp\frac{10M_\odot}{M}\rp\sim 10^{21}H_0\lp\frac{10M_\odot}{M}\rp\,.
\ee
Thus, for any signal of this kind there will be a great difference between the time scales of the problem, having $\fGW\gg H_0$. This motivates solving equation (\ref{eq:generalequation}) using an adiabatic or WKB approximation. For that, we introduce a dimensionless, small parameter $\eps$ suppressing the time derivatives
\be
\lb\eps^2\frac{\d^2}{\d\eta^2} + \eps\,\nM\frac{\d}{\d\eta}+\cM k^2 +\pM k +\mM\rb\vPhi=0
\ee
and enhancing the phase of the wave
\be
\vPhi=\eM e^{\frac{i}{\epsilon}\int\hat{\theta}d\eta}\lp\vPhi_0+\epsilon\vPhi_1+\cdots\rp\,,
\ee
where $\vPhi$ stands for $\vPhi=\bpm h \\ t \epm$.
Since we are in a multidimensional problem, we are expanding the solution around the basis determined by the matrix $\hat{E}$ solving the constant-parameter case. The amplitude is expanded in different orders of $\eps$ and $\hat{\theta}$ is the diagonal phase matrix. 
Defining $\hat{G}\equiv e^{\frac{i}{\epsilon}\int\hat{\theta}d\eta}$, then we have the following equations at increasing order in $\eps$
\begin{align}
&\eps^0:  & &\lb \lp\cM k^2+\pM k+\mM\rp\eM-\eM\tM^2+i\nM\eM\tM\rb \gM\vPhi_0=0\,, \label{eq:WKBeps0}\\
&\eps^1:  & &\lp2\eM\tM-i\nM\eM\rp\gM\vPhi'_0+\lp\eM\tM'+2\eM'\tM-i\nM\eM'\rp\gM\vPhi_0=0\,, \label{eq:WKBeps1}\\
&\eps^2: & &\eM\gM\vPhi_0''+2\eM'\gM\vPhi'_0+\eM''\hat{G}\vPhi_0= \label{eq:WKBeps2}\\
& & &~~~~~~~-i\lp2\eM\tM-i\nM\eM\rp\gM\vPhi'_1-i\lp\eM\tM'+2\eM'\tM-i\nM\eM'\rp\gM\vPhi_1\,, \nonumber\\
&\cdots & &\cdots \nonumber\\
&\eps^{n+1}: & &\eM\gM\vPhi_{n-1}''+2\eM'\gM\vPhi'_{n-1}+\eM''\hat{G}\vPhi_{n-1}= \label{eq:WKBepsN}\\
& & &~~~~~~~-i\lp2\eM\tM-i\nM\eM\rp\gM\vPhi'_n-i\lp\eM\tM'+2\eM'\tM-i\nM\eM'\rp\gM\vPhi_n\,.\nonumber 
\end{align} 
To solve the leading order equation, which gives the exact solution when the coefficients are constant, we have to find the roots of the quartic equation
\be \label{eq:determinant}
\det\lb \cM k^2+\pM k+\mM-\iM\theta^2+i\nM\theta\rb=0\,,
\ee
with $\iM$ the identity matrix. The matrix $\eM$ is then
\be
\hat{E}=\bpm 1 & -\frac{\hat{W}_{12}+i\nM_{12}\theta_i}{\hat{W}_{11}-\theta_i^2+i\nM_{11}\theta_i} \\ -\frac{\hat{W}_{21}+i\nM_{21}\theta_j}{\hat{W}_{22}-\theta_j^2+i\nM_{22}\theta_j} & 1\epm\,, 
\ee
where for shortness we have defined $\hat{W}\equiv\cM k^2+\pM k+\mM$. Note, that since we have dropped the indices for the two polarizations, the matrices are all $2\times2$ matrices.
Here, $\theta_i$ and $\theta_j$ correspond to two different solutions of Eq. (\ref{eq:determinant}). 
 
At next to leading order, $\Od\lp\eps^1\rp$, we solve $\vPhi_0$ from the first order differential equation (\ref{eq:WKBeps1}). In general, this is a system of first order ordinary differential equations with time dependent coefficients without analytic solutions.\footnote{Although a \emph{formal} analytical solution could be written in terms of a time-ordered matrix exponential.} 
However, within the WKB expansion the matrix exponential is a good approximate solution.\footnote{To be an exact solution, the matrix in the exponent should commute with itself at any two instants of time. Because the parameters vary slowly in the WKB compared to the GW frequency this is a good approximation.} Accordingly, we can solve $\vPhi_0$ as
\be
\vPhi_0=\tM^{-1/2}e^{-\int\Awkb\,\d\eta}\,\vec{C}_0\,,
\ee 
where the matrix in the exponent corresponds to 
\be \label{eq:matrix_exp}
\Awkb=\gM^{-1}\tM^{1/2}\lp2\eM\tM-i\nM\eM\rp^{-1}\lp2\eM'\tM-i\nM\eM'+\frac{i}{2}\nM\eM\tM'\tM^{-1}\rp\gM	\tM^{-1/2}\,,
\ee
and $\vec{C}_0$ is a vector of constant coefficients to be fixed with the initial conditions. Here one should recall that $\tM$ is a diagonal matrix and thus the term $\tM^{-1/2}$ in front is just the usual WKB scaling $1/\sqrt{\theta_i}$ of the one-dimensional problem (see appendix \ref{app:1D}). If there is time dependence, there can be corrections to this scaling, which corresponds to the matrix exponential. 

At next to next to leading order, the first correction to the amplitude $\vPhi_1$ can be computed from (\ref{eq:WKBeps2}), which is analogous to (\ref{eq:WKBeps1}) but with a non-homogeneous term. In fact, the solution of the $n$-th correction will have the same structure given by the iterative solution
\be
\vPhi_{n}=\tM^{-1/2}e^{-\int\Awkb \d\eta}\lp\vec{C}_n+i\int e^{\int\Awkb \d\eta}\,\Bwkb^{-1}\,\vec{F}^{\text{wkb}}_{n-1}\, \d\eta\rp\,,
\ee
where 
\begin{align}
\Bwkb&=\lp2\eM\tM-i\nM\eM\rp\gM\tM^{-1/2}\,,\\
\vec{F}^{\text{wkb}}_{n-1}&=\lp\eM\gM\vPhi_{n-1}''+2\eM'\gM\vPhi'_{n-1}+\eM''\hat{G}\vPhi_{n-1}\rp\,,
\end{align}
and $\vec{C}_n$ is a constant vector. 
In this way we have solved the problem up to order $\epsilon^{n+1}$.

The above general solution can be simplified in some cases. For instance, when $\nM$ commutes with $\eM$ and $\tM$, the friction matrix $\nM$ may be absorbed by defining
\be
\left.\vPhi\right\vert_{[\eM,\nM]=0}=e^{-\frac{1}{2}\int\nM \d\eta}\eM e^{\frac{i}{\epsilon}\int\hat{\theta}\d\eta}\lp\vPhi_0+\epsilon\vPhi_1+\cdots\rp\,,
\ee
where again we are using the matrix exponential as an approximate solution whenever $\nM$ is non-diagonal. 
Then (\ref{eq:determinant}) becomes a quadratic equation for $\theta^2$,
\be \label{eq:eigenvalues}
\theta^2_{1,2}=\frac{1}{2}\lp\tr{\hat{W}}\pm\sqrt{4\hat{W}_{12}\hat{W}_{21}+\lp\hat{W}_{11}-\hat{W}_{22}\rp^2}\rp\,,
\ee
and $\eM$ is the associated matrix of eigenvectors of $\hat{W}$.

Altogether, we can decompose the general solution in each of the components, obtaining
\begin{align} \label{eq:hwkb1}
h(\eta)&=\left[
 c_1\Phi_h(\eta)+c_2 \eM_{12}(\eta)\Phi_t(\eta)e^{i\int\delta\theta(\eta) \d\eta}\right] e^{i\int\theta_1(\eta) \d\eta}\,, \\
t(\eta)&=\left[
 c_2\Phi_t(\eta)+c_1\eM_{21}(\eta)\Phi_h(\eta)e^{-i\int\delta\theta(\eta) \d\eta}\right] e^{i\int\theta_2 (\eta) \d\eta}\,. \label{eq:twkb1}
\end{align}
Here, we have denoted the difference in the phases as $\delta\theta=\theta_2-\theta_1$, and $\Phi_{h,t}$ incorporate all the corrections from $\sum_n\vPhi_n$ to the amplitude of $h$ and $t$, respectively. One should note that the above expressions (\ref{eq:hwkb1}-\ref{eq:twkb1}) correspond only to the contribution of two distinct (in absolute value) phases $\theta_{1,2}$. Whenever there are four independent roots of (\ref{eq:determinant}), one should add to (\ref{eq:hwkb1}-\ref{eq:twkb1}) the equivalent terms depending on $\theta_{3,4}$. 
Finally, we can fix the constants $c_{1,2}$ using the initial conditions at the time of emission $\eta_e$. Imposing that initially only one of the tensor perturbations is excited with an amplitude $h_0$ dictated by GR, i.e. $h(\eta_e)=h_0$ and $t(\eta_e)=0$, we find
\begin{align}
c_1&=\frac{h_0}{\Phi_h(\eta_e)(1-\eM_{12}(\eta_e)\eM_{21}(\eta_e))}\,, \\
c_2&=-\frac{h_0\eM_{21}(\eta_e)}{\Phi_h(\eta_e)(1-\eM_{12}(\eta_e)\eM_{21}(\eta_e))}\,.
\end{align}

\subsubsection{Large-$k$ expansion}
\label{sec:largek}

In addition to the hierarchy between the time variation of the parameters of the theory and the frequency of the GWs, it could be the case that the parameters themselves are small compared to the wavenumber $k$. Accordingly, one could make a large-$k$ or shortwave expansion (also known as eikonal approximation \cite{misner2017gravitation}), which is a more restrictive approximation compared to the WKB. Using the same ansatz for $\vPhi$, the system of equations is however different, i.e.
\be
\lb\frac{\d^2}{\d\eta^2} + \nM\frac{\d}{\d\eta}+\eps^{-2}\cM k^2 +\eps^{-1}\pM k +\mM\rb\vPhi=0.
\ee
Splitting in the different orders, we find
\begin{align}
&\eps^{-2}: & &\lp \cM \eM k^2-\eM\tM^2\rp \gM\vPhi_0=0\,, \\
&\eps^{-1}: & &2\eM\tM\gM\vPhi'_0+\lp\eM\tM'+2\eM'\tM+\nM\eM\tM-i\pM\eM k\rp\gM\vPhi_0=0\,, \\
&\eps^{0}: & &\eM\gM\vPhi_0''+\lp2\eM'+\nM\eM\rp\gM\vPhi'_0+\lp\eM''+\nM\eM'+\mM\eM\rp\hat{G}\vPhi_0= \\
& & &~~~~~~~~~-2i\eM\tM\gM\vPhi'_1-i\lp\eM\tM'+2\eM'\tM+\nM\eM\tM-i\pM\eM k\rp\gM\vPhi_1\,, \nonumber \\
&\cdots & &\cdots \nonumber \\
&\eps^{n-1}: & &\eM\gM\vPhi_{n-1}''+\lp2\eM'+\nM\eM\rp\gM\vPhi'_{n-1}+\lp\eM''+\nM\eM'+\mM\eM\rp\hat{G}\vPhi_{n-1}= \\
& & &~~~~~~~~~-2i\eM\tM\gM\vPhi'_n-i\lp\eM\tM'+2\eM'\tM+\nM\eM\tM-i\pM\eM k\rp\gM\vPhi_n\,. \nonumber 
\end{align} 
To solve the leading order equation, we take $\theta^2$ as the eigenvalues of $\cM$ (cf. (\ref{eq:eigenvalues}))
and $\eM$ the matrix of eigenvectors
\be
\hat{E}=\bpm 1 & -\frac{\cM_{12}}{\cM_{11}-\theta_2^2} \\ -\frac{\cM_{21}}{\cM_{22}-\theta_1^2} & 1\epm\,. 
\ee
In the case in which the velocity matrix $\cM$ is diagonal, which is the most common case, then the matrix of eigenvectors becomes the identity matrix $\eM=\iM$.
 
At next order, we obtain the amplitude again using an approximate matrix exponential solution
\be \label{eq:largek_amplitude}
\vPhi_0=\tM^{-1/2}e^{-\int\Ak\,\d\eta}\,\vec{C}_0\,,
\ee 
but now with a matrix in the exponent 
\be
\Ak=\frac{1}{2}\gM^{-1}\tM^{-1/2}\eM^{-1}\lp2\eM'+\nM\eM-i\pM\eM\tM^{-1}k\rp\tM^{1/2}\gM
\ee
different to (\ref{eq:matrix_exp}). 
The higher-order corrections to the amplitude can be computed as before,
\be
\vPhi_{n}=\tM^{-1/2}e^{-\int\Ak \d\eta}\lp\vec{C}_n+i\int e^{\int\Ak \d\eta}\,\Bk^{-1}\,\vec{F}^{\text{large-k}}_{n-1}\, \d\eta\rp\,,
\ee
with
\begin{align}
\Bk&=2\eM\gM\tM^{1/2}\,,\\
\vec{F}^{\text{large-k}}_{n-1}&=\eM\gM\vPhi_{n-1}''+\lp2\eM'+\nM\eM\rp\gM\vPhi'_{n-1}+\lp\eM''+\nM\eM'+\mM\eM\rp\hat{G}\vPhi_{n-1}\,,
\end{align}
In this way we have solved the problem up to order $\epsilon^{n-1}$.

If we focus on the leading order amplitude, we could rewrite the previous formula (\ref{eq:largek_amplitude}) as
\be \label{eq:phi0_largek}
\vPhi_0=\tM^{-1/2}\,e^{-\frac{1}{2}\tr{\bA}}\bpm\cos\omega+\frac{\Delta\bA}{2\omega}\sin\omega & -\frac{\bA_{12}}{\omega}\sin\omega \\ -\frac{\bA_{21}}{\omega}\sin\omega & \cos\omega-\frac{\Delta\bA}{2\omega}\sin\omega \epm\bpm c_1 \\ c_2 \epm
\ee
by denoting the integral of the matrix in the exponent $\bA_{ij}=\int_{\eta_e}^\eta\hat{A}_{ij}d\eta$, defining the difference of the diagonal entries $\Delta\bA=\bA_{22}-\bA_{11}$ and introducing a frequency
\begin{align}
\omega^2&=-\bA_{12}\bA_{21}-\Delta\bA^2/4\,.
\end{align}
In the case in which the velocity matrix is diagonal, the mixing of the modes is controlled by $\omega$. This is explicit when we compute each tensor perturbation
\begin{align} \label{eq:hlargek}
h(\eta)&=\frac{e^{-\frac{1}{2}\tr{\bA}}}{\sqrt{\theta_1}}\left[
 c_1\lp\cos\omega+\frac{\Delta\bA}{2\omega}\sin\omega\rp-c_2 \frac{\bA_{12}}{\omega}\sin\omega\right] e^{i\int\theta_1(\eta) \d\eta}\,, \\
t(\eta)&=\frac{e^{-\frac{1}{2}\tr{\bA}}}{\sqrt{\theta_2}}\left[
 c_2\lp\cos\omega-\frac{\Delta\bA}{2\omega}\sin\omega \rp-c_1 \frac{\bA_{21}}{\omega}\sin\omega\right] e^{i\int\theta_2 (\eta) \d\eta}\,. \label{eq:tlargek}
\end{align}
when we impose the initial conditions $h(\eta_e)=h_0$ and $t(\eta_e)=0$, this expression simplifies further to (note that $\bA_{ij}(\eta_e)=0$)
\begin{align} 
h(\eta)&=h_0\,e^{-\frac{1}{2}\tr{\bA}}
 \lp\cos\omega+\frac{\Delta\bA}{2\omega}\sin\omega\rp \frac{\sqrt{\theta_1(\eta_e)}}{\sqrt{\theta_1(\eta)}}\,e^{i\int\theta_1(\eta) \d\eta}\,, \\
t(\eta)&=-h_0\,e^{-\frac{1}{2}\tr{\bA}}\frac{\bA_{21}}{\omega}\sin\omega \frac{\sqrt{\theta_1(\eta_e)}}{\sqrt{\theta_2(\eta)}}\,e^{i\int\theta_2 (\eta) \d\eta}\,. 
\end{align}
From this expression we can also see that there will be an overall damping determined by $\tr{\bA}$.
\subsection{Particular cases}

In order to gain insights from the general, approximate, analytical solutions that we have found, let us consider some particular cases. It is important to note that in general there will be degeneracies between different parameters. For this reason, we also consider representative examples separately.

\subsubsection{Mixing through the mass matrix}
\label{sec:mass_mixing}

In analogy with neutrinos, if the mass matrix of the tensor perturbations is non-diagonal, the propagation and mass eigenstates are different, implying that they will mix while traveling. In the following we consider $h$ and $t$ propagating at different speeds and interacting through the mass matrix $\mM$,\footnote{One should note that if the fields are normalized canonically and we assume that these equations of motion come from a Lagrangian then $m_{ht}=m_{th}$ and all the expressions in the following discussion simplify.}
\be
\lb \frac{\d^2}{\d\eta^2} +\bpm c_h^2 & 0 \\ 0 &  c_t^2\epm k^2 + \bpm m^2_h & m^2_{ht} \\ m^2_{th} &  m^2_t\epm\rb \bpm h \\ t \epm =0\,. 
\ee
The associated eigenvalues are
\be
\theta_{1,2}^{2}=\lp c_h^2+\frac{1}{2}\dc^2\rp k^2+\frac{1}{2}M^2\mp\frac{1}{2}\sqrt{\dc^4 k^4+2\dc^2\dm^2k^2+M^4(1+\Delta)^2}\,, \label{eq:t12_mass}
\ee
where we have defined the difference in the speeds $\dc^2\equiv c_t^2-c_h^2$, the sum of the square masses $M^2\equiv m_h^2+m_t^2$, their difference $\dm^2=m_t^2-m_h^2$ and the parameter $\Delta\equiv\sqrt{1-4\det[\mM]/M^4}-1$, which vanishes when $\mM$ is degenerate. 
The eigenvectors are
\be \label{eq:eMeigenvectors}
\hat{E}=\bpm 1 & -\frac{m_{ht}^2}{c_h^2k^2+m_h^2-\theta_2^2} \\ -\frac{m_{th}^2}{c_t^2k^2+m_t^2-\theta_1^2} & 1\epm\,. 
\ee

\paragraph{High-$k$ limit:}
It is interesting to study first the high-$k$ limit. The phases are
\begin{align}
&\theta_1^2=c_h^2k^2+m_h^2+\Od(k^{-2})\,, \\
&\theta_2^2=(c_h^2+\dc^2) k^2+m_t^2+\Od(k^{-2})\,,
\end{align} 
and the matrix of eigenvectors scales as
\be \label{eq:eM_wkb_mass}
\hat{E}=\bpm 1 & \frac{m_{ht}^2}{\dc^2k^2+\dm^2+\Od(k^{-2})} \\ -\frac{m_{th}^2}{\dc^2k^2+\dm^2+\Od(k^{-2})} & 1\epm=\iM+\Od\lp \frac{m_{ij}^2}{\dc^2k^2}\rp\,. 
\ee
Therefore, if $h$ and $t$ propagate at different speeds, $\dc\neq0$, and the wavenumber $k$ is much larger than the matrix elements of $\mM$, the mixing will be suppressed, with $\eM$ approaching the identity matrix. 

\paragraph{Small-$\dc$ limit:} 
Since we are interested in studying the regime in which $k$ is large, let us consider the limit in which the difference in the speeds is small, $\dc^2\ll1$, and the mixing is not suppressed. In this limit, the phases simplify to 
\begin{align}
&\theta_1^2=\lb c_h^2+\frac{1}{2}\dc^2\lp1-\frac{\dm^2}{M^2(1+\Delta)}\rp\rb k^2-\frac{1}{2}M^2\Delta+\Od(\dc^4)\,, \label{eq:dclimit}\\
&\theta_2^2=\lb c_h^2+\frac{1}{2}\dc^2\lp1+\frac{\dm^2}{M^2(1+\Delta)}\rp\rb k^2+M^2\lp1+\frac{1}{2}\Delta\rp+\Od(\dc^4)\,,
\end{align}
and the eigenvectors to
\be
\hat{E}=\bpm 1 & \frac{m_{ht}^2}{m_t^2+M^2\Delta/2}\lp1-\frac{\dc^2k^2}{M^2(1+\Delta)}+\Od(\dc^4)\rp \\ -\frac{m_{th}^2}{m_t^2+M^2\Delta/2}\lp1-\frac{\dc^2k^2}{M^2(1+\Delta)}+\Od(\dc^4)\rp & 1\epm\,. 
\ee

\begin{figure}[t]
\centering 
\includegraphics[width=.49\textwidth]{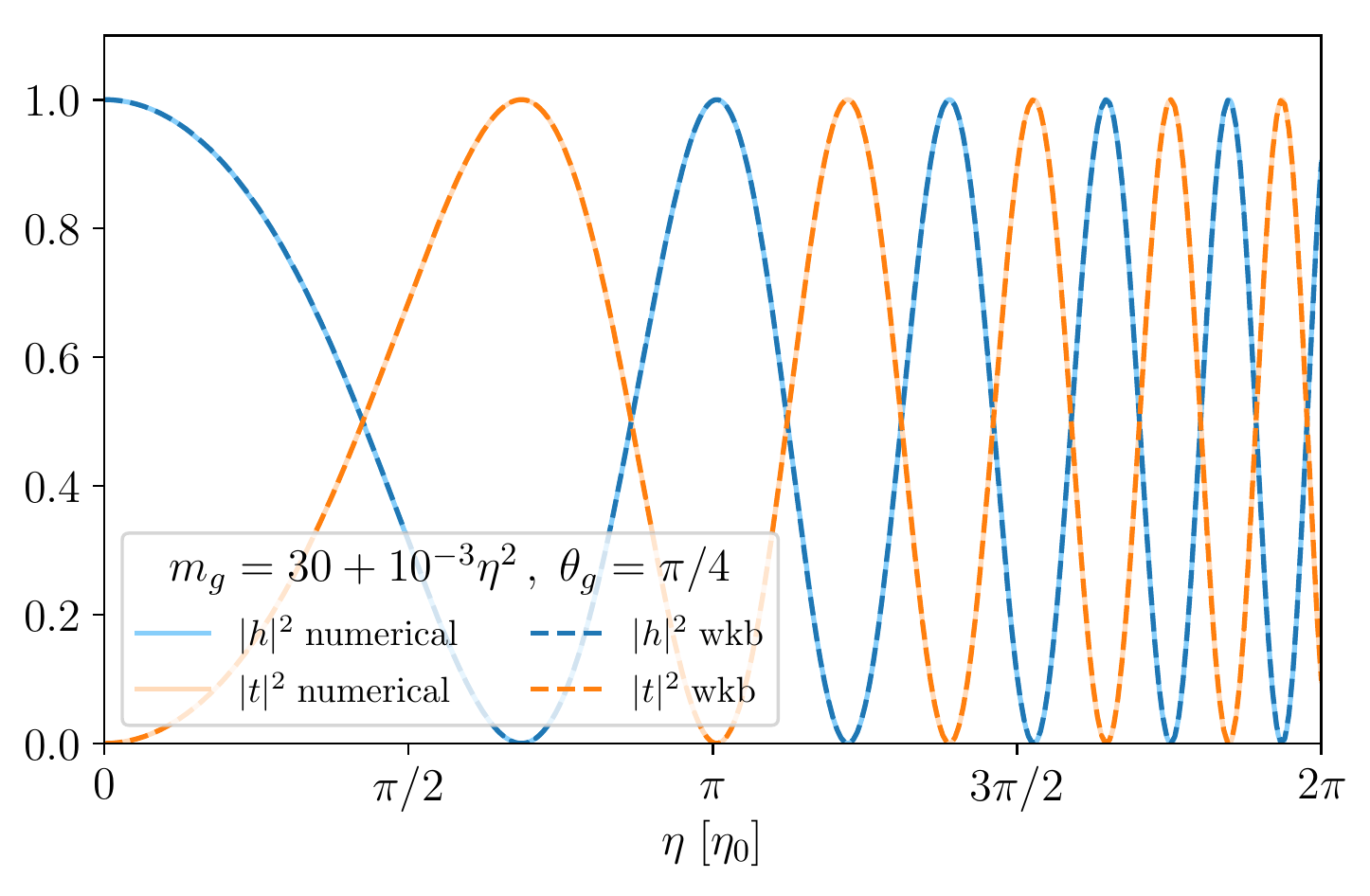}
\includegraphics[width=.49\textwidth]{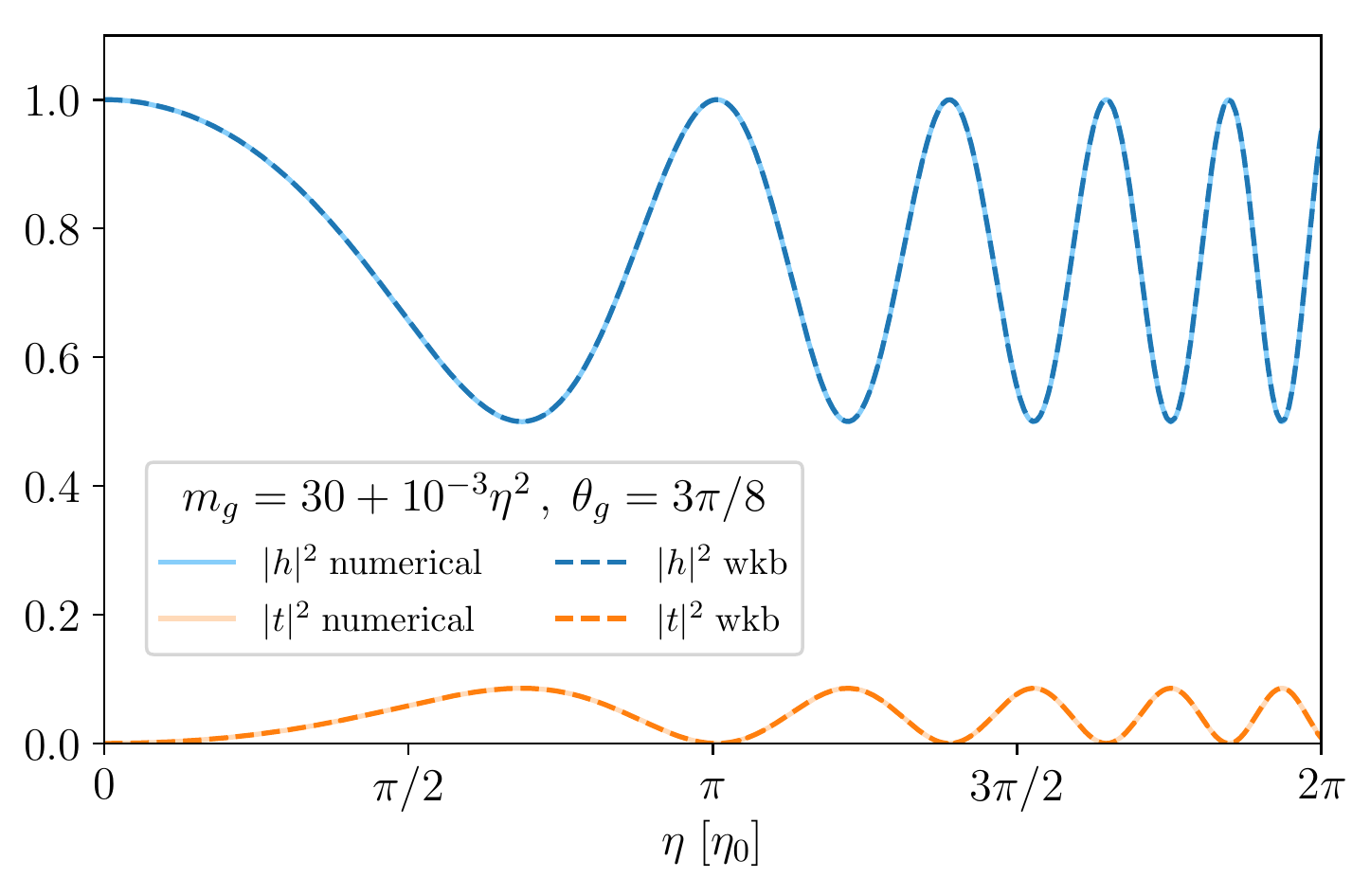}
\includegraphics[width=.49\textwidth]{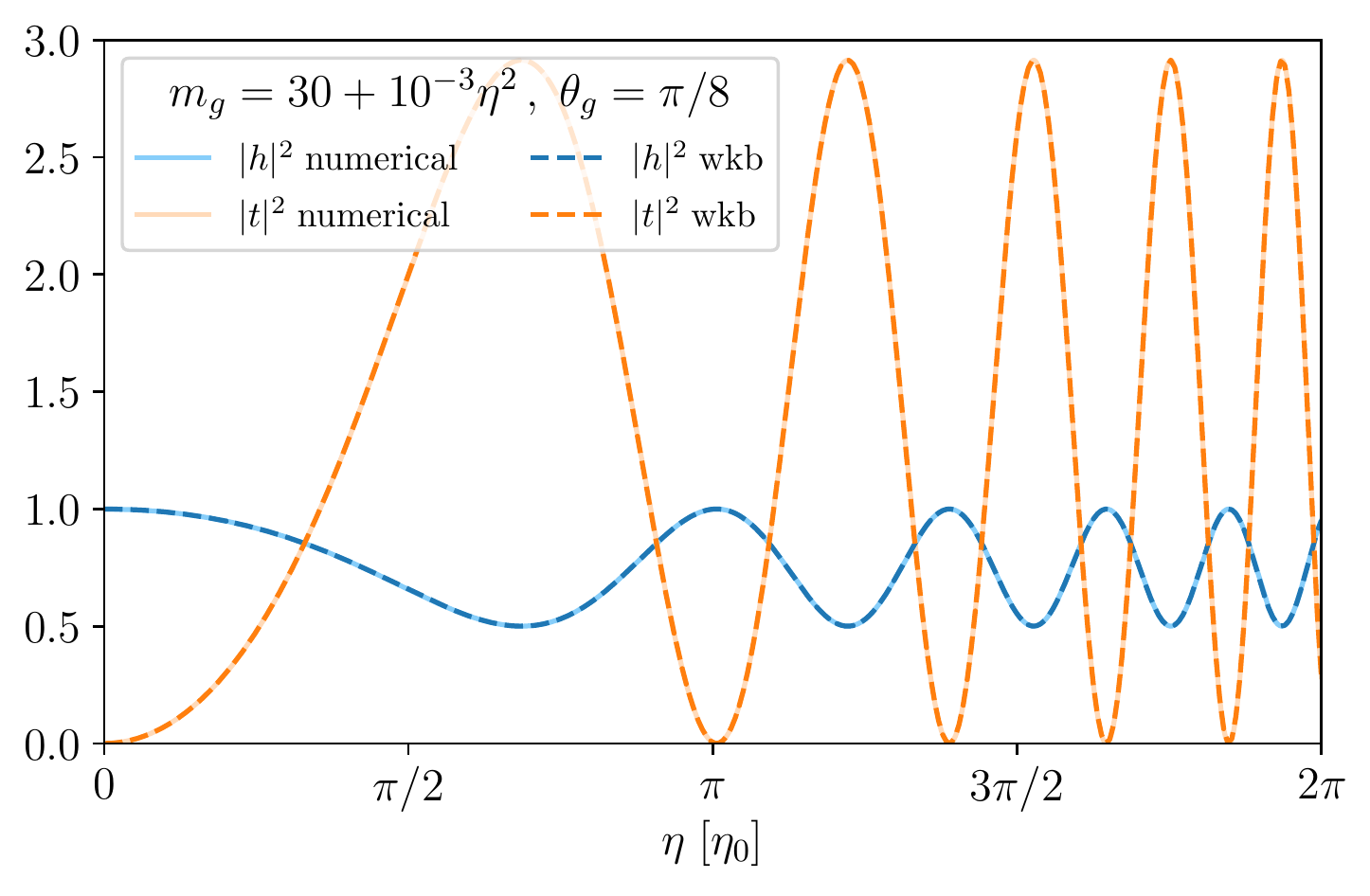}
\includegraphics[width=.49\textwidth]{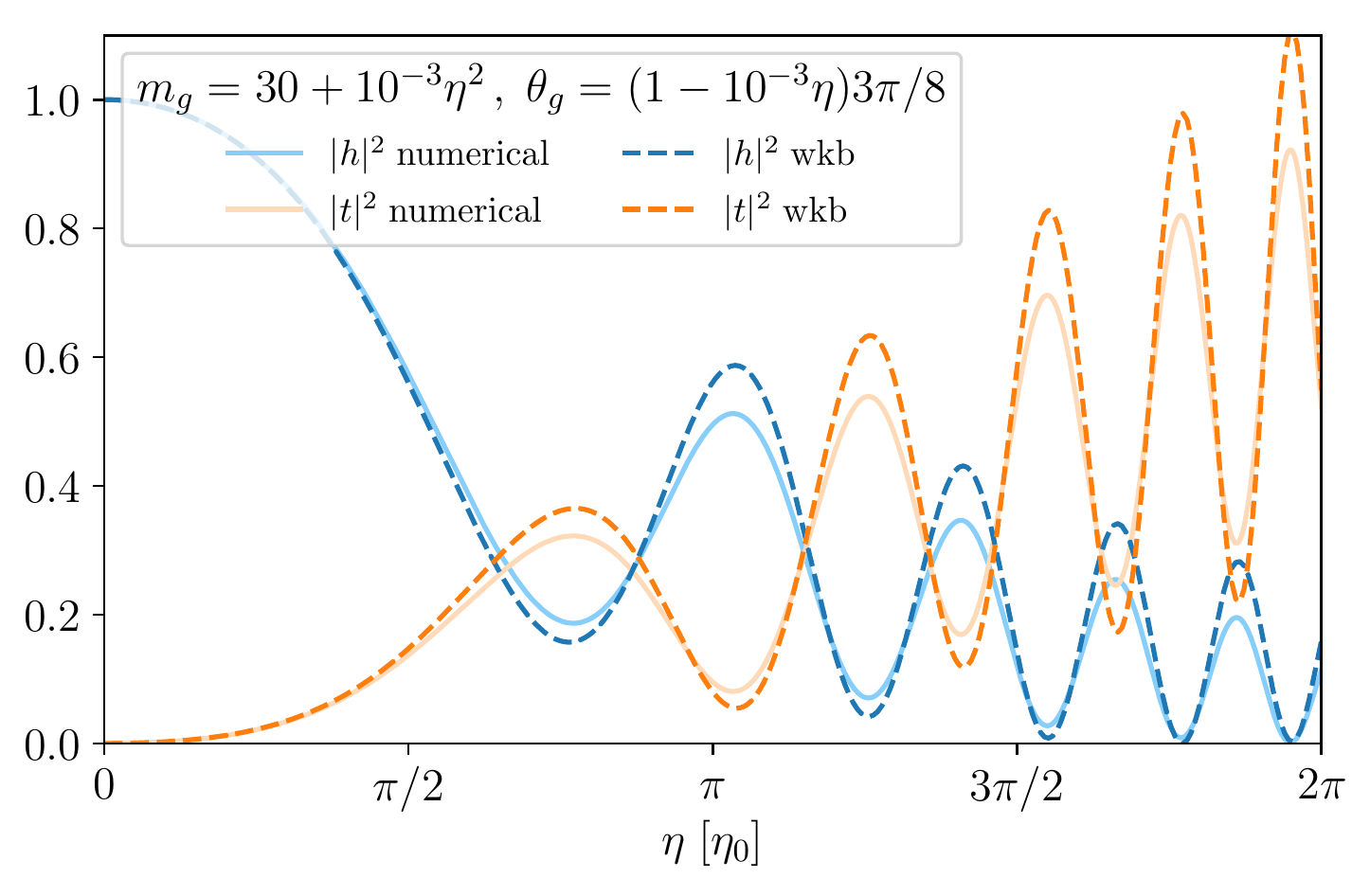}
 \caption{Oscillation of the GW amplitude $\vert h\vert$ and the tensor perturbation $\vert t\vert$ due to a mass mixing. We choose the same time dependent effective mass $m_g$ in all the panels. 
The mixing angle $\theta_g$ corresponds to $\pi/4$ (upper left), $3\pi/8$ (upper right), $\pi/8$ (lower left) and a function of time (lower right). 
We compare the numerical solution (solid) with the WKB expansion (dashed) fixing $\Delta c=0$, $k=10^{3}$, $h_0=1$ and normalizing time w.r.t. the initial period $\eta_0$.}
 \label{fig:mass_mixing}
\end{figure}

The frequency of oscillation due to the mixing is governed by the difference in the eigenfrequencies
\be
\delta\theta\equiv\theta_2-\theta_1=\frac{M^2(1+\Delta)}{2c_hk}+\Od(k^{-3},\dc^2)\,,
\ee
which suggests the introduction of the effective mass
\be \label{eq:effective_mass}
m_g\equiv M\sqrt{1+\Delta}\,.
\ee
In the same manner, whenever the components of the mass matrix scale similarly with time, e.g. $m_{ij}^2\propto a(\eta)^2$, the matrix of eigenvectors becomes approximately constant, i.e. $\eM={\rm const.}+\Od(\Delta c^2k^2/M^2)$. The WKB solutions are then
\begin{align} 
h(\eta)&=\left[
 \frac{c_1}{\sqrt{\theta_1(\eta)}}+\frac{c_2}{\sqrt{\theta_2(\eta)}} \eM_{12}(\eta)e^{i\int\delta\theta(\eta) \d\eta}\right] e^{i\int\theta_1(\eta) \d\eta}\,, \\
t(\eta)&=\left[
 \frac{c_2}{\sqrt{\theta_2(\eta)}}+\frac{c_1}{\sqrt{\theta_1(\eta)}}\eM_{21}(\eta)e^{-i\int\delta\theta(\eta) \d\eta}\right] e^{i\int\theta_2 (\eta) \d\eta}\,. \label{eq:twkb}
\end{align}
This suggests the definition of a mixing angle $\theta_g$,
\be \label{eq:mixing_angle}
\tan^2\theta_g=-\eM_{12}\eM_{21}=\frac{m_{ht}^2m_{th}^2}{(m_t^2+M^2\Delta/2)^2}+\Od\lp\frac{\dc^2k^2}{M^2}\rp\,,
\ee
so that, after imposing the initial conditions $h(\eta_e)=h_0$ and $t(\eta_e)=0$, the amplitude of $h$ becomes
\be
\vert h(\eta)\vert^2=h_0^2\cos^4\theta_g\lp\frac{\theta_1(\eta_e)}{\theta_1(\eta)}+\frac{\theta_2(\eta_e)}{\theta_2(\eta)}\tan^4\theta_g+2\frac{\sqrt{\theta_1(\eta_e)\theta_2(\eta_e)}}{\sqrt{\theta_1(\eta)\theta_2(\eta)}}\tan^2\theta_g\,\cos\lb\int_{\eta_e}^\eta\delta\theta(\eta') \d\eta'\rb\rp\,,
\ee
and the one of $t$ reads
\be
\vert t(\eta)\vert^2=h_0^2\cos^4\theta_g\frac{m_{th}^4}{(m_t^2+M^2\Delta/2)^2}\lp\frac{\theta_1(\eta_e)}{\theta_1(\eta)}+\frac{\theta_2(\eta_e)}{\theta_2(\eta)}-2\frac{\sqrt{\theta_1(\eta_e)\theta_2(\eta_e)}}{\sqrt{\theta_1(\eta)\theta_2(\eta)}}\,\cos\lb\int_{\eta_e}^\eta\delta\theta(\eta') \d\eta'\rb\rp\,.
\ee
Therefore, the amplitude of the GW signal $\vert h\vert^2$ detected will oscillate with a frequency given by $\delta\theta=m_g^2/(2c_hk)$ and an amplitude controlled by the mixing angle $\theta_g$. This type of mixing precisely occurs in bigravity \cite{Belgacem:2019pkk}.

In Fig. \ref{fig:mass_mixing} we have plotted the evolution of the amplitudes of $h$ and $t$ when there is a mass mixing, in the limit in which the propagation speeds are the same $\dc=0$. We normalize time with respect to the initial period of oscillation $\eta_0\equiv 1/\delta\theta(\eta_e)$. Although the effective mass $m_g$ varies with time, as can be seen from the change in the frequency of oscillation, the WKB solution (dashed lines) is a very good approximation for the exact numerical result (solid lines). We notice that the mixing angle determines the amplitude of the second tensor. When $\theta_g=\pi/4$, as in the upper left panel, there is a complete conversion of $h$ into $t$. As we will see later in section \ref{sec:pheno}, this configuration will maximize the detectability of the GW oscillations. When $\theta_g>\pi/4$, as in the upper right panel, there is not a complete conversion and $\vert t\vert$ is smaller than $h_0$. In the opposite case when $\theta_g<\pi/4$, as in the lower left panel, $\vert t\vert$ can be larger than $h_0$. Finally, if $\theta_g$ varies in time, as in the lower right panel, the amplitude of $\vert t \vert$ will change accordingly in time. In this case, due to the rapid time variation both in $m_g(\eta)$ and $\theta_g(\eta)$ and the choice of $k=10^3$, the leading order WKB solution does not fully capture the dynamical behavior. For a larger wavenumber, the agreement improves. In this respect, one should remember that for astrophysical sources of GWs and cosmologically varying parameters, the hierarchy in $k$ is many orders of magnitude larger than the one presented in these examples.

\begin{figure}[t]
\centering 
 \includegraphics[width=.49\textwidth]{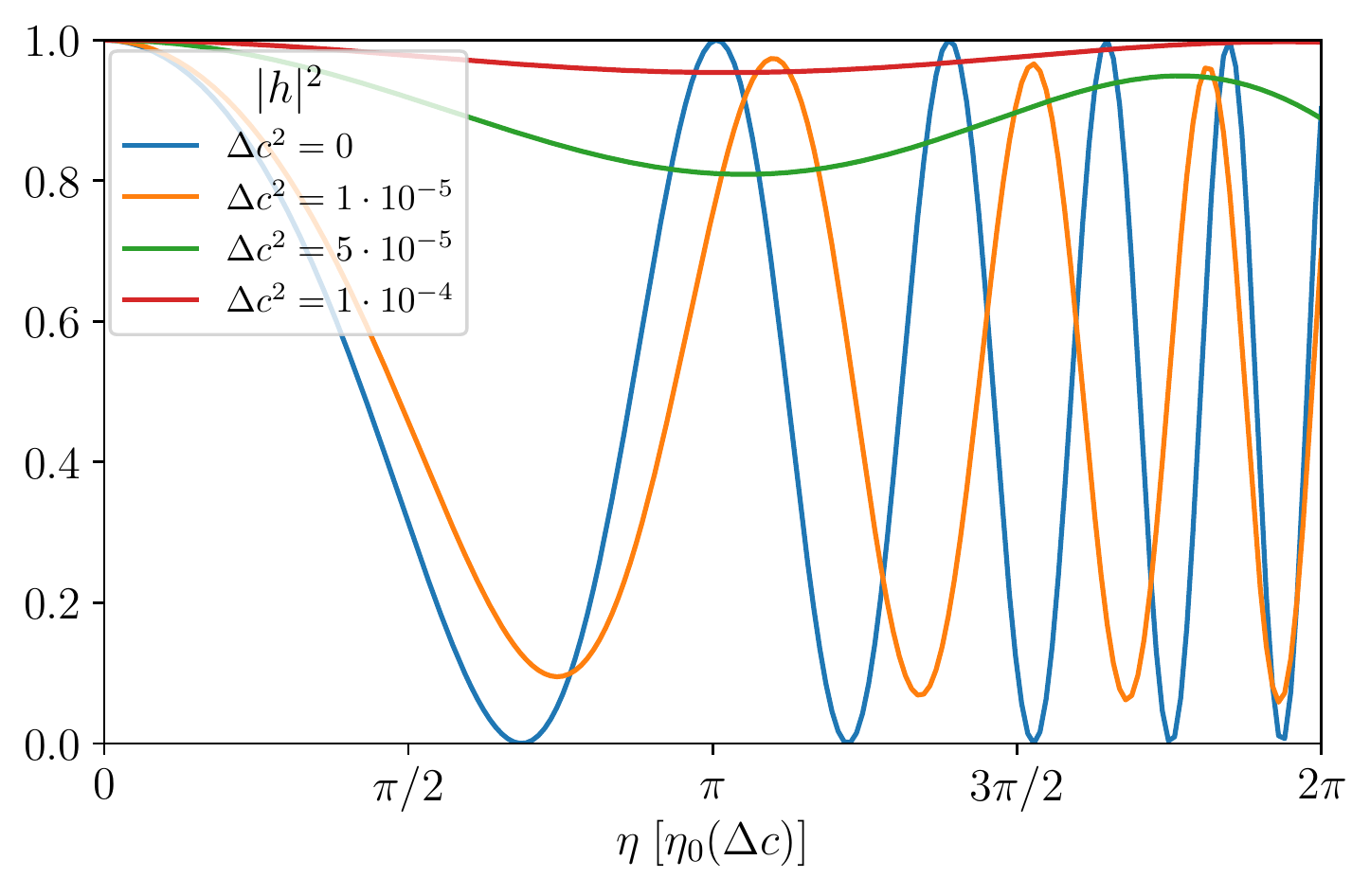}
  \includegraphics[width=.49\textwidth]{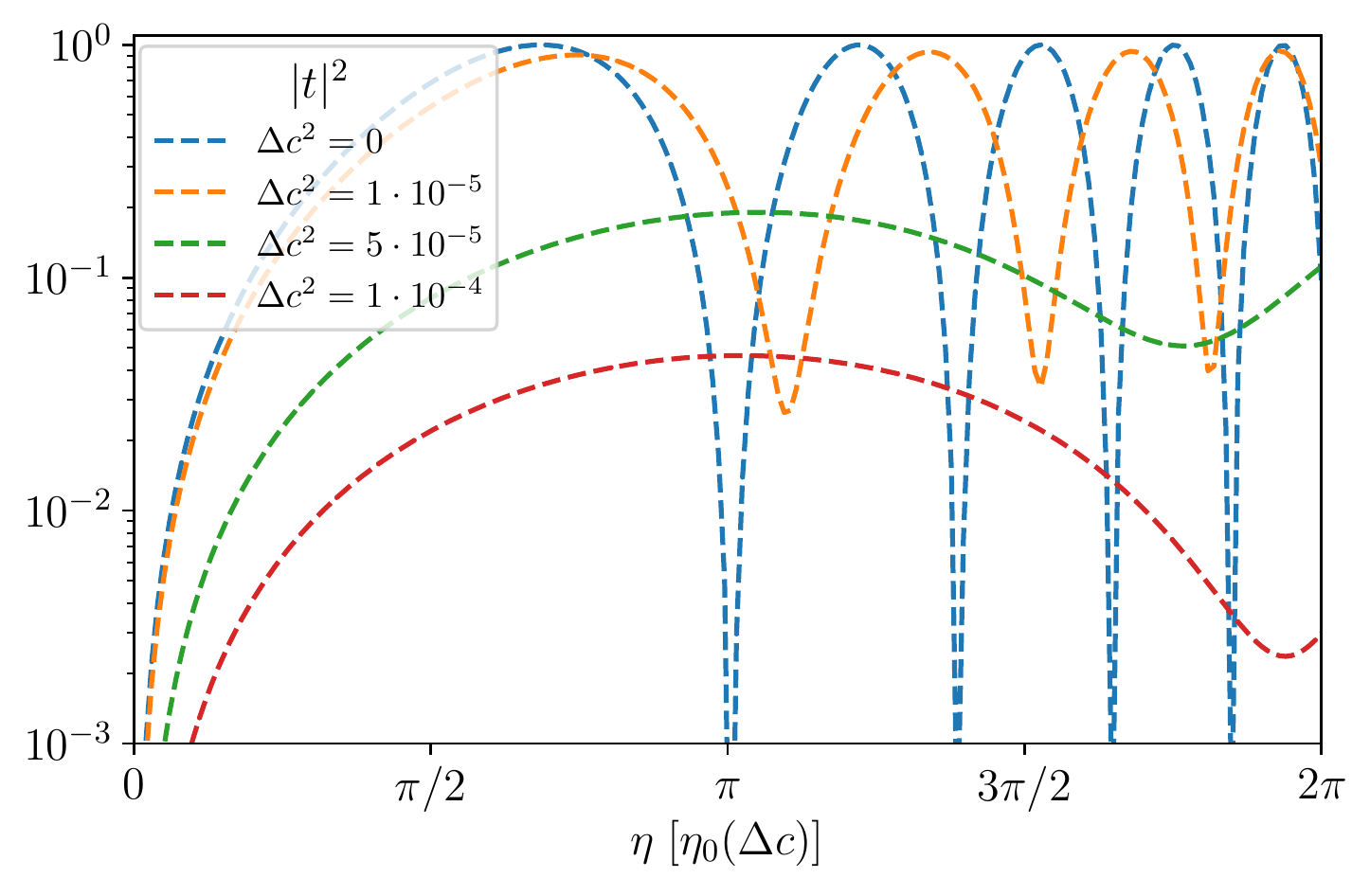}
 \caption{Mass mixing with different propagation speeds. We plot the amplitudes $\vert h\vert^2$ (left) and $\vert t\vert^2$ (right) for different values of $\dc^2=c_t^2-c_h^2$. We have chosen $\theta_g=\pi/4$, $k=10^{3}$, $h_0=1$ and the same time dependent effective mass $m_g$ of Fig. \ref{fig:mass_mixing}.
For each $\Delta c$, we have normalized the time w.r.t. its initial period $\eta_0(\Delta c)$.}
 \label{fig:mass_mixing_suppression}
\end{figure}

In the case in which $\dc\neq0$, there is suppression of the amplitude of $t$ w.r.t. $h$ determined by $m_{ht}^2/(\dc^2k^2)$, recall \eqref{eq:eM_wkb_mass}. This suppression can be observed in Fig. \ref{fig:mass_mixing_suppression} where we plot $\vert h\vert$ (left) and $\vert t \vert$ (right) for different values of $\dc$. For comparison, we use the same parameters as in Fig. \ref{fig:mass_mixing}. As the difference in the speeds increases, the amplitude of $h$ approaches the initial value $h_0$ and the second tensor $t$ reduces. Since we have chosen $k=10^3$ and $m_{ht}\sim10$, one needs $\dc<10^{-4}$ not to get a negligible amplitude of $t$. In practice, for large hierarchies between $k$ and $m_{ij}$, one needs to have $\dc\sim0$ in order to have observable GW oscillations.

Finally let us remark that not only the amplitude of the GW will differ w.r.t. GR, also the phase will change. From (\ref{eq:dclimit}) we learn something important: even if we set the speed of $h$ equal to the speed of light, $c_h=c$, when there is a mixing and the second tensor propagates at a different speed, the speed of GWs can differ from $c$. If we define $\cgw$ from the leading $k^2$ term in (\ref{eq:dclimit}), we can parameterize the anomalous speed through 
\be\label{eq:agw_mass}
\agw=\frac{\cgw^2}{c^2}-1=\frac{1}{2}\frac{\dc^2}{c^2}\lb1-\frac{\dm^2}{M^2(1+\Delta)}\rb\,.
\ee 
Note that when there is no mixing, $m_{th}$ or $m_{ht}$ vanish, then $M^2(1+\Delta)=\dm^2$ and $\agw=0$. This implies that $\agw$ will be degenerate in $\dc$ and $m_{ht}$, meaning that $\agw\simeq0$ whenever $\dc\ll1$ or $m_{ht}\ll1$.

It may be convenient to clarify that by anomalous propagation speed we refer to the high frequency regime, although the actual propagation speed of GWs can  differ from $c$ even for $\agw=0$ due to the presence of mass terms. It is interesting however to notice that, precisely because of the induced mixing of the off-diagonal terms of the mass matrix, these will be more tightly constrained than the diagonal ones.

\subsubsection{Mixing through the friction matrix}
\label{sec:friction_mixing}

If the friction is non-diagonal, there will also be GW oscillations. Our starting ansatz is
\be
\lb \frac{\d^2}{\d\eta^2} + \bpm 0 & -2\alpha \\ 2\alpha & 4\dnu  \epm\frac{\d}{\d\eta} +\bpm c_h^2 & 0 \\ 0 &  c_t^2\epm k^2\rb \bpm h \\ t \epm =0\,, 
\ee
where $\alpha$ is the parameter controlling the mixing and we have defined $4\dnu=\nu_t-\nu_h$. Note that we make this choice because it is always possible to absorb the part of the friction matrix proportional to the identity via a field redefinition $\vPhi=e^{-\frac{1}{2}\int\nu_1 \d\eta}\tilde{\Phi}$. One might be tempted to proceed similarly and absorb the whole friction matrix with a matrix exponential. However, this is not consistent with the WKB expansion unless this matrix commutes with the matrix of eigenvectors and eigenfrequencies, i.e. $[\eM,\nM]=[\tM,\nM]=0$.

\paragraph{WKB approximation:} 

Therefore, in general, we will have to solve the quartic equation (\ref{eq:determinant}) for the eigenfrequencies. Although analytically solvable, the solutions themselves are not very illuminating. In the following, for simplicity, we restrict to $\dc=0$.\footnote{In the situation in which $\dc\neq0$, the mixing will be suppressed in analogy to the mass mixing case (\ref{eq:eM_wkb_mass}), although with less strength: now $\sim 1/k$ instead of $\sim1/k^2$.} 
In that case, the eigenfrequencies become
\begin{align}
\theta_{1,2}&=\sqrt{c_h^2k^2+(\onu\pm i\dnu)^2}\pm\onu+i\dnu\,, \label{eq:t1_friction}\\
\theta_{3,4}&=-\sqrt{c_h^2k^2+(\onu\pm i\dnu)^2}\pm\onu+i\dnu\,, \label{eq:t2_friction}
\end{align}
where we have defined the oscillation frequency $\onu^2\equiv \alpha^2-\dnu^2$ associated to the friction mixing. 

\begin{figure}[t]
\centering 
\includegraphics[width=.49\textwidth]{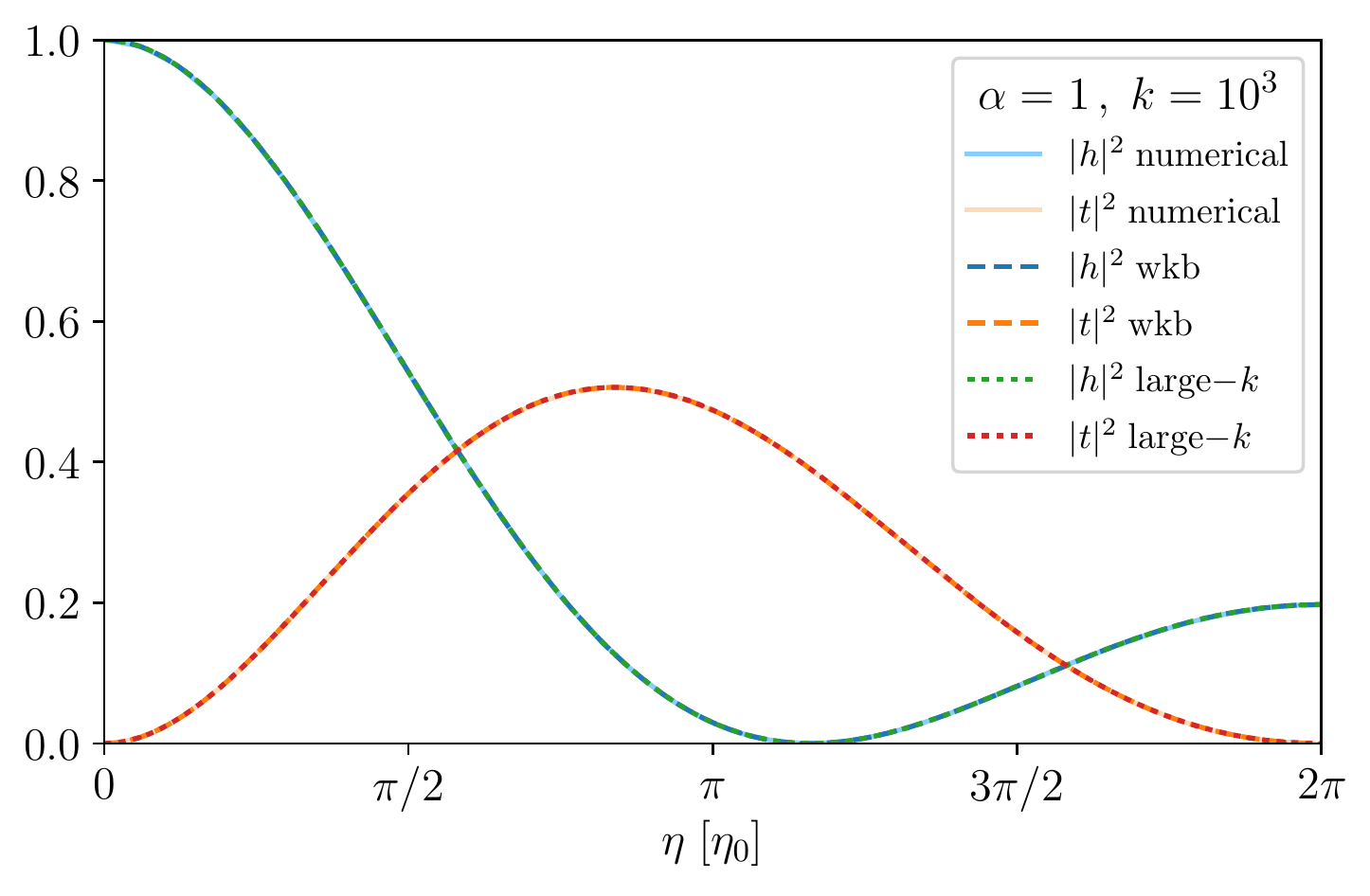}
\includegraphics[width=.49\textwidth]{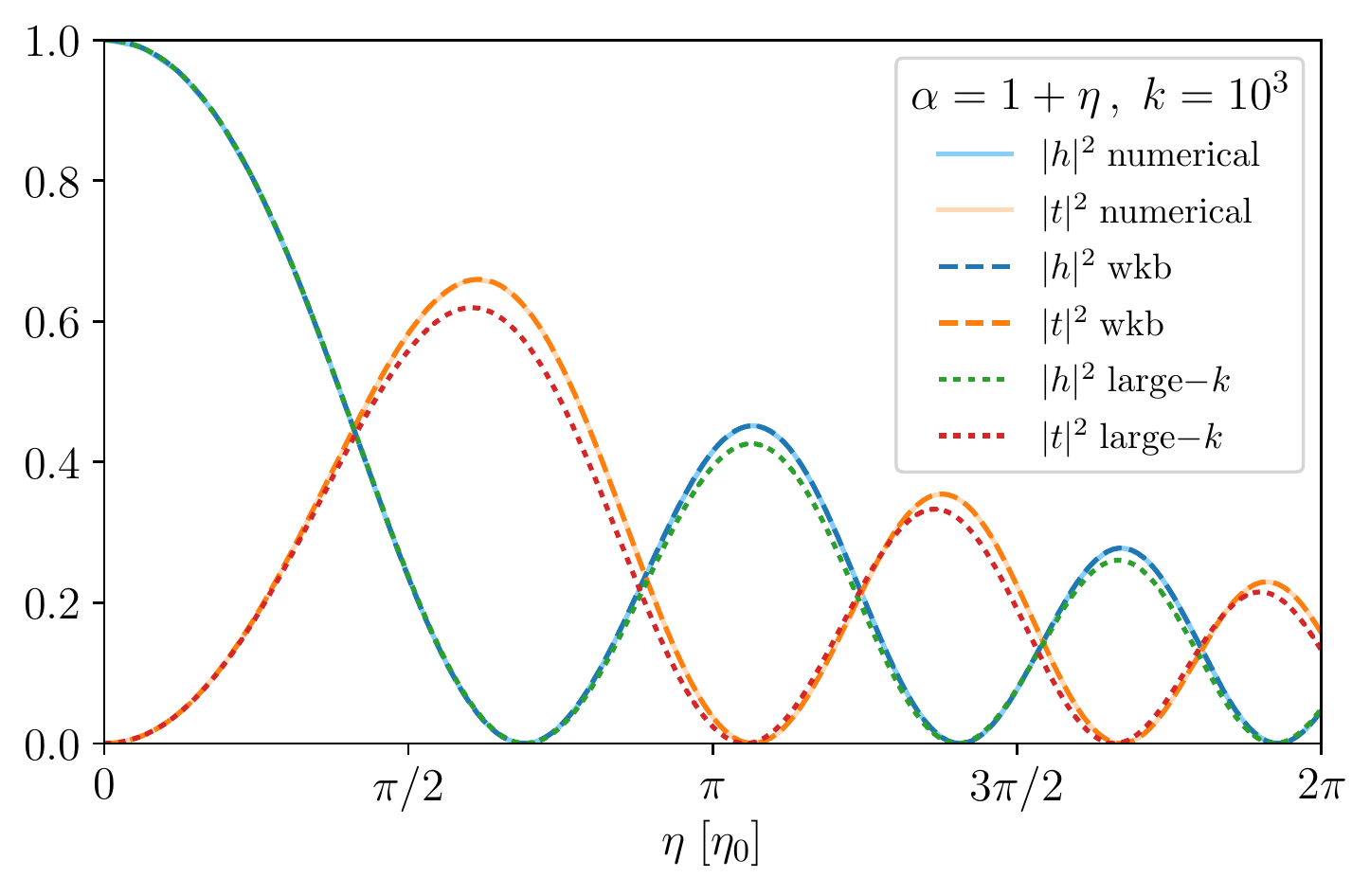}
\includegraphics[width=.49\textwidth]{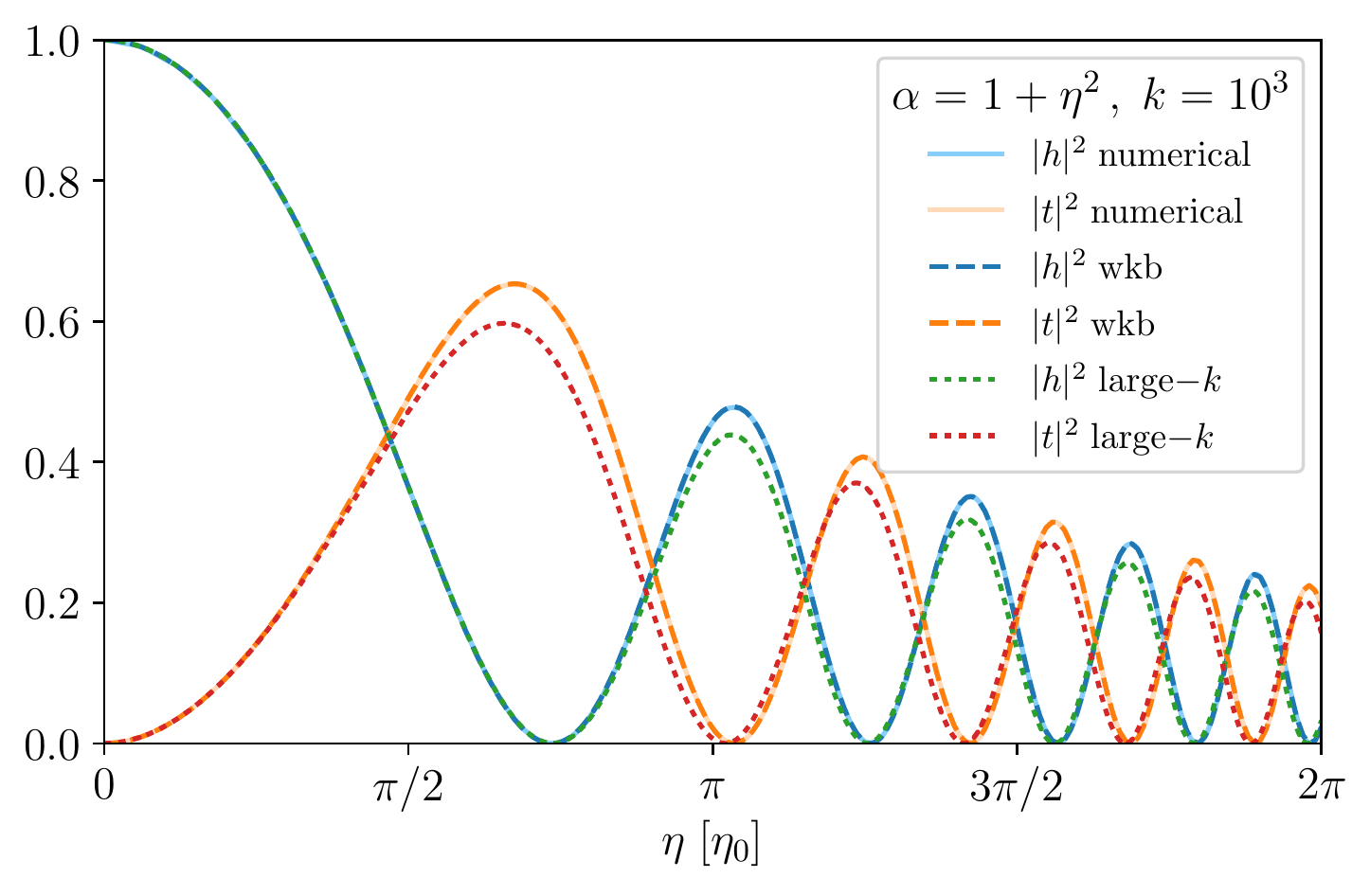}
\includegraphics[width=.49\textwidth]{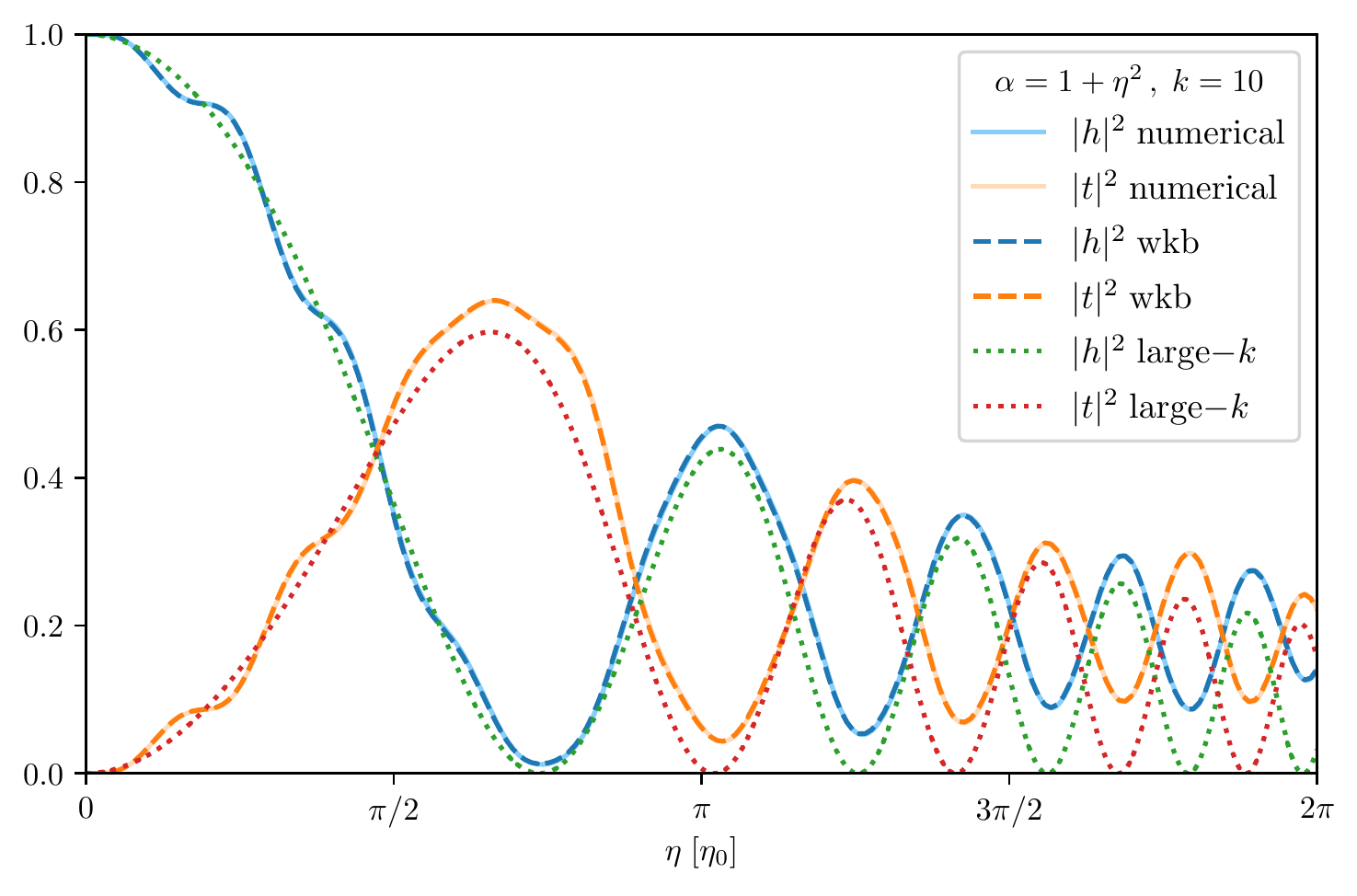}
\caption{Oscillation of the GW amplitude $\vert h\vert$ and the tensor perturbation $\vert t\vert$ due to a friction mixing. 
We vary the time dependence of the non-diagonal matrix element $\alpha$ from constant (upper left), to linear (upper right), to quadratic (lower panels).  
We compare the numerical solution (solid) with the WKB expansion (dashed) and the large-$k$ approximation (dotted) fixing $\Delta c=0$, $h_0=1$ and $k=10^{3}$, except in the lower right plot where $k=10$. We normalize time w.r.t. the initial period $\eta_0$.}
 \label{fig:friction_mixing}
\end{figure}

\paragraph{Large-$k$ approximation:} 

If we were in a situation in which the parameters are small themselves compared to $k$, we could apply the large-$k$ approximation. In the $\dc=0$ case, the phases are simply $\theta_{1,2}^2=c_h^2k^2$ and $\eM=\iM$. Thus, all the mixing information is contained in the amplitude (\ref{eq:phi0_largek}). Particularizing for the case under consideration, it becomes
\be
\begin{split}
\vPhi_0&={\frac{\sqrt{c_h(\eta_e)}}{\sqrt{c_h(\eta)}}}e^{-\frac{1}{2}\int\nM d\eta}\bpm c_1 \\ c_2 \epm \\
&={\frac{\sqrt{c_h(\eta_e)}}{\sqrt{c_h(\eta)}}} e^{-\bdnu} \bpm \cos\bonu+\frac{\bdnu}{\bonu}\sin\bonu & \frac{\bar{\alpha}}{\bonu}\sin\bonu \\ -\frac{\bar{\alpha}}{\bonu}\sin\bonu & \cos\bonu-\frac{\bdnu}{\bonu}\sin\bonu \epm \bpm c_1 \\ c_2 \epm\,,
\end{split}
\ee
where we have defined the integrals
\begin{align} \label{eq:friction_integrals}
&\bdnu=\int_{\eta_e}^\eta\dnu \d\eta\,, & &\bar{\alpha}=\int_{\eta_e}^\eta\alpha \d\eta\,, & &\bonu^2=\bar{\alpha}^2-\bdnu^2\,, 
\end{align}
from the time of emission $\eta_e$ to a given instant $\eta$. Then, imposing the initial conditions, $h(\eta_e)=h_0$ and $t(\eta_e)=0$, we get
\begin{align}
\vert h(\eta)\vert^2&=h_0^2\frac{c_h(\eta_e)}{c_h(\eta)}e^{-2\bdnu}\lp\cos\bonu+\frac{\bdnu}{\bonu}\sin\bonu\rp^2\,, \\ 
\vert t (\eta)\vert^2&=h_0^2\frac{c_h(\eta_e)}{c_h(\eta)}e^{-2\bdnu}\frac{\bar{\alpha}^2}{\bonu^2}\sin^2\bonu\,.
\end{align}
Therefore, the frequency of oscillation is controlled by $\bonu$ and the damping of the signal by~$\bdnu$. 

In the simplest case in which the friction matrix $\nM$ is constant, $\alpha=\alpha_0$ and  $\dnu=\dnu_0$, the integrals (\ref{eq:friction_integrals}) simplify to $\bdnu=\dnu_0(\eta-\eta_e)$, $\bar{\alpha}=\alpha_0(\eta-\eta_e)$ and $\bonu=\sqrt{\alpha_0^2-\dnu_0^2}(\eta-\eta_e)=\omega_0(\eta-\eta_e)$. Accordingly, only the terms in the sines and cosines, and the global damping depend on time
\begin{align} \label{eq:amplitude_constant_friction}
\vert h(\eta)\vert^2&=h_0^2\frac{c_h(\eta_e)}{c_h(\eta)}e^{-2\dnu_0\bar{\eta}}\lp\cos\lb\omega_0\bar{\eta}\rb+\frac{\dnu_0}{\omega_0}\sin\lb\omega_0\bar{\eta}\rb\rp^2\,, \\ 
\vert t (\eta)\vert^2&=h_0^2\frac{c_h(\eta_e)}{c_h(\eta)}e^{-2\dnu_0\bar{\eta}}\,\frac{\alpha_0^2}{\omega_0^2}\sin^2\lb\omega_0\bar{\eta}\rb\,,
\end{align}
where $\bar{\eta}=(\eta-\eta_e)$. This solution resembles the model of GW-gauge field oscillations studied in Ref. \cite{Caldwell:2016sut,Caldwell:2018feo}.

In Fig. \ref{fig:friction_mixing} we plot different examples of the oscillations and damping in the amplitude of the tensor perturbations $h$ and $t$ induced by the friction mixing. In order to compare the numerical solution (solid lines) with the WKB expansion (dashed lines) and the large-$k$ approximation (dotted lines), we consider different time dependences of the mixing parameter $\alpha$ and different wave-numbers $k$. For $k=10^3$, we observe that the leading WKB solution gives an excellent approximation of the numerical result for a constant, linear and quadratic dependence in time of the mixing $\alpha$ (upper left, upper right and lower left panels respectively). On the contrary, for this wavenumber, the large-$k$ expansion does not match perfectly the numerical result when there is a time dependence. Even when we lower the wavenumber to $k=10$ (lower right panel), for this choice of parameters and time interval, the WKB follows nicely the numerical solution. This serves to exemplify that the WKB is a better approximation in general than the large-$k$ expansion since it expands over the variation of the parameters with respect to the frequency and not the parameters themselves. However, when there is a large value of $k$, both approximations tends to converge and the large-$k$ expansion becomes more useful since the analytical expressions are simpler.

\subsubsection{Mixing through the velocity matrix}
\label{sec:velocity_mixing}

\begin{figure}[t]
\centering 
 \includegraphics[width=.49\textwidth]{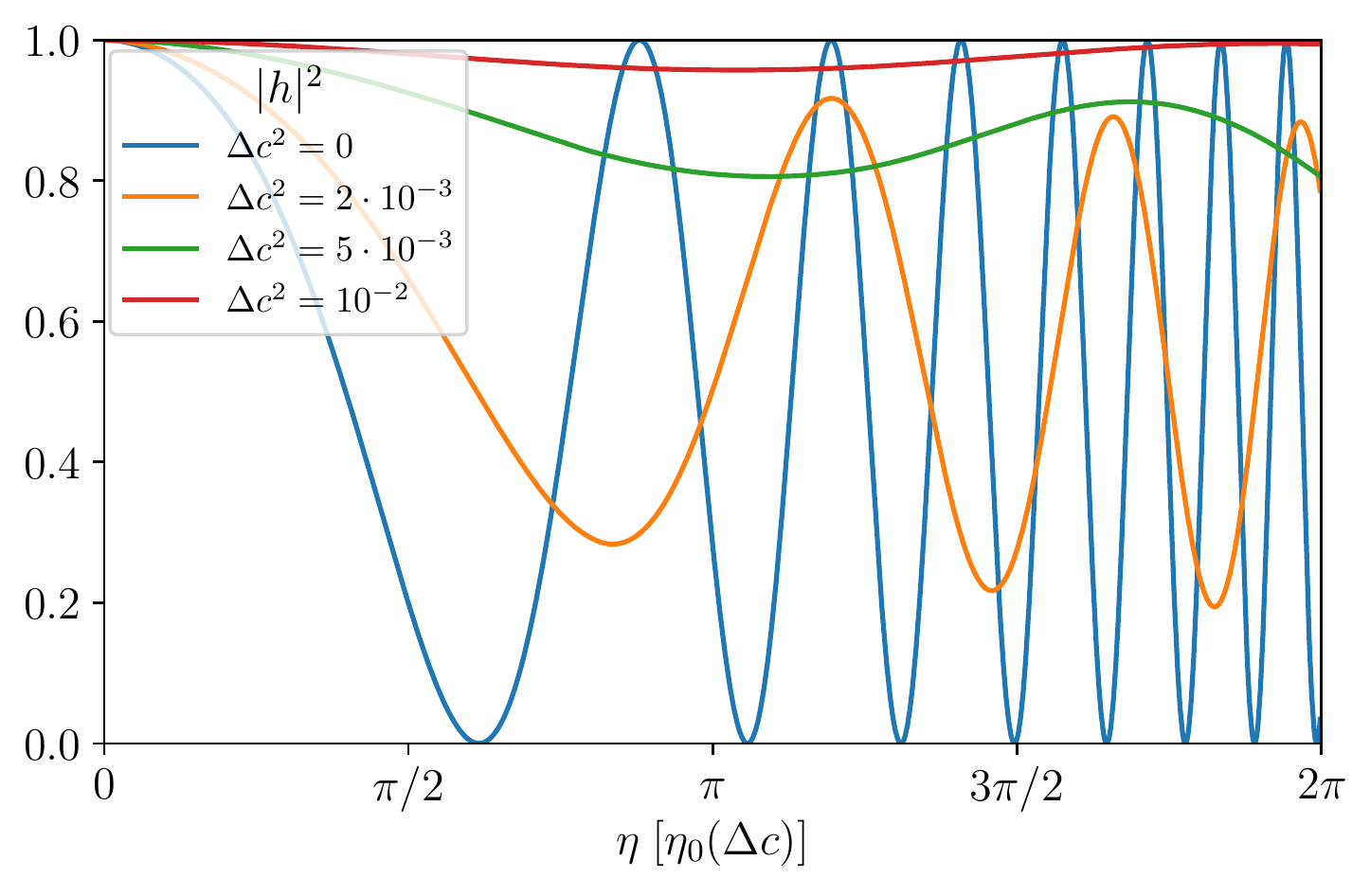}
  \includegraphics[width=.49\textwidth]{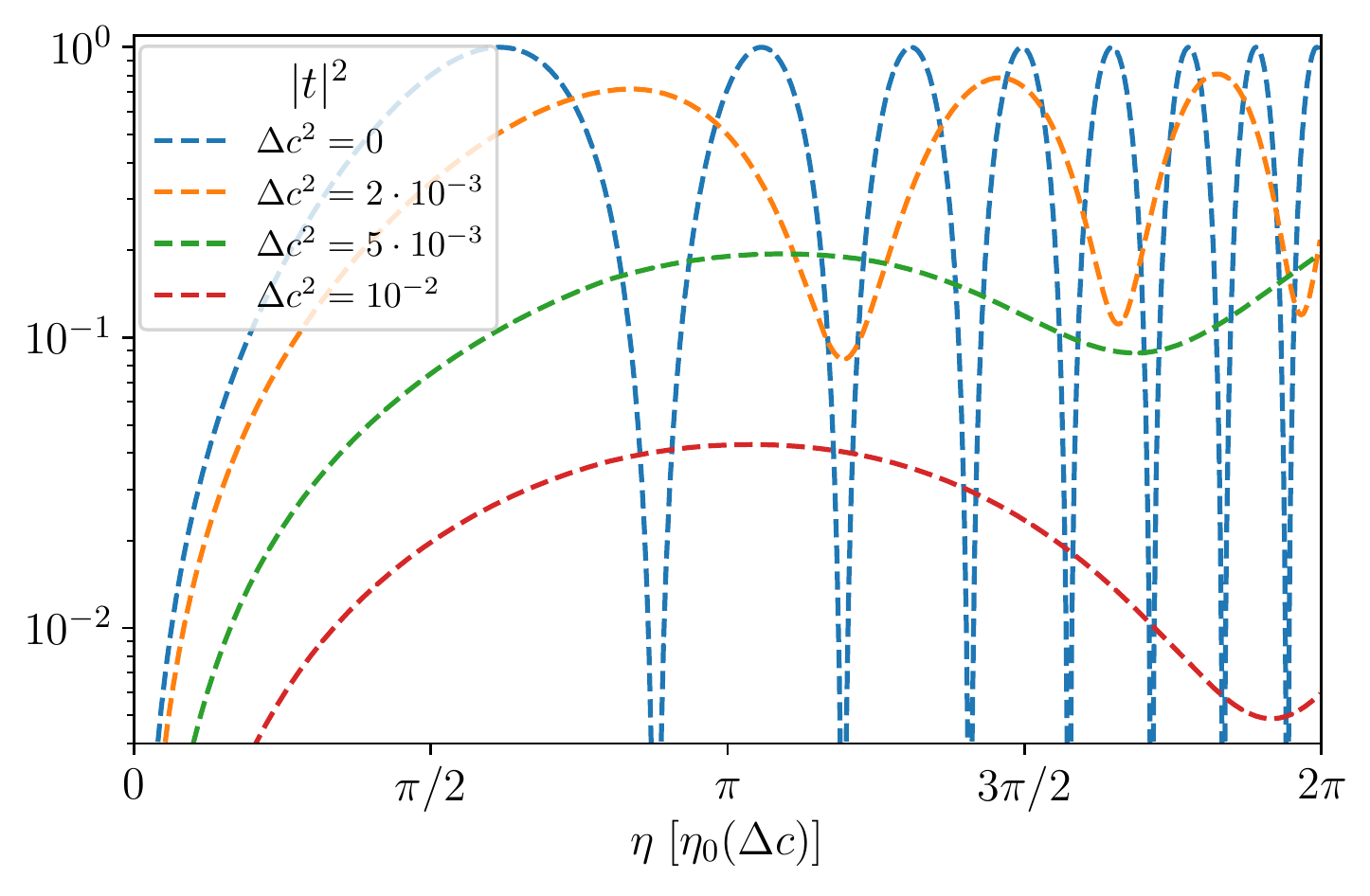}
\caption{Oscillation of the GW amplitude $\vert h\vert$ (left) and the tensor perturbation $\vert t\vert$ (right) due to a velocity mixing for different values of $\dc^2=c_t^2-c_h^2$. We have chosen $k=10^{5}$, $h_0=1$ and the mixing $c_{ht}$ quadratic in time. 
For each $\Delta c$, we have normalized the time w.r.t. its initial period $\eta_0(\Delta c)$.}
 \label{fig:velocity_mixing}
\end{figure}

As we will show later, certain operators introduce also a non-diagonal velocity matrix. Then, a mixing occurs at leading order in both WKB and large-$k$ expansions. Focusing only on this source of mixing,
\be
\lb \frac{\d^2}{\d\eta^2} +\bpm c_h^2 & c_{ht}^2 \\ c_{ht}^2 &  c_t^2\epm k^2\rb \bpm h \\ t \epm =0\,, 
\ee
we can easily solve the propagation. Using the notation introduce above $\dc^2=c_t^2-c_h^2$, the eigenfrequencies are given by
\be
\theta^2_{1,2}= \lp c_h^2+\frac{1}{2}\dc^2\mp\frac{1}{2}\sqrt{4c_{ht}^4+\dc^4}\rp k^2\,, \label{eq:t12_velocity}
\ee
and the matrix of eigenvectors by
\be
\hat{E}=\bpm 1 & -\frac{2c_{ht}^2}{\dc^2+\sqrt{4c_{ht}^4+\dc^4}} \\ \frac{2c_{ht}^2}{\dc^2+\sqrt{4c_{ht}^4+\dc^4}} & 1\epm\,. 
\ee
Then, the mixing in the amplitude is determined by the non-diagonal entry $c_{ht}$. 
Note here that although a different propagation speed $\dc\neq0$ tends to suppress the mixing, this is not enhanced by $k$ as in previous cases. For instance, for the mass mixing case (\ref{eq:eM_wkb_mass}), the non-diagonal terms were suppressed by $\sim1/k^2$. This can be seen in Fig. \ref{fig:velocity_mixing}, where we show the oscillation in $h$ and $t$ for different values of $\dc$ in the left and right panels  respectively. The larger $\dc$ becomes, the more $\vert t\vert$ is suppressed and the more $\vert h\vert$ approaches the initial value~$h_0$. This plot is analogous to the mass mixing case presented in Fig. \ref{fig:mass_mixing_suppression}. However, for the velocity mixing the suppression is not enhanced by $k$ and, thus, $\dc$ can be larger. This different behavior becomes more pronounced as $k$ grows.

As in the previous cases, even if $c_h=c$, there can be an anomalous speed $\cgw\neq c$ whenever there is a mixing via $c_{ht}\neq0$ and the second tensor $t$ has a non-luminal propagation speed $c_t\neq c$, as it can be easily deduced from
\be\label{eq:agw_velocity}
\agw=\frac{1}{2}\frac{\dc^2}{c^2}\lp1-\sqrt{1+4\frac{c_{ht}^4}{\dc^4}}\rp\,.
\ee
The anomalous speed $\agw$ is thus degenerate in the difference of the speeds $\dc$ and the mixing term $c_{ht}$.

\subsubsection{Chiral mixing}
\label{sec:chiral_mixing}

In the presence of parity violating terms it is convenient to work in the left- and right-circular polarizations basis, which we assume in the following. In the simplest set-up, there is only the parity violating matrix $\pM$ linear in $k$ and the velocity matrix,
\be
\lb \frac{\d^2}{\d\eta^2}  +\bpm c_h^2 & 0 \\ 0 &  c_t^2\epm k^2\pm\bpm \mu_h  & \gamma \\ \gamma &  \mu_t \epm k \rb \bpm h_{L,R} \\ t_{L,R} \epm =0\,. 
\ee
Due to the $\pm$ in front of the $\pM$ matrix, the $L$ and $R$ polarizations evolve differently. As in the previous cases, we compare different approximate solutions of these coupled differential equations. We will see that the WKB analysis will be similar to the mass mixing studied in section \ref{sec:mass_mixing} and the large-$k$ similar to the friction mixing studied in section \ref{sec:friction_mixing}.

\begin{figure}[t]
\centering 
 \includegraphics[width=.49\textwidth]{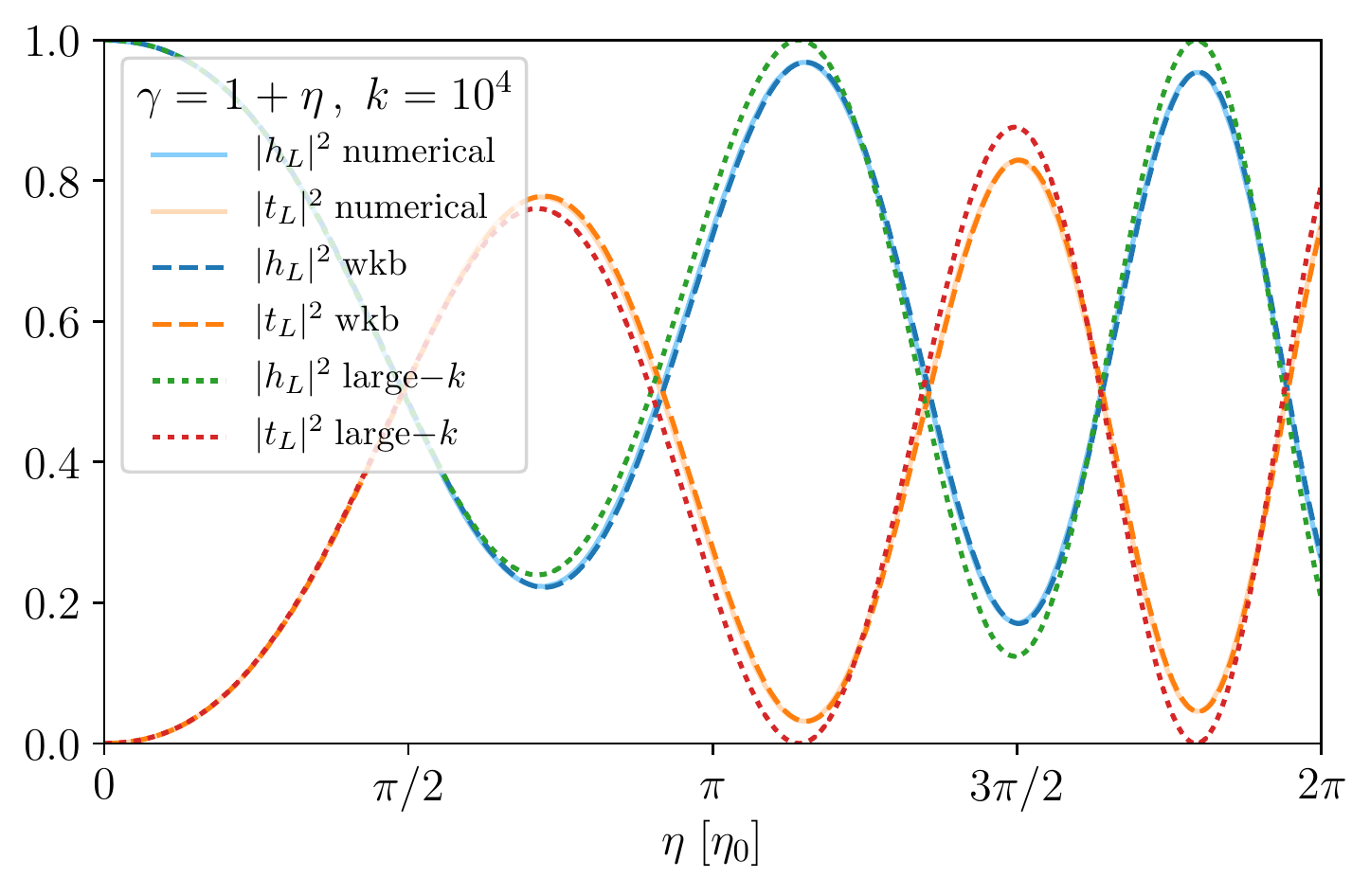}
  \includegraphics[width=.49\textwidth]{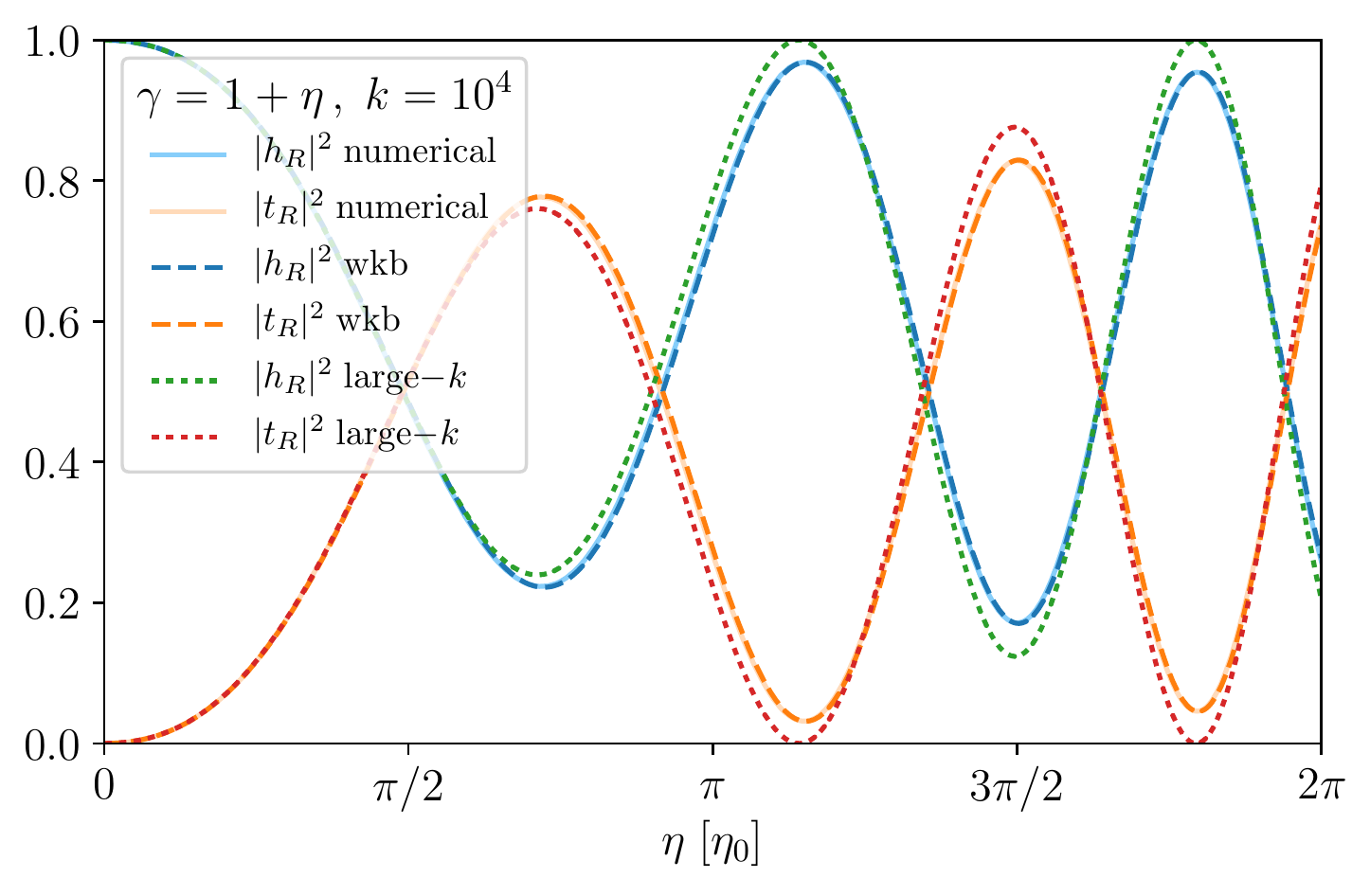}
 \caption{Oscillation of the circular polarizations (left and right) of the GW amplitude $\vert h_{L,R}\vert$ and the tensor perturbation $\vert t_{L,R}\vert$ due to a chiral mixing. 
We choose the mixing $\gamma$ to vary linearly in time. 
We compare the numerical solution (solid) with the WKB expansion (dashed) and the large-$k$ approximation (dotted) fixing $\Delta c=0$, $h_0=1$, $k=10^{4}$ and normalizing time w.r.t. the initial period $\eta_0$.}
 \label{fig:chiral_mixing}
\end{figure}

\paragraph{WKB approximation:} at leading order in the WKB we obtain the phase of the wave by solving the algebraic equation \eqref{eq:determinant}. 
The corresponding phases for each polarization are
\begin{align}
(\theta_{L,R;\,1})^2&=c_h^2k^2+\frac{\dc^2 k^2}{2}\pm\frac{\mtot k}{2}-\frac{k}{2}\sqrt{4\gamma^2+(\dmu\mp k\dc^2)^2}\,, \label{eq:t1_chiral}\\
(\theta_{L,R;\,2})^2&=c_h^2k^2+\frac{\dc^2 k^2}{2}\pm\frac{\mtot k}{2}+\frac{k}{2}\sqrt{4\gamma^2+(\dmu\mp k\dc^2)^2}\, \label{eq:t2_chiral},
\end{align}
where we have defined $\mtot=\mu_h+\mu_t$ and $\dmu=\mu_t-\mu_h$. The matrix of eigenvectors is analogous to the mass mixing case (\ref{eq:eMeigenvectors}) substituing $\mM$ for $\pM\,k$ and accounting for the different sign of the parameters of each polarization
\be \label{eq:eMeigenvectors_chiral}
\hat{E}_{L,R}=\bpm 1 & \mp\frac{\gamma}{c_h^2k^2\pm\mu_h-\theta_2^2} \\ \mp\frac{\gamma}{c_t^2k^2\pm\mu_t-\theta_1^2} & 1\epm\,. 
\ee

With these expressions one can proceed and analyze how a GW signal will be modified. There will be both a modification of the amplitude and the phase due to the GW oscillations. These modifications will depend on the polarization. In fact we can already anticipate from the matrix \eqref{eq:eMeigenvectors_chiral} that there will be a chiral effect in the amplitude. We will discuss in section \ref{sec:chirality} how to probe this chirality. For the moment, let us focus on the phase. In the limit in which the difference in the propagation speeds $\dc$ is small, we can extract the GW speed from the leading $k^2$ term. Although $h$ propagates at the speed of light, the non-luminal speed of $t$ together with the mixing $\gamma$ induces an anomalous speed for the GWs, parametrized by $\agw=\cgw^2/c^2-1$. We obtain
\be\label{eq:agw_chiral}
(\agw)_{L,R}=\frac{1}{2}\frac{\dc^2}{c^2}\lp1\mp\frac{\dmu}{4\omu}\rp\,,
\ee 
where, for later convenience, we have introduced the frequency $16\omu^2=4\gamma^2+\dmu^2$.

\begin{figure}[t]
\centering 
 \includegraphics[width=.49\textwidth]{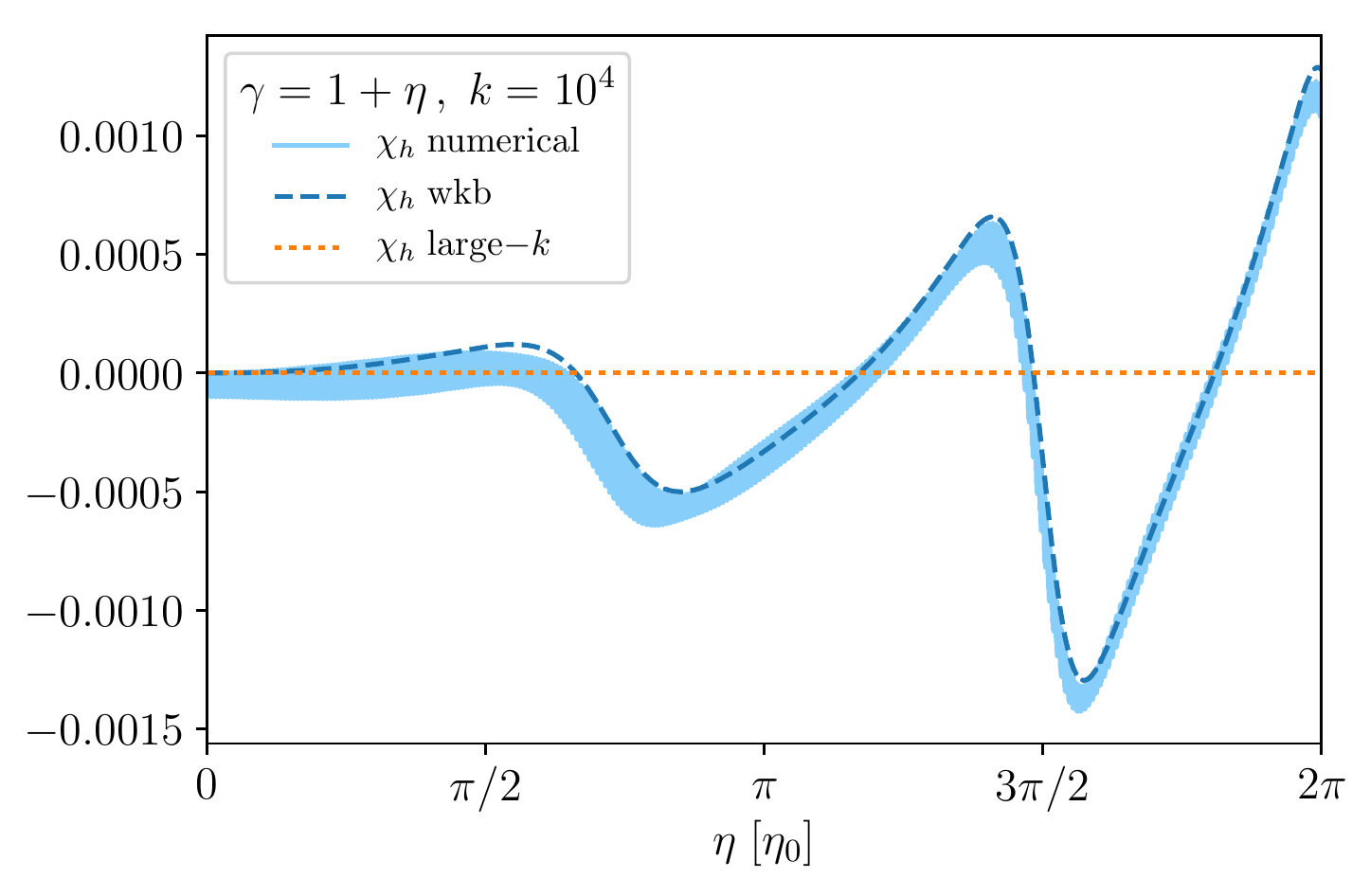}
  \includegraphics[width=.49\textwidth]{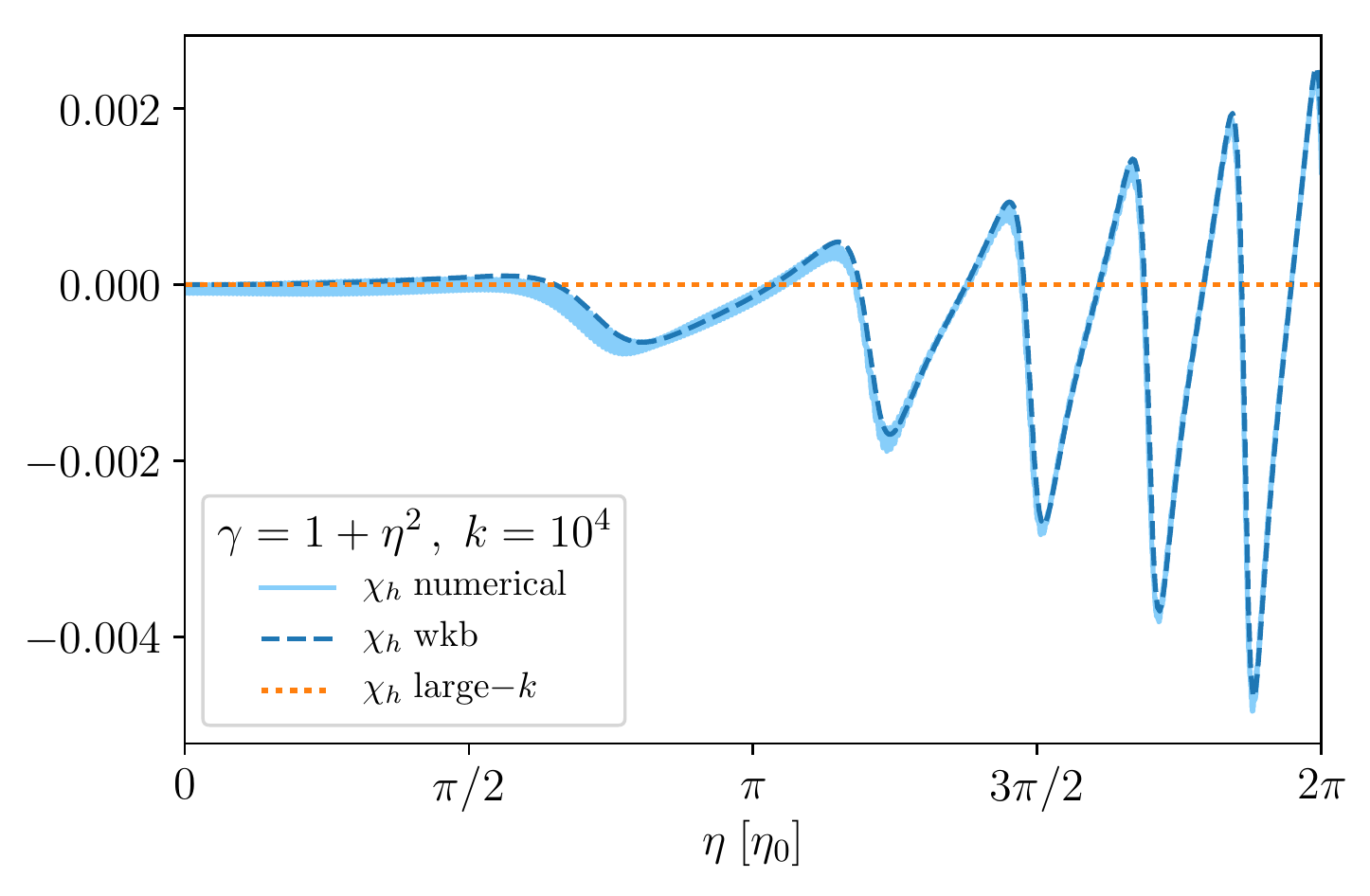}
 \caption{Chirality induced by GW oscillations. We compare the value of $\chi_h$, see \eqref{eq:chirality}, for the numerical (solid), WKB (dashed) and large-$k$ (dotted) solutions. We plot a linear and a quadratic time dependent mixing $\gamma$ in the left and right panels respectively, fixing $\Delta c=0$, $h_0=1$, $k=10^{4}$ as in Fig. \ref{fig:chiral_mixing}.}
 \label{fig:chirality}
\end{figure}

\paragraph{Large-$k$ approximation:} Let us now consider the limit in which both perturbations propagate at the same speed, $\dc=0$. 
Then, similarly to the friction mixing, in the large-$k$ expansion one obtains $\theta_{1,2}^2=c_h^2k^2$ and $\eM=\iM$ for both polarizations. Using (\ref{eq:largek_amplitude}), we obtain the leading order amplitude $\vPhi_0$. One may notice that the situation is equivalent to the friction mixing if we exchange $\nM\rightarrow -i\,k\pM\tM^{-1}$. Thus we get
\be
\begin{split}
\lp\vPhi_0\rp_{L,R}&={\frac{\sqrt{c_h(\eta_e)}}{\sqrt{c_h(\eta)}}}e^{\pm\frac{i}{2}\int\pM\,c^{-1}_h d\eta}\bpm c_1 \\ c_2 \epm \\
&={\frac{\sqrt{c_h(\eta_e)}}{\sqrt{c_h(\eta)}}} e^{\pm\frac{i}{4}\bmtot} \bpm \cos\bomu\mp i\frac{\bdmu}{4\bomu}\sin\bomu & \pm i\frac{\bar{\gamma}}{2\bomu}\sin\bomu \\ \pm i\frac{\bar{\gamma}}{2\bomu}\sin\bomu & \cos\bomu\pm i\frac{\bdmu}{4\bomu}\sin\bomu \epm \bpm c_1 \\ c_2 \epm\,,
\end{split}
\ee
where we have defined the average values as
\begin{align}
&\bmtot=\int_{\eta_e}^\eta\frac{\mtot}{c_h} \d\eta\,, & &\bdmu=\int_{\eta_e}^\eta\frac{\dmu}{c_h} \d\eta\,, & &\bar{\gamma}=\int_{\eta_e}^\eta\frac{\gamma}{c_h} \d\eta\,, & &16\bomu^2=4\bar{\gamma}^2+\bdmu^2\,. 
\end{align}
Imposing the initial conditions, $h(\eta_e)=h_0$ and $t(\eta_e)=0$, we obtain the amplitude of the tensor perturbations
\begin{align}
\vert h_{L,R}\vert^2&=h_0^2\frac{c_h(\eta_e)}{c_h(\eta)}\lp\cos^2\bomu+\frac{\bdmu^2}{16\bomu^2}\sin^2\bomu\rp\,, \\ 
\vert t_{L,R}\vert^2&=h_0^2\frac{c_h(\eta_e)}{c_h(\eta)}\frac{\bar{\gamma}^2}{4\bomu^2}\sin^2\bomu\,.
\end{align}
Noticeably, the amplitude is the same for both polarizations. This means that in this setup, at leading order in the large-$k$ expansion, there is no chiral effect. Note however that if instead of starting only with a chiral matrix $\pM$, we include a friction matrix $\nM$, there will be chiral effects. This is because the matrix in the exponent of \eqref{eq:largek_amplitude} will then be $\Ak=\nM-i\,k\pM\tM^{-1}$. Another point to highlight is that, differently to the friction mixing, now there is not a global damping because the $e^{\pm\frac{i}{4}\bmtot}$ term only contributes to the phase.

In Fig. \ref{fig:chiral_mixing} we plot the amplitude of the two polarizations (left and right panels respectively) of the perturbations of $h$ and $t$. We choose a mixing parameter that varies linearly in time. As for the friction mixing, the WKB solution (dashed lines) provides a better approximation of the numerical result (solid lines) than the large-$k$ expansion (dotted lines). Moreover we observe that both the left and right polarizations evolve qualitatively in the same manner. To quantify the difference in the evolution, we introduce the chirality parameter $\chi$ that, for the GW amplitude, we define as
\be \label{eq:chirality}
\chi_h=\frac{\vert h_L\vert^2-\vert h_R\vert^2}{\vert h_L\vert^2+\vert h_R\vert^2}\,.
\ee
In Fig. \ref{fig:chirality} we plot $\chi_h$ for the numerical, WKB and large-$k$ solutions. As discussed, at leading order in the large-$k$ expansion, there is no chirality. On the other hand, the WKB gives a good agreement with the numerical result. 
An important point is that the chirality can grow along the propagation, as exemplified in the plots. Thus, a sizable difference between the left and right polarization can be induced due to the modified propagation. One should note however that the particular time evolution is subject to the specific time dependence of the parameters. Moreover, by definition, the chirality is bounded by $\vert\chi_h(\eta)\vert<1$. We compare the growth of the chirality for a linear (left panel) and quadratic (right panel) time dependence of the chiral mixing $\gamma$. We will discuss the detectability of this effect in section~\ref{sec:chirality}.

\section{Theoretical landscape}\label{sec:theorLandscape}

In the precedent sections we have developed a general framework for cosmologies containing two helicity-2 modes. We will now discuss different theoretical scenarios that can give rise to these cosmologies. It is clear that the usual GWs will be provided by the GR sector, which is precisely the theory for a massless spin-2 field. The appearance of a second tensor mode thus requires the presence of additional fields propagating at least two degrees of freedom as to conform the two polarizations of the extra tensor mode. However, the precise nature of the additional helicity-2 mode can have different physical origins depending on the underlying mechanism that generates it. Below we will explain in some detail different scenarios featuring a second tensor mode that can be broadly classified as follows:

\begin{itemize}

\item Additional spin-2 field. This is the most straightforward way to have an extra tensor perturbation that directly originates from the helicity-2 mode of the additional spin-2 field.

\item Non-trivial realisations of the Cosmological Principle. The additional tensor mode arises in a less direct manner within these scenarios and it is a consequence of realising the homogeneity and isotropy of the FLRW universes by combining spatial rotations and translations with some internal symmetries.

\end{itemize}

The first scenario with a second spin-2 field is not difficult to understand so we will not explain it any further here and we will give all the necessary details in Sec. \ref{subsec:bigravity}. A cautionary comment should be stated however due to the delicate nature of spin-2 field interactions. It is well-known that higher order curvature theories that contain arbitrary powers of the Riemann tensor propagate an additional massive spin-2 field besides the usual massless spin-2 that describes GWs. Crucially, this second mode is associated to the higher order nature of the corresponding field equations and, hence, the theory possess an instability in the form of an Ostrogradski ghost. For this reason we will not consider these theories within our theoretical landscape. 

The cosmologies based on non-trivial realizations of the Cosmological Principle can be more subtle so we will discuss the general mechanism. The basic idea is to have some internal symmetry group $G$ (that can be either global or local) in the matter sector besides the Poincar\'e group so we have $G\times ISO(3,1)$. The matter fields then adopt a background configuration that breaks both the internal and the Poincar\'e groups\footnote{For a thorough classification of systems whose vacuum state break spacetime Poincar\'e symmetries, while preserving some kind of translations and/or rotations see \cite{Nicolis:2015sra}.}. However, for appropriate internal groups (e.g. $SO(3)$ or $SU(2)$) the fields configuration can in turn leave some linear combinations of internal and Poincar\'e generators unbroken and, provided the generators correspond to some translations and rotations, there can be a residual $ISO(3)$ symmetry as required by the Cosmological Principle. Let us notice that this $ISO(3)$ is not the corresponding subgroup of the original Poincar\'e group (as it is the case in standard realizations). In other words, the symmetry breaking pattern is $G\times ISO(3,1)\rightarrow ISO(3)_{\text{diagonal}}$ where the generators of the unbroken $ISO(3)_{\text{diagonal}}$ group are not those in the original $ISO(3,1)$. The importance of having this symmetry group for the background is that the perturbations can be classified according to it and, provided the matter sector is the appropriate one, some perturbations can arrange themselves into an additional helicity-2 mode, even if there are no extra spin-2 fields in the theory. Let us emphasize that the usual decomposition theorem for the perturbations into scalar, vector and tensor modes is still valid, but one must choose the appropriate group for their classification. In the case at hand, decomposing the perturbations into irreducible representations of the purely spatial rotations would mix the different helicity modes, since that is no longer a symmetry of the background configuration. The decoupling will however occur when decomposing the perturbations according to the diagonal rotational group that combines both spatial and internal transformations\footnote{As an example of the importance of appropriately decomposing the perturbations we can mention \cite{ArmendarizPicon:2004pm}, where a violation of the decomposition theorem is claimed to occur. However, this is precisely because the perturbations were decomposed according to the purely spatial rotations. Had the unbroken diagonal group been used to classify the perturbations, the decomposition theorem would have been found to hold.}. The possibility of having linear combinations of internal and external generators that combine to produce the unbroken $ISO(3)_{\rm diagonal}$ can be guaranteed if the internal group contains $SO(3)$ and/or $SU(2)$ as subgroups. Furthermore, since we want to have perturbations that arrange into a helicity-2 mode with respect to the unbroken diagonal group, we need to consider the internal symmetry realized with fields that transform non-trivially under the Lorentz group, i.e., carrying some Lorentz indices. Perhaps the simplest realizations are those based on spin-1 fields and for that reason we will give explicit examples with vector fields below. 

Lastly, let us point out that we are only considering scenarios with one additional tensor mode. Multiple extra tensor modes could be straightforwardly accommodated within the presented formalism by just appropriately enlarging the dimensionality of the matrices in the equations of motion (\ref{eq:generalequation}). In this case, new possibilities arise since the new tensor modes can arise from a mixture of the two mechanisms discussed above. In particular, we could have several spin-2 fields such as multi-gravity \cite{Hinterbichler:2012cn}, non-trivial realizations of the cosmological principle involving more fields\footnote{A particular realization of this possibility was discussed in \cite{Caldwell:2016sut} by using the independent $SU(2)$ subgroups of some large $SU(N)$ group. The case of $SU(4)$ with two independent $SU(2)$ subgroups giving rise to two tensor modes was explicitly worked out.}   or combinations of both mechanisms. An interesting example of the former would be tri-gravity with some internal $SO(3)$ symmetry. We will not explore these interesting scenarios here and we will leave them for future work. 

\subsection{Bigravity}\label{subsec:bigravity}

The quest to promote a massless spin-2 particle to a massive one enforces the introduction of a second metric, the fiducial metric $f_{\mu\nu}$. This is easy to understand, since any contraction of the metric with itself will only contribute a cosmological constant $\Lambda_1=g_{\mu\nu}g^{\mu\nu}$ and $\Lambda_2=g_{\mu}{}^{\mu}g_{\nu}{}^{\nu}$. Hence, we need to introduce the fiducial metric in order to build a mass term or a general potential term. The fiducial metric can be made dynamical by including an explicit kinetic term for it. This gives rise to ghost-free bigravity, whose Lagrangian is given by
\begin{equation}\label{action_MG_dRGT}
\mathcal{S} = \int \mathrm{d}^4x  \left(\frac{\mpl^2}{2} \sqrt{-g}R[g]+\frac{M_{\rm f}^2}{2} \sqrt{-f}R[f]-\frac{m^2M_{\rm eff}^2}{2}\sum_n \beta_n e_n(\sqrt{g^{-1}f})  +\mathcal{L}_{\rm matter} \right)\,,
\end{equation}
where the corresponding potential interactions ${\cal U}_n$ read \cite{deRham:2010kj}
\begin{eqnarray}
e_0&=& \mathcal{E}^{\mu\nu\rho\sigma}  \mathcal{E}_{\mu \nu\rho\sigma}\nonumber\\
e_1&=& \mathcal{E}^{\mu\nu\rho\sigma}  \mathcal{E}^{\alpha}_{\;\;\;\nu\rho\sigma} \mathcal{S}_{\mu\alpha} \nonumber\\
e_2&=& \mathcal{E}^{\mu\nu\rho\sigma}  \mathcal{E}^{\alpha\beta}_{\;\;\;\;\; \rho\sigma} \mathcal{S}_{\mu\alpha} \mathcal{S}_{\nu\beta} ,
\nonumber\\
e_3&=& \mathcal{E}^{\mu\nu\rho\sigma}  \mathcal{E}^{\alpha\beta\kappa}_{\;\;\;\;\;\;\; \sigma} \mathcal{S}_{\mu\alpha} \mathcal{S}_{\nu\beta}  \mathcal{S}_{\rho\kappa},
\nonumber\\
e_4&=& \mathcal{E}^{\mu\nu\rho\sigma}  \mathcal{E}^{\alpha\beta\kappa\gamma} \mathcal{S}_{\mu\alpha} \mathcal{S}_{\nu\beta}  \mathcal{S}_{\rho\kappa}  \mathcal{S}_{\sigma\gamma} ,
\end{eqnarray}
with the Levi-Cevita tensor $\mathcal{E}^{\mu\nu\rho\sigma}$. The fundamental tensor of the theory $\mathcal{S}^\mu{}_\nu$ must have the specific form
$\mathcal{S}^\mu{} _\nu = \left(\sqrt{g^{-1}f}\right)^\mu{}_\nu$. Note that it is assumed that the matter fields only couple to $g_{\mu\nu}$.
The equations of motion are obtained by varying the action with respect to $g_{\mu\nu}$ and $f_{\mu\nu}$, respectively
\begin{eqnarray}
&&G_{\mu\nu}^g+V_{\mu\nu}=\frac{T_{\mu\nu}}{\mpl^2} \nonumber\\
&&G_{\mu\nu}^f+Y_{\mu\nu}=0,
\end{eqnarray}
where $V$ and $Y$ arise as the respective variation of the potential with respect to $g$ and $f$. For the FLRW cosmological background compatible with homogeneity and isotropy we can assume the most general ansatz 
\be
\d s^2_g=a^2(\eta)\Big(-\d\eta^2+\delta_{ij}\d x^i\d x^j\Big)\quad\text{and} \quad\d s^2_f=b^2(\eta)\Big(-\tilde{c}^2\d\eta^2+\delta_{ij}\d x^i\d x^j\Big).
\ee 
Considering small tensor perturbations for both metrics yields the following coupled differential equations
\be \label{eq:bigravity}
\lb \frac{\d^2}{\d\eta^2} +2\mH \frac{\d}{\d\eta}  +\bpm c_h^2 & 0 \\ 0 &  c_t^2\epm k^2+m^2\bpm m_1  & -m_1 \\ -m_2 &  m_2 \epm \rb \bpm h \\ t \epm =0\,,
\ee
with the propagation speeds $c_t=1=c_h$ for backgrounds with $\tilde{c}=1$ and $\partial_\eta(b/a)=0$. Introducing the following linear combinations $u_{\mu\nu}=M_{\rm eff}(h_{\mu\nu}/M_{\rm f}+t_{\mu\nu}/\mpl)$ and $v_{\mu\nu}=M_{\rm eff}(h_{\mu\nu}/\mpl-t_{\mu\nu}/M_{\rm f})$, one immediately observes that one massless and one massive tensor perturbation propagate and one of the equations decouples
\be
\lb \frac{\d^2}{\d\eta^2} +2\mH \frac{\d}{\d\eta}  + k^2+m^2\bpm 0  & 0 \\ -\tilde{m}_2 &  \tilde{m}_2 \epm \rb \bpm u \\ v \epm =0\,,
\ee
The mixing of the two tensor modes via the mass term gives rise to oscillatory behavior of gravitational waves.
In the presence of a non-diagonal coupling of the two modes to matter results in modulations of the strain that could in principle be observable. In principle, the two tensor modes could propagate at different speeds. This is the case for backgrounds with $\tilde{c}\ne1$. They could also have different friction terms for backgrounds with $\partial_\eta(b/a)\ne0$. For more details we refer the reader to \cite{Belgacem:2019pkk} where the propagation of GWs outside of the de Sitter branch was studied.

\subsection{Yang-Mills theories}\label{subsec_YM}
In this section we will consider theories for a non-Abelian gauge field $A^a{}_\mu$ that can support homogeneous and isotropic solutions (see e.g. \cite{Galtsov:1991un,Zhao:2005bu,Maleknejad:2011jw,Adshead:2012kp,Maleknejad:2012fw} for some cosmological applications). The simplest realisation of this scenario utilizes an internal $SU(2)$ gauge group. The background field configuration compatible with the required symmetries has vanishing temporal components and three spatial vector fields of the same magnitude and mutually orthogonal as follows:
\be
\bar{A}^a{}_i=A(t)\delta^a{}_i,\quad \bar{A}^a{}_0=0
\label{triad}
\ee
where we have aligned the three vector fields with the corresponding coordinate axis. This configuration breaks both the internal $SU(2)$ and the external rotations, but there are  combinations of generators that remain unbroken, leaving a diagonal $SO(3)$ symmetry unbroken. This can be easily understood by noticing that the effect of an internal $SU(2)$ transformation can be compensated by performing a spatial rotation. Since the background only depends on time, homogeneity is trivially realized.

The Yang-Mills theories that we will consider will be constructed in terms of Lorentz and $SU(2)$ invariants  that can be formed with the field strengths defined as:
\be
\F^a{}_{\mu\nu}=\partial_\mu A^a{}_{\nu}-\partial_\nu A^a{}_{\mu}-g\epsilon^{abc}A^b{}_{\mu}A^c{}_{\nu}
\label{eq:defFnA}
\ee
with $g$ the gauge coupling constant and $\epsilon^{abc}$ the completely antisymmetric tensor in colour space. The independent invariants can be built in terms of the bilinear Lorentz invariants\footnote{Our construction follows the analogous one performed in \cite{Piazza:2017bsd}.}
\be
X^{ab}=\F^a{}_{\mu\nu}\F^{b\mu\nu},\quad \Xt^{ab}=\F^a{}_{\mu\nu}\tilde{\F}^{b\mu\nu}
\ee
and the following Lorentz invariant trilinears
\be
Y^{abc}=\F^{a\mu}{}_\nu \F^{b\nu}{}_\rho \F^{c\rho}{}_\mu,\quad \Yt^{abc}=\F^{a\mu}{}_\nu \F^{b\nu}{}_\rho \tilde{\F}^{c\rho}{}_\mu.
\ee
Then, the independent scalars can be chosen to be
\begin{align}\label{SU2scalars}
&X_0=[X],\quad X_1=[\Xt],
\nonumber\\
&X_2=[X^2],\quad 
X_3=[\Xt^2],\quad 
X_4=[X\Xt],\nonumber\\
&X_5=[X^3],\quad X_6=[X^2\Xt],\quad X_7=[\Xt^3],\nonumber\\
&X_8=[X^3\Xt],\quad X_9=\epsilon_{abc} Y^{abc},\quad X_{10}=\epsilon_{abd} \Yt^{abc}\,,
\end{align}
where the square brackets stand for the trace over the internal $SU(2)$ space. Obviously, analyzing the general theory depending on all the eleven invariants is a very arduous task clearly beyond the scope of the present work. However, in order to provide a concrete illustration for the implications of gravitational waves oscillations discussed in the above section, we shall focus on a very simple model described by the following action
\be
\mS_{\rm YM}=\int\d^4x\sqrt{-g} \,f(X_0).
\ee
For this class of theories, we have two tensor modes, namely, the usual ones corresponding to the GWs of GR plus a tensor perturbation of the gauge field. More explicitly, we have the helicity-2 modes defined by
\be
h_{ij}=\frac{\mpl}{a^2}\delta g_{ij}\quad{\text {and}}\quad t_{ij}=\delta^a_{(i}\delta A^a{}_{j)}
\label{deftensorA}
\ee
with the traceless and transverse conditions $\delta^{ij}h_{ij}=\delta^{ij}t_{ij}=0$ and $\partial_ih_{ij}=\partial_it_{ij}=0$ respectively. The equation of motion for the tensor modes in conformal time take the form 
\begin{eqnarray}\label{eqTen_YM}
\left[ \bpm 1  & 0 \\ 0 &-16f_X   \epm\frac{\d^2}{\d\eta^2} +\bpm 2\mH  & -16A'f_X/(a\mpl) \\ 16A'f_X/(a\mpl) &\nu_t   \epm \frac{\d}{\d\eta}  +\bpm 1  & 0 \\ 0 &-16f_X   \epm k^2\right. \nonumber\\
\left.\pm32gAf_X\bpm 0  & -A/(2a\mpl) \\ -A/(2a\mpl) &1   \epm k+\bpm m_1  & m_2 \\ m_3 &  m_4 \epm \right] \bpm h \\ t \epm_{R,L} =0\,,
\end{eqnarray}
where $\pm$ correspond to the $R$ and $L$ polarizations respectively, we have defined
\be
\nu_t\equiv-32\mH f_X+\frac{192f_{XX}}{a^4}\Big[2g^2A^3\big(A\mH-A'\big)+A'\big(A''-2\mH A'\big)\Big]
\ee
 with $f_X=\partial_{X_0}f$, $f_{XX}=\partial_{X_0X_0}f$ and the components of the mass matrix are given by
\begin{eqnarray}
m_1&=&2\big(\mH^2+2\mH'\big)+2\frac{a^4f+4f_X\big(2A'^2-3g^2A^4\big)}{a^2\mpl^2}, \nonumber\\
m_2&=&\frac{16f_X\Big(g^2A^3-\mH A'\Big)}{a\mpl}, \nonumber\\
m_3&=&\frac{16}{a\mpl}\left[f_X\Big(g^2A^3+A''\Big)-\frac{12f_{XX}A'}{a^4}\Big(2g^2A^3\big(A\mH-A'\big)+A'\big(A''-2\mH A'\big)\Big)\right], \nonumber\\
m_4&=&-16\big(\mH^2+\mH'\big)f_X+\frac{192\mH f_{XX}A'}{a^4}\Big[2g^2A^3\big(A\mH-A'\big)+A'\big(A''-2\mH A'\big)\Big]\,.
\end{eqnarray}
The field equations for the tensor modes in this simple scenario already show non-trivial mixings in the contributions of friction, velocity, chirality and mass. One remarkable property of these models is that, as we already explained above, they can give a chiral mixing even when the original theory is parity-preserving. It is easy to understand that the origin for this chiral effect is precisely the Levi-Civita symbol in colour space in the non-Abelian piece of the field strengths. In fact, the chiral effect is proportional to the gauge coupling constant in that case. For this very simple model, the propagation speeds for the high frequency modes of the two tensor modes are the same and equal to the speed of light. However, this is an accident of only including $X_0$. By including higher contributions $X_n$, with $n>1$, from \eqref{SU2scalars} the propagation speed of $t_{ij}$ can be modified. As an example, the second order expansions of $X_2$ and $X_3$ contain terms like
\bea
X_2&\supset&\frac{32}{a^8}\left[\Big(3A'^2-g^2A^4\Big)\vert t'\vert^2-k^2\Big(A'^2-3g^2A^4\Big)\vert t\vert^2\right]\\
X_3&\supset&\frac{64}{a^8}\Big(g^2A^4\vert t'\vert^2+k^2A'^2\vert t\vert^2\Big)
\eea
that will clearly gives rise to propagation speed for $t_{ij}$ different from the speed of light. In order to modify the propagation speed of GWs one would need more contrived operators involving non-minimal couplings. Some of those operators also allow for the coupling mediated by $\tilde{\mathcal{N}}^{ab}$ in \eqref{ActionTensorModes}. For that we need to resort to the Horndeski Yang-Mills interaction given by $L^{\mu\nu\rho\sigma} \F^a{}_{\mu\nu} \F^{a}{}_{\rho\sigma}$, with $L^{\mu\nu\rho\sigma}=\frac14 \epsilon^{\mu\nu\alpha\beta}\epsilon^{\rho\sigma\gamma\delta} R_{\alpha\beta\gamma\delta}$ the double dual Riemann tensor, whose property of being divergenceless guarantees the absence of higher than second order derivatives in the field equations. However, the generation of the mixing mediated by $\tilde{\mathcal{N}}^{ab}$ from this interaction will come in accompanied by an anomalous $c_T^2$ (see e.g. \cite{BeltranJimenez:2017cbn}), thus jeopardizing the tight constraints on the propagation speed of GWs.

\subsection{Multi-Proca theories}

While the Yang-Mills theories of the precedent section already provide very interesting effects, the requirement of the gauge symmetry constrains the allowed operators. By abandoning the gauge symmetry, we can enlarge the theory space because a larger number of operators besides those given in \eqref{SU2scalars} are allowed. For instance, an arbitrary mass matrix can be present now, while in the Yang-Mills theories the mass terms can only arise from the background configuration. In other words, the helicity-2 sector in the Yang-Mills theories is subject to gauge symmetry constraints that will be reflected in special properties of the matrices governing the evolution and mixing of the tensor modes. Moreover, due to the lack of any gauge symmetry, we are no longer forced to use \eqref{eq:defFnA} to describe derivative operators for the vector fields. Thus, in this section we will use the usual $U(1)$ gauge invariant field strengths defined as
\be
F^a{}_{\mu\nu}=\partial_\mu A^a{}_\nu-\partial_\nu A^a{}_\mu.
\ee
Another important consequence of abandoning gauge symmetries is the appearance of new propagating degrees of freedom. However, in order to avoid ghostly degrees of freedom, we will assume that all the non-gauge invariant interactions are compatible with having, at most, three degrees of freedom per vector field. In general, giving up on the gauge symmetry allows the following three distinctive type of new operators
\begin{itemize}
\item {\it Non-derivative operators}. These correspond to ultra-local operators that describe non-derivative self-interactions such as e.g. $(A^a{}_\mu A^{a\mu})^n$.

\item {\it Partially gauge-invariant derivative operators}. By this we mean interactions that involve derivatives of the vector fields but only through their gauge-invariant field strengths as e.g. $(A^a{}_\mu A^{a\mu}) (F^b{}_{\mu\nu}F^{b\mu\nu})$. Obviously, these interactions are not gauge invariant, but they have the property that they identically vanish for purely longitudinal modes of the form $A^a{}_\mu=\partial_\mu\varphi^a$.

\item {\it Non-gauge invariant derivative operators}. These operators contain derivative self-interactions that are not required to respect any gauge symmetry not even for the derivatives of the vector fields. Examples of these operators are e.g. 
\be
(A^a{}_\mu A^{a\mu})^2\Big(\nabla_\mu A^{b\mu}\nabla_\nu A^{b\nu}-\nabla_\mu A^b{}_\nu\nabla^\mu A^{b\nu}\Big).
\ee
This type of interactions is the  most delicate case because they require very precise structures to avoid introducing ghost-like degrees of freedom (see e.g. \cite{Allys:2016kbq,Jimenez:2016upj,ErrastiDiez:2019ttn,ErrastiDiez:2019trb,Jimenez:2019hpl}).

\end{itemize}

In order to guarantee the existence of homogeneous and isotropic solutions with the field configuration given in \eqref{triad} we still need to impose an internal global $SO(3)$ symmetry, that is nevertheless a much weaker condition than requiring a gauge symmetry. We will consider again a very simple model where the oscillation takes place for illustrative purposes. Thus, let us consider  a Lagrangian of the form 
\be
\lag=f(Y,X)
\ee
with $Y\equiv A^a_\mu A^{b\mu}\delta_{ab}$ and $X=\delta_{ab}F^a{}_{\mu\nu} F^{b\mu\nu}$. The cosmological configuration for the fields will be again
\be
\bar{A}^a_\mu=A(t)\delta^a_\mu.
\ee
The evolution of the tensor modes for this case can be obtained from \eqref{deftensorA} by taking $g$ and adding the contributions from the $Y-$dependence. The equations now read
\begin{eqnarray}\label{eqTen_YM}
\left[ \bpm 1  & 0 \\ 0 &-16f_X   \epm\frac{\d^2}{\d\eta^2} +\bpm 2\mH  & -16A'f_X/(a\mpl) \\ 16A'f_X/(a\mpl) &\nu_t   \epm \frac{\d}{\d\eta}  +\bpm 1  & 0 \\ 0 &-16f_0   \epm k^2\right. \nonumber\\
\left.+\bpm m_1  & m_2 \\ m_3 &  m_4 \epm \right] \bpm h \\ t \epm =0\,,
\end{eqnarray}
with
\be
\nu_t=-32\mH f_X+\frac{192f_{XX}}{a^4}A'\big(A''-2\mH A'\big)-\frac{96f_{YX}}{a^4}(A'-\mH A)
\ee
and the components of the mass matrix are given by
\begin{eqnarray}
m_1&=&2\big(\mH^2+2\mH'\big)+2\frac{a^4f+8f_XA'^2-4a^2f_YA^2}{a^2\mpl^2}, \\
m_2&=&8\frac{-2f_X\mH A'+a^2f_Y A}{a\mpl }, \nonumber\\
m_3&=&\frac{8}{a\mpl}\left[2f_XA''-\frac{24f_{XX}A'^2}{a^4}\big(A''-2\mH A'\big)+a^2 Af_Y+\frac{12}{a^2}f_{YX}A A'\big(A'-\mH A\big)\right], \nonumber\\
m_4&=&-16\big(\mH^2+\mH'\big)f_X+\frac{192\mH f_{XX}A'^2}{a^4}\big(A''-2\mH A'\big)-8a^2f_Y-\frac{96A\mH f_{YZ}}{a^2}(A'-\mH A)\,.\nonumber
\end{eqnarray}
We see again a non-trivial mixing that will result in GW oscillations. Similarly to the Yang-Mills case, the apparently limited contributions is due to the extremely simple form of the chosen Lagrangian. More general interactions will evidently result in a much richer structure for the equations.

\subsection{Vanishing gauge coupling limit: Gaugids}
In section \ref{subsec_YM} we have considered models based on an $SU(2)$ symmetry and studied the explicit form of the equations of motion for the tensor modes with their characteristic terms. Interesting models also arise in the limit $g\to0$ of the gauge field coupling constant, where the initial $SU(2)$ symmetry factorizes into three copies of $U(1)$ gauge symmetries times a global $SO(3)$ invariance. In this case we can similarly construct the scalar quantities given in equations \eqref{SU2scalars} where the field strength becomes $F^a{}_{\mu\nu}=\partial_\mu A^a_{\nu}-\partial_\nu A^a_{\mu}$. This allows to introduce a different type of background field configurations than in section \ref{subsec_YM}. We can distinguish between two fundamentally different field configurations. 

First of all, in the triad configuration the background gives rise to the electric gaugid model with $F^a{}_{0i}=A'\delta_i^a$ in which case the tensor perturbations equations are the simplified version of those in the Yang-Mills theories with $g=0$. Since not much is gained for this configuration as compared to the Yang-Mills theories, we will not discuss it any further.

A more interesting field configuration that realizes the cosmological principle in a different way is the so-called magnetic gaugid, used in \cite{Piazza:2017bsd} to develop an interesting inflationary model. This configuration uses the three copies of $U(1)$ gauge symmetries in order to allow for an inhomogeneous field configuration of the form $A^a{}_\mu=\frac12B\epsilon^a{}_{i\mu}x^i$, with $B$ some constant, that gives the purely magnetic field strength $F^a{}_{ij}=B\epsilon^a{}_{ij}$. Since this background field configuration is fundamentally different from the triad configuration, the tensor perturbations will acquire very distinctive features and will not be related to the Yang-Mills models even in the $g\to0$ limit. A remarkable and distinctive property of the magnetic gaugid is that the background field does not evolve in time. As an illustrative example we will reproduce the results already computed in \cite{Piazza:2017bsd} for the particular action
\begin{equation}
\mS=\int\d^4x\sqrt{-g}\left[\frac{\mpl^2R}{2}-P(X_0)-(27M_1^4+18M_2^4)\frac{X_2}{X_1^2}+\frac{72M_2^4}{4}\Big(1+\frac{X_2}{X_1^2}+\frac{X_3}{X_1^2}\Big)\right]\,,
\end{equation}
where the quantities $X_n$ are given by the expressions in \eqref{SU2scalars} with $g=0$. This particular model was chosen in order to guarantee the stability of all the perturbations. The equations for the helicity-2 sector can be written as \cite{Piazza:2017bsd}:
\be
\lb \frac{\d^2}{\d\eta^2}  +\bpm 1 & 0 \\ 0 &  c_t^2\epm k^2\pm\bpm \mu_h  & \gamma \\ \gamma &  \mu_t \epm k +\bpm m_1  & 0 \\ 0 &  m_2 \epm \rb \bpm h \\ t \epm_{R,L} =0\,. 
\ee
where the explicit expressions for the coefficients of the matrices are given in \cite{Piazza:2017bsd} where it can be seen that they depend on the background evolution. We see again a chiral effect even if the action is not parity violating analogous to the Yang-Mills case. The origin of the chirality this time is the Levi-Civita tensor in the background configuration. Furthermore, we see that $t_{ij}$ features a propagation speed $c_t^2\neq1$.

After reviewing some scenarios giving rise to oscillations of GWs, we will now proceed to analyzing the phenomenology and potential observational signatures of such scenarios.

\section{Phenomenology}
\label{sec:pheno} 

So far we have shown that the coupled evolution of two tensor perturbations (\ref{eq:generalequation}) can modify their propagation in several different ways: 
\begin{enumerate}[\itshape(i)]
\item their amplitudes mix whenever there are non-diagonal terms in the mass, friction, velocity or chiral matrices;
\item their amplitude can get damped due to the friction matrix;
\item the propagation speed of the perturbation coupled to matter could be anomalous whenever the second perturbation has a non-luminal propagation and there is a mixing;
\item each polarization propagates differently when there is a chiral matrix. 
\end{enumerate}
Moreover, we have seen that this type of propagation equations arises in cosmological set-ups with multiple vector fields and in many classes of modified gravity theories. 
For instance, bigravity leads to a mass mixing while cosmological gauge fields yields to a friction and chiral mixing. More sophisticated vector-tensor theories can even induce a velocity mixing. 

In this section we are going to investigate the phenomenological implications of these effects on GW observations. In order to discuss each possible observable, we will consider representative examples. Here, we do not aim at setting firm constraints on particular theories but rather show the potential of GW oscillations to test certain classes of models. A detailed analysis solving the full cosmological evolution together with GW propagations for each particular example would be necessary for that task and is left for the interested readers to test their favorite theory.

The mixing of the amplitude of the different tensor perturbations has clear consequences for the GW signals. Even if we start only with perturbations of one class at emission, we will generically have both of them excited at detection. Since only one of the perturbations, $h$ in our convention, couples to matter, the excitation of the other perturbation $t$ would be seen in the detector as a deficit of $h$ signal. If the conversion of $h$ into $t$ continues periodically, this will induce an oscillation of the GW wave-form. We will study this characteristic effect in section \ref{sec:oscillation_wave-form}. But, if the amplitude detected is lower, this would be interpreted as the source being further away. Therefore, there will be also a modification of the GW luminosity distance that we analyze in section \ref{sec:modified_dL}. Now, if $t$ propagates at a speed different from the speed of light $c$, the mixing causes that the net propagation speed of $h$ becomes anomalous. This can be constrained with multi-messenger detections as we discuss in section \ref{sec:anomalous_speed}. Finally, for the case in which the mixing is chiral, the two tensor polarizations $h_{+,\times}$ evolve differently leaving an imprint that could be distinguished with a network of ground based detectors. We study this imprint in section \ref{sec:chirality}.

In addition to these characteristic phenomena associated to the GW oscillations, it is important to note that there will be also modifications to the dispersion relation.\footnote{The modification of the dispersion relation can be read directly from the eigenfrequencies that we have computed in each particular case: (\ref{eq:t12_mass}) for a mass mixing, (\ref{eq:t1_friction}-\ref{eq:t2_friction}) for a friction mixing, (\ref{eq:t12_velocity}) for a velocity mixing and (\ref{eq:t1_chiral}-\ref{eq:t2_chiral}) for a chiral mixing.} Such modified dispersion relation can be constrained as usual with the time evolution of the frequency of the GW. Because this does not constitute a unique probe of GW oscillations we do not discuss it further here, but they should not be forgotten when considering the detectability of a given model.

\subsection{Oscillations of the wave-form}
\label{sec:oscillation_wave-form}

The mixing of the tensor perturbations causes that the GW strain of the signal emitted is modified during the propagation. This modification will depend on the particular theory and on the location of the source. To exemplify this effect, we are going to consider two representative examples. 

On the one hand, we are going to investigate a scenario in which there is a \emph{mass mixing} whose time dependence is proportional to the square of the scale factor, i.e. $\mM\propto a(\eta)^2\mM_0$. We will work in the high-$k$ limit in which $\theta_1\simeq\theta_2\simeq k$. Accordingly, we parametrize the problem with an effective mass $m_g$ and a mixing angle $\theta_g$ which are constants, recall (\ref{eq:effective_mass}) and (\ref{eq:mixing_angle}) respectively. This example resembles bigravity theory in the large mass limit. See Section \ref{subsec:bigravity} and \cite{Belgacem:2019pkk}. 
The transfer function between the initial GR emission $\vert h_{_{GR}}\vert$ and the signal detected $\vert h(z)\vert$ is given by
\be
\frac{\vert h(z,k)\vert}{\vert h_{_{GR}}\vert}=\cos^2\theta_g\left(1+\tan^4\theta_g+2\tan^2\theta_g\,\cos\left[\frac{m_g^2}{2k}\int_{0}^z\frac{dz}{(1+z)^2H(z)}\right]\right)^{1/2}\,.
\ee
Importantly, the transfer function depends on the parameters of the model, $m_g$ and $\theta_g$, the redshift $z$ and the frequency $k$. In addition, the modified amplitude is also sensitive to the cosmic expansion history through $H(z)$. As observations suggest and for simplicity, we have assumed that the background cosmology is $\Lambda$CDM.

On the other hand, we will work with an example with a \emph{friction mixing}. In this case, we consider that the friction matrix $\nM$ is constant, which corresponds to (\ref{eq:amplitude_constant_friction}). The transfer function
\be
\frac{\vert h(z)\vert}{\vert h_{_{GR}}\vert}=(1+z)^{-\Delta\nu}\left(\cos\lb\omega_\nu\,\log(1+z)\rb+\frac{\Delta\nu}{\omega_\nu}\sin\lb\omega_\nu\,\log(1+z)\rb\right)
\ee
is then controlled by the frequency of oscillation $\onu$ and the damping factor $\dnu$, where $\onu$ is a function of both the non-diagonal entry $\alpha$ and $\dnu$. Noticeably, now the amplitude does not depend on the frequency of the GW.

\begin{figure}[t!]
\centering 
\includegraphics[width=.49\textwidth]{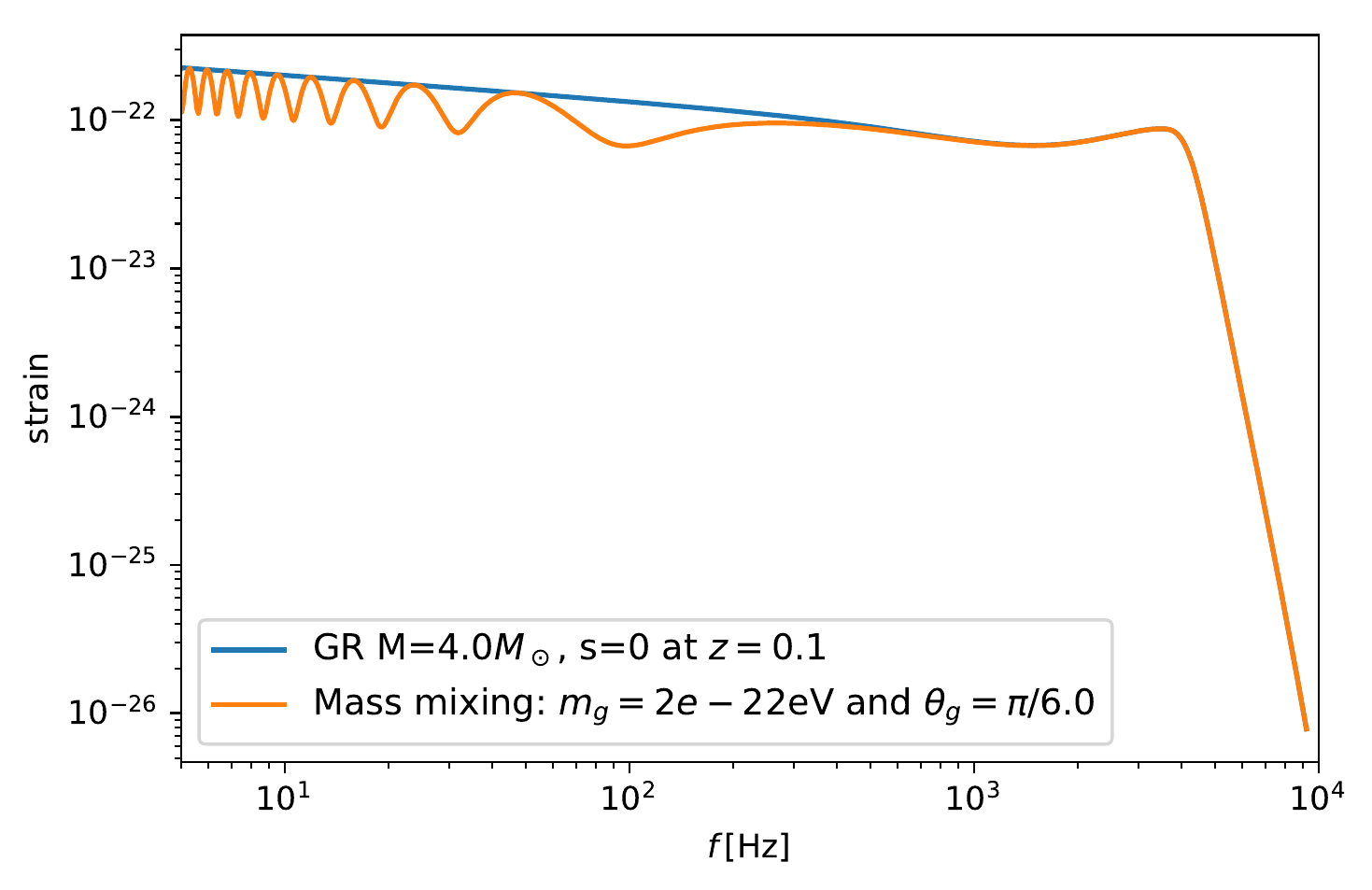}
\includegraphics[width=.49\textwidth]{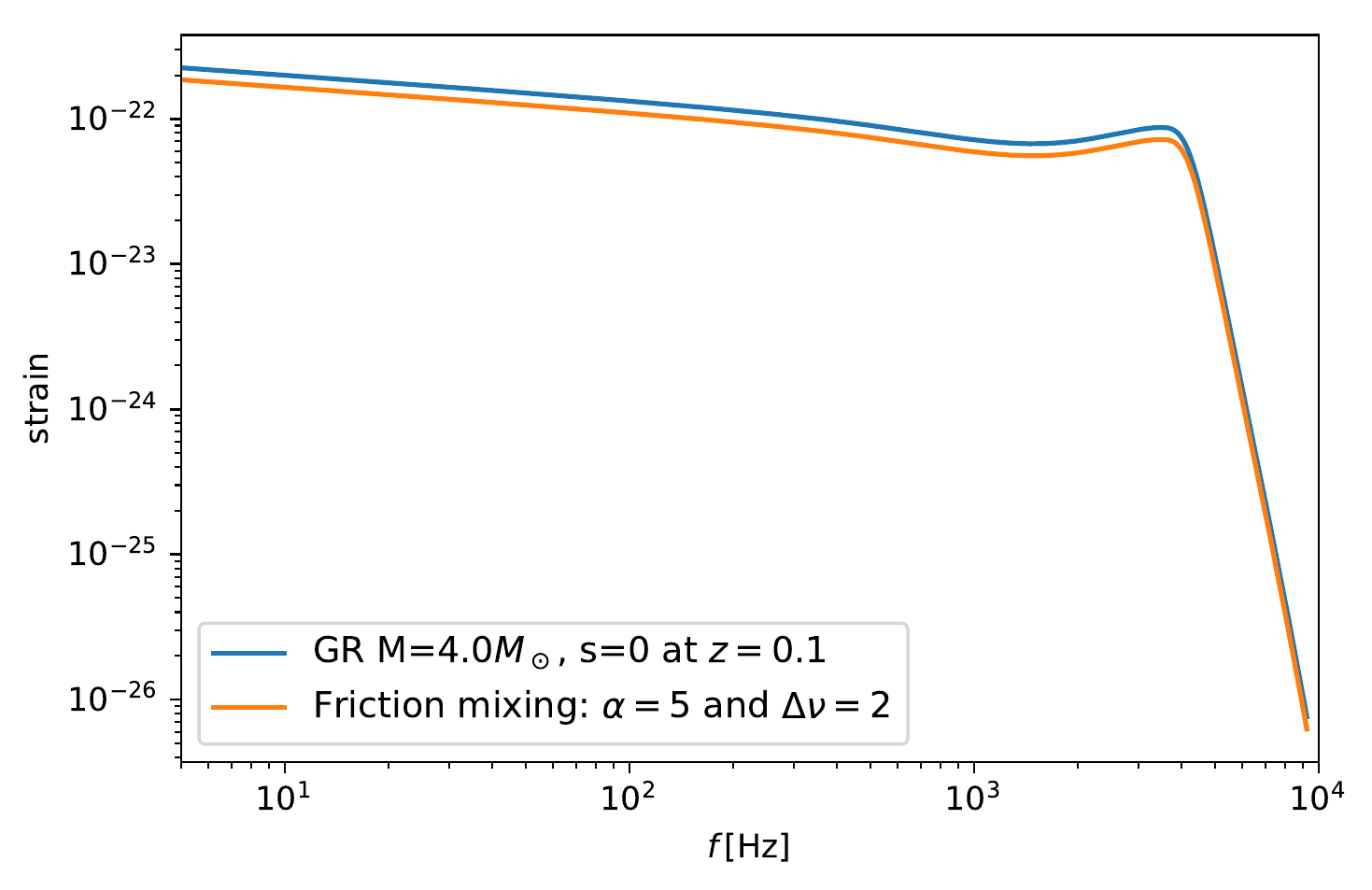}
\includegraphics[width=.49\textwidth]{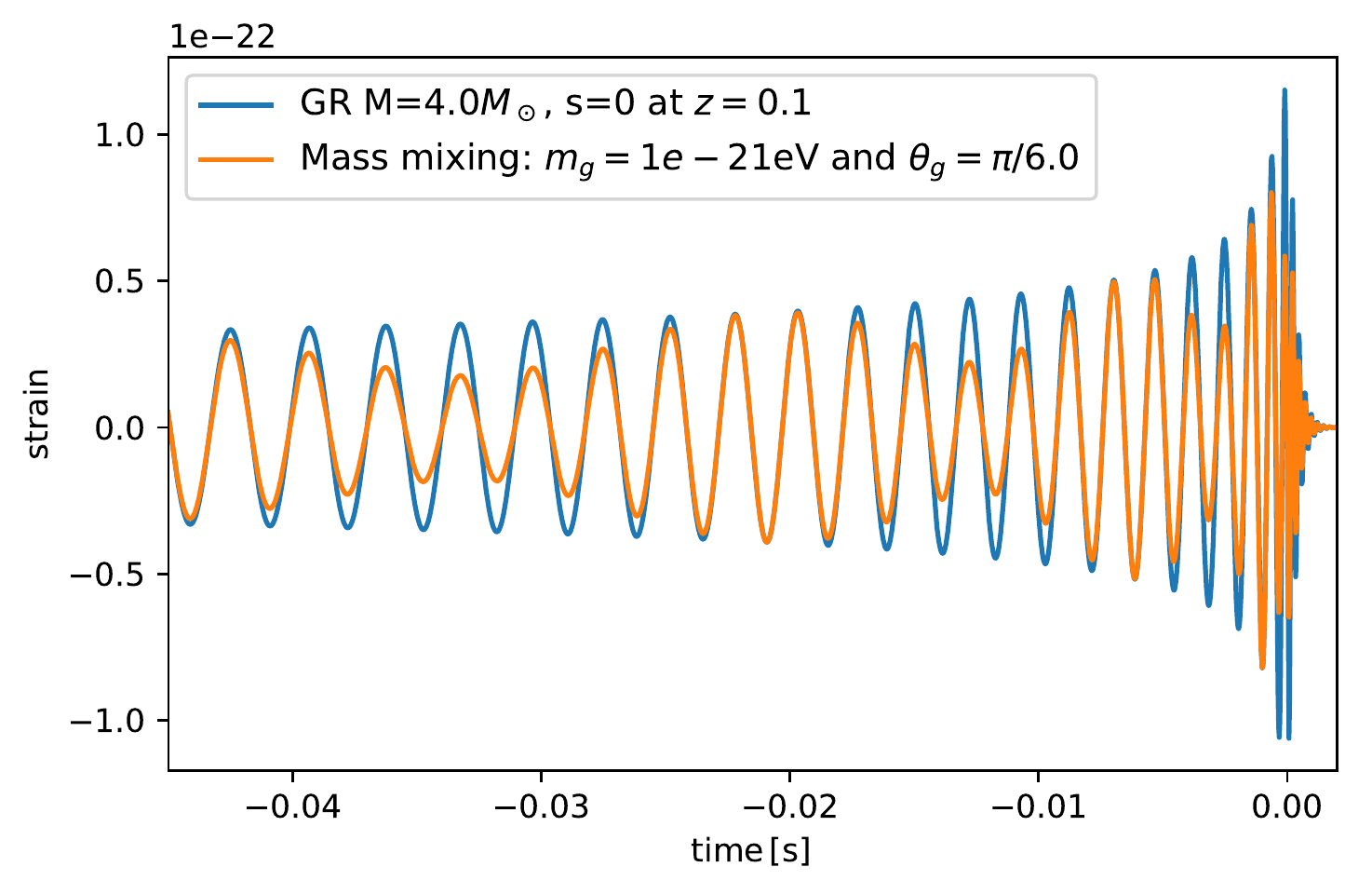}
\includegraphics[width=.49\textwidth]{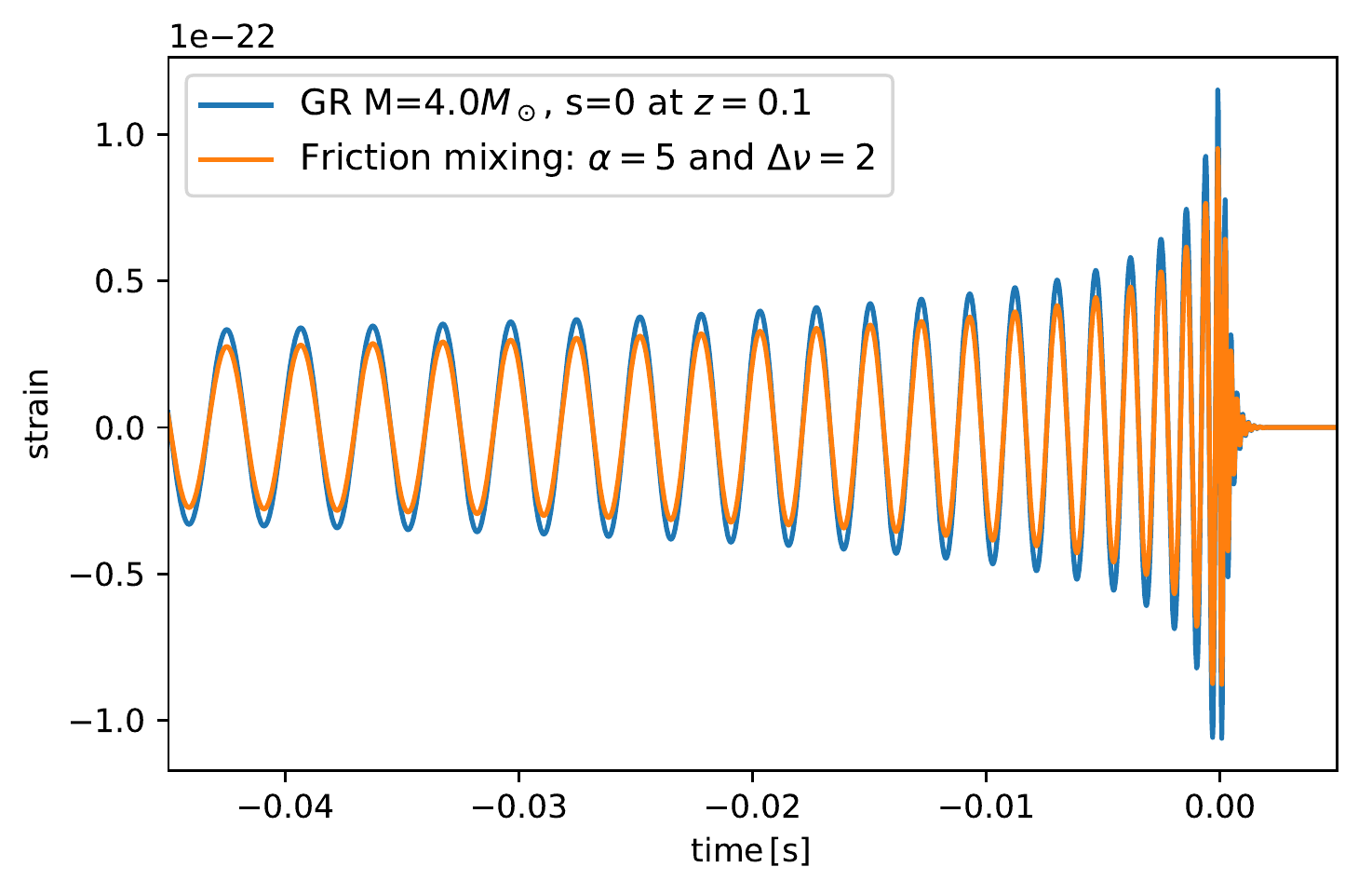}
 \caption{Examples of modified wave-forms in the frequency (top) and time (bottom) domain for theories with a mass mixing (left) or a friction mixing (right). The emitted GR signal (blue) corresponds to an equal mass, non-spinning BBH merger of $M=4M_{\odot}$ at $z=0.1$.}
 \label{fig:modified_strain}
\end{figure}

In Fig. \ref{fig:modified_strain} we compare the modification of the GW strain for the mass and friction mixing scenarios. We plot the modified strain on top of the original GR signal both in the frequency and time domain. As it can be clearly seen, the fact that the mass mixing transfer function depends on the frequency makes the strain to oscillate leaving a very distinct wave-form. On the contrary, for the friction mixing there is only a dimming of the signal that globally rescales the amplitude. This effect is completely degenerate with the distance to the source and could not be distinguished through wave-form modeling. Still, since the transfer function depends on redshift, a friction mixing leaves a measurable imprint in the GW luminosity distance as we will see in section \ref{sec:modified_dL}.

Focusing on the mass mixing case, the oscillatory pattern in the GW strain could be used to constrain the parameters of the theory by comparing this wave-form with the observed ones. In this sense, the best target will be a signal with a long inspiral part, allowing to constrain the strain over many oscillations. In the context of present ground-based detectors, long binary neutron star signals like GW170817 have more constraining power than short binary black-hole detections like GW150914. Moreover, with the future space-based detector LISA, we could be sensitive to much lower frequencies of oscillations. We will also benefit from very long signals that could last months or years. Eventually, a multi-band GW detection could be as well a very powerful test of this type of mixing.

Lastly, let us emphasize the importance of searching for possible astrophysical degeneracies that could mimic this fundamental oscillatory pattern. In particular, we note that binaries with precessing spins also lead to an oscillation of the wave-form \cite{Apostolatos:1994mx}. The oscillation effect is enhanced when there is a hierarchy in the masses of the compact objects. Thus, this will be more relevant to LISA sources. This possible degeneracy could be broken by observing several events covering different masses and spins. This is because the precessing spin effect is linked to the characteristics of the binary while the GW oscillation modulation is universal for a given cosmological distance.

\subsection{Modified GW luminosity distance}
\label{sec:modified_dL}

If part of the initial GW signal is converted into the second tensor or diluted by the friction term, this would be interpreted as the source being further away since the received signal would be dimmer. In other words, the GW luminosity distance $\dLgw$ would be modified.\footnote{Modifications of the GW luminosity distance $\dLgw$ are ubiquitous in the landscape of modify gravity and not restricted to scenarios with multiple tensor modes. In fact, modifications in the cosmological background where GWs propagate can make $\dLgw$ to differ from the GR prediction. For recent analysis of these class of models see Refs. \cite{Saltas:2014dha,Lombriser:2015sxa,Nishizawa:2017nef,Belgacem:2017ihm,Belgacem:2018lbp,Belgacem:2019pkk} (or \cite{Ezquiaga:2018btd} for a review).} 
We can obtain the GW luminosity distance from the inverse of the amplitude 
\be
\vert h_{+,\times}\vert=\frac{\mathcal{M}_c^{5/3}f^{2/3}}{\dLgw}F_{+,\times}\,,
\ee
where $\mathcal{M}_c$ is the chirp mass, $f$ the frequency and $F_{+,\times}$ is a polarization dependent function of the inclination angle. Recall that in GR the GW luminosity distance is equal to the EM one and determined by the Hubble parameter
\be
d_{L}^{\text{GR}}=\dLem=(1+z)\int_0^z\frac{c}{H(z)}dz\,.
\ee
Because the amount of damping of the signal depends on the distance travelled, this effect would be different for GWs emitted at different redshift. Then, it is necessary to have multiple detections to constrain this modification of the propagation. 

There are two ways in which we can test it. With GWs alone, we could use the merger rate of compact binaries as a function of redshift $R(z)$. If there is an additional friction term (induced by the diagonal terms of $\nM$), the amplitude of the GWs at the detector will be lower and, as a consequence, less events would be detected. Comparing the observed rate with the theoretical prediction from a given astrophysical model could constrain this modification in the propagation. However, given the intrinsic uncertainty in the theoretical modeling of the merger rate, this is not very promising. Moreover, the effect of the additional friction term would be degenerate with $H_0$ \cite{Lagos:2019kds}. 
On the contrary, when there is a mixing of the perturbations, for given redshifts periodically separated, the number of detections will be much smaller than predicted. This distinct pattern could be very well distinguished from an astrophysical effect. In the extreme case in which there is a complete conversion of $h$ into $t$ (see for instance Fig. \ref{fig:mass_mixing}), this implies that $R(z)$ at certain redshift bins would be zero. Having a large population of compact binaries over a wide range of redshift could bound the mixing of the GWs with other tensor modes.

\begin{figure}[t!]
\centering 
\includegraphics[width=.49\textwidth]{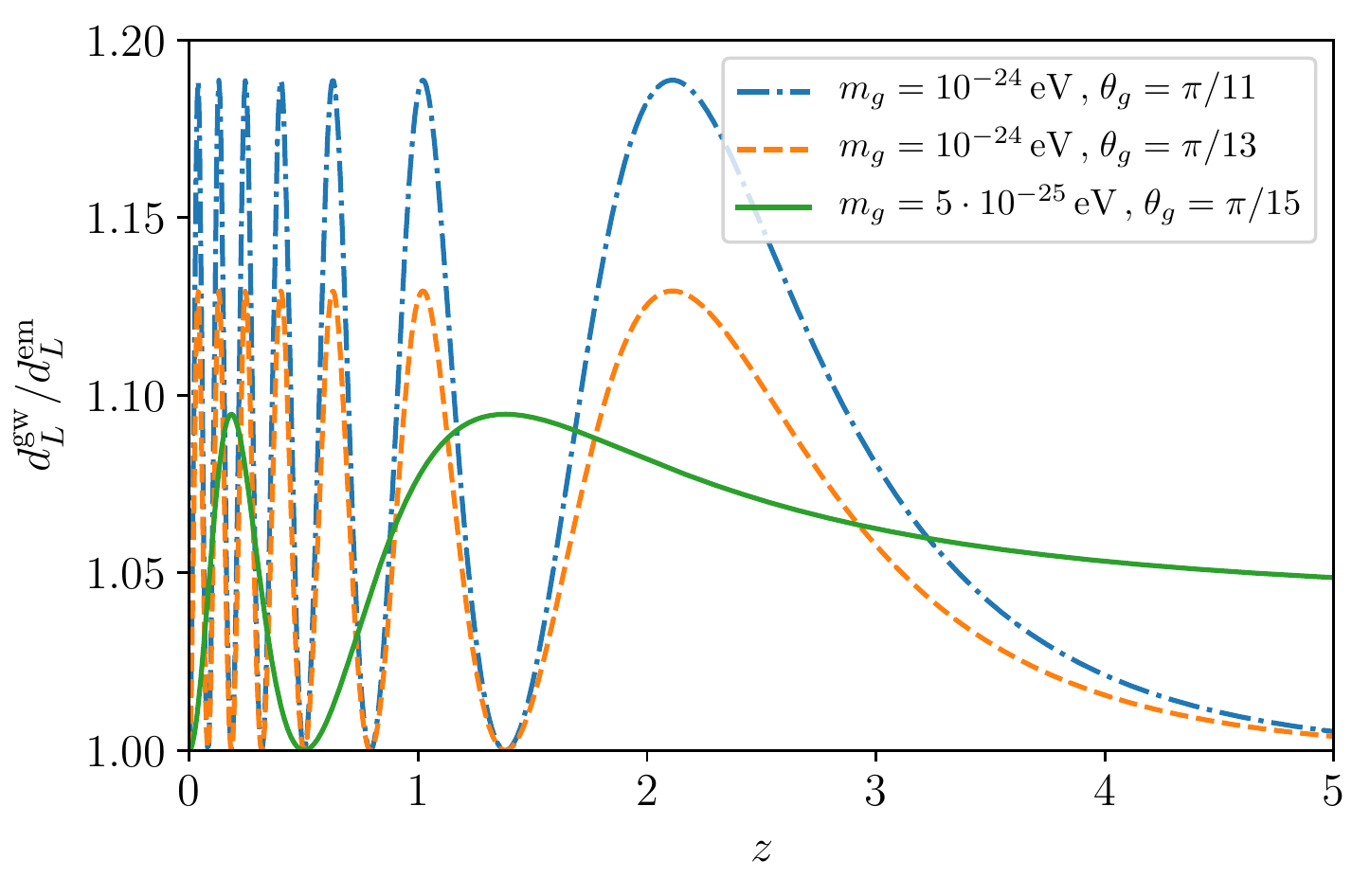}
\includegraphics[width=.49\textwidth]{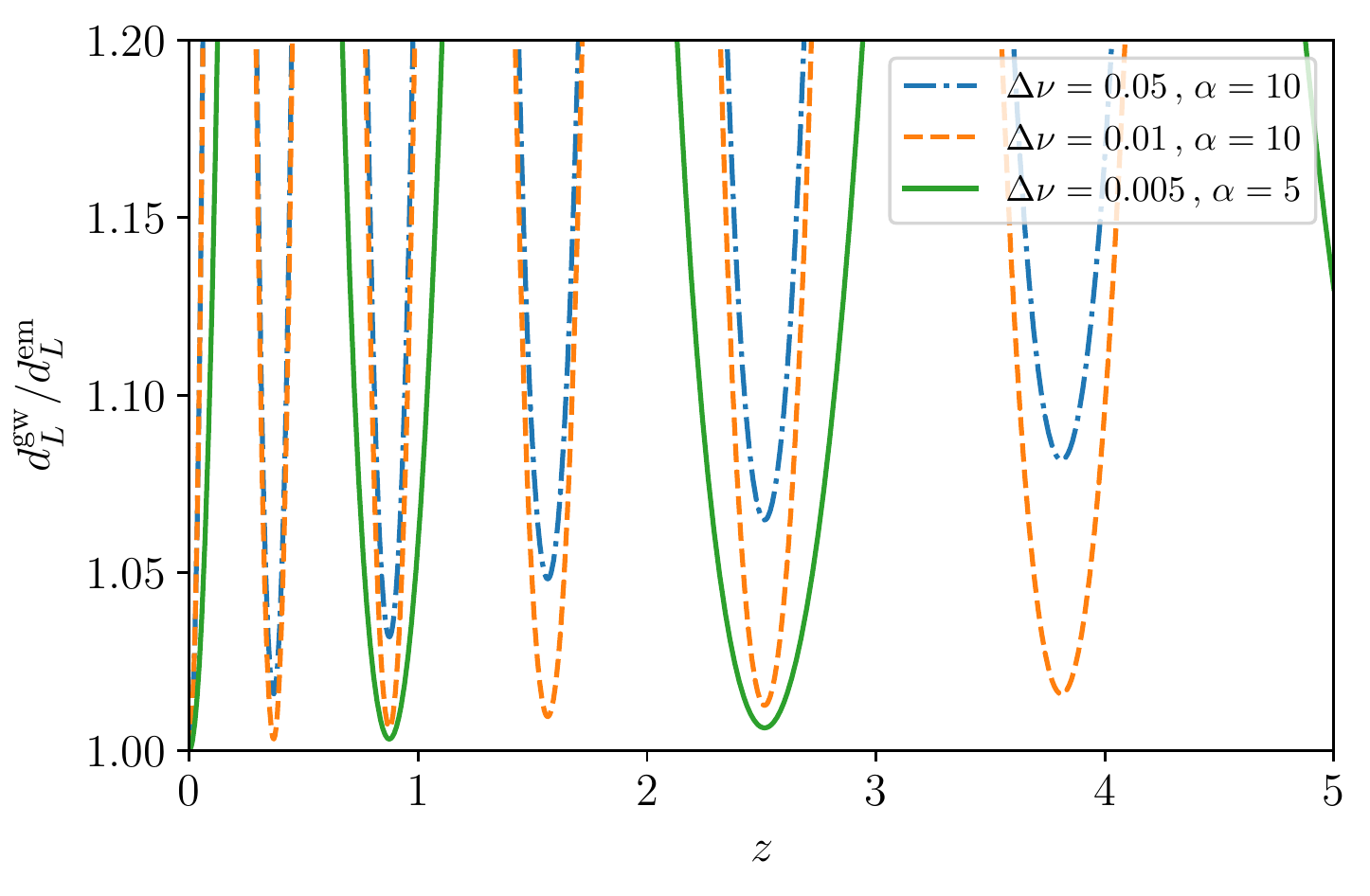}
 \caption{Modified GW luminosity distance as a function of redshift for theories with a mass mixing (left) or a friction mixing (right). We plot the ratio $\dLgw/\dLem$ for different values of the effective mass $m_g$ and mixing angle $\theta_g$; and damping factor $\dnu$ and friction mixing $\alpha$.}
 \label{fig:modified_dL}
\end{figure}

With multi-messenger events, we can constrain this effect further. Either by a direct EM counterpart or a statistical analysis, if we can determine the EM luminosity distance of the source, we can then test the ratio $\dLgw/\dLem$. Any deviation of this ratio from being 1 would be a smoking gun for physics beyond GR in the standard model of cosmology $\Lambda$CDM. Directly from the transfer function in the GW amplitude we can compute the ratio of luminosity distances. For the mass mixing described in the previous section it would be given by
\be
\frac{\dLgw}{\dLem}=\frac{1}{\cos^2\theta_g}\left(1+\tan^4\theta_g+2\tan^2\theta_g\,\cos\left[\frac{m_g^2}{2k}\int_{0}^z\frac{dz}{(1+z)^2H(z)}\right]\right)^{-1/2}\,.
\ee
Possible mixing angles span from 0 to $\pi/2$, having the maximum mixing at $\theta_g=\pi/4$. At this value, a complete conversion of $h$ into $t$ can occur. As a consequence, the amplitude of $h$ vanishes and $\dLgw$ diverges at 
\be
\frac{m_g^2}{2k}\int_{0}^z\frac{dz}{(1+z)^2H(z)}=\frac{\pi}{2}+2\pi n \quad \text{for} \quad n=0,1,2,\cdots
\ee
Note also that the parameter space is symmetric around $\pi/4$. The frequency of oscillation is controlled by the effective mass and the frequency of the GW through $m_g^2/k$. To have a frequency of oscillation of order 1 at low redshift, this ratio should be of order $H_0$ to compensate the Hubble parameter in the denominator.\footnote{The reader should remember that $H_0\sim10^{-33}\text{eV}\sim10^{-18}\text{Hz}$.} This implies that with present ground-based interferometer, $f_{\text{LIGO}}\sim100$Hz, we can test $m_g\sim 10^{-23}$eV.\footnote{This is the same order of magnitude that present LIGO constraints on a mass term in the dispersion relation $w(k)^2=c^2k^2+m_g^2$ \cite{LIGOScientific:2019fpa}.}  In the same manner, the future space-based detector LISA, $f_{\text{LISA}}\sim10$mHz, will be sensitive to $m_g\sim 10^{-25}$eV.

For the friction mixing example, the ratio of the GW and EM luminosity distance becomes
\be
\frac{\dLgw}{\dLem}=(1+z)^{\Delta\nu}\left(\cos\lb\omega_\nu\,\log(1+z)\rb+\frac{\Delta\nu}{\omega_\nu}\sin\lb\omega_\nu\,\log(1+z)\rb\right)^{-1}\,.
\ee
The first term introduces a global friction term that increases $\dLgw$, while the rest makes the luminosity distance to oscillate at a rate determined by $\onu=\sqrt{\alpha^2-\dnu^2}$, where $\alpha$ is the non-diagonal term producing the mixing and $\dnu$ is the difference between the friction term of $t$ and $h$. Note that since we are taking the ratio of $\dLgw$ over $\dLem$, the standard damping due to the cosmic expansion does not appear in this expression. Moreover, in the limiting case in which there is no mixing, $\alpha\rightarrow0$, we recover $\dLgw/\dLem\rightarrow1$. Contrary to the mass mixing case, there is always a complete conversion of one type of perturbation into the other happening at
\be
\tan\lb\omega_\nu\,\log(1+z)\rb=-\frac{\omega_\nu}{\Delta\nu}\,.
\ee
This extreme feature in $\dLgw/\dLem$ makes it easier to probe.

In Fig. \ref{fig:modified_dL} we have plotted the ratio $\dLgw/\dLem$ for both the mass mixing and friction mixing scenarios as a function of redshift. In the left panel one can observe how the GW luminosity distance varies with $\theta_g$ and $m_g$. As the mixing angle approaches to $\pi/4$, the amplitude of the oscillation increases. Accordingly, the frequency of oscillation increases with $m_g$. On the right panel we present the corresponding plot for the friction mixing. Noticeably, the ratio diverges periodically due to the complete conversion of the original signal into tensor perturbations not coupled to matter. Also, the global friction term $(1+z)^{\dnu}$ makes the minimum values of $\dLgw/\dLem$ to increase away from 1 as $\dnu$ increase. On the other hand, the mixing parameter $\alpha$ controls the frequency of oscillation. 

Detecting standard sirens over different redshift ranges allows to cover a larger patch of the parameter space. Second generation ground-based interferometers are sensitive to BNS up to $z\sim0.05$ and BBH up to $z\sim0.5$. This range will be much increased with third generation detectors such as Einstein Telescope, reaching possibly $z\sim2$ for BNS and $z\sim15$ for BBH. LISA from space could also hear up to very high redshifts. Because of the precision and high redshift signal that future detectors will observe, lensing of GWs will be a relevant factor for parameter estimation, potentially affecting the inference of any modification of the GW luminosity distance. However, for the distinct, oscillatory features that we are considering here, we expect that the determination of the mixing parameters will not be affected significantly by lensing.

Fig. \ref{fig:modified_dL} corresponds to the expected redshift range, $z\sim2-6$, and sensitivity, $\Delta d_L/d_L\sim10\%$, where LISA could detect standard sirens from super massive BHs with EM counterparts. In fact, the capability of LISA to detect modifications in $\dLgw$ was recently studied in detail in \cite{Belgacem:2019pkk}. For the case of bigravity, it was shown LISA could probe masses of $m_g\gtrsim 2\cdot10^{-25}$eV with mixing angles $0.05\pi\lesssim\theta_g\lesssim0.45\pi$. Here, we show that LISA could also probe scenarios with a friction mixing.

\subsection{Anomalous GW speed}
\label{sec:anomalous_speed}

So far, we have focused on modifications of the amplitude of GWs due to the mixing of the tensor modes coupled to matter with other cosmological tensor fields. Nonetheless, GW oscillations can also modify the phase of the GW, which is indeed much better constrained with interferometers than the amplitude. One of the results of our analysis was to demonstrate that even if the tensor mode coupled to matter $h$ propagates at the speed of light $c_h=c$, if there is a mixing and the second tensor propagates at a different speed, the effective velocity of the GW could be non-luminal, $\cgw\neq c$. We have shown this explicitly for the mass-mixing (\ref{eq:agw_mass}) and velocity mixing (\ref{eq:agw_velocity}) scenarios.

An anomalous speed, which can be parametrised by $\agw=\cgw^2/c^2-1$, yields to a delay between the GW and any other EM counterpart $\Delta t$. After GW170817, which was followed by GRB170817A just $1.74\pm0.05$s later \cite{Monitor:2017mdv}, we know that $\agw$ is constrained to the level of $10^{-15}$ at LIGO frequencies. Assuming that $h$ propagates at the speed of light, this multi-messenger event constrains the possible deviation from $c$ of the second tensor whenever there is mixing. Suppressing the mixing could also be a way to avoid this limit but then there will be no other GW oscillation effects in the amplitude. 

Present constraints on the propagation speed of GWs could be improved in the future by observing more distant events (remember that GW170817 was only at about 40Mpc). The increase in the sensitivity and distance reach will be more significant when moving from second to third generation interferometers. Also promising candidates are the SMBH standard siren at high redshift that LISA target to detect, although it is still not clear if prompt emission could be observed for such distant sources. Interestingly, LISA also provides the opportunity to test the propagation speed of GWs at a different frequency range, which is relevant to constrain a possible frequency dependence in $\cgw$ \cite{deRham:2018red,Ezquiaga:2018btd}. A secure test of the speed of GWs at mHz is to measure the phase lag between GW and EM radiation of LISA verification binaries \cite{Finn:2013toa,Bettoni:2016mij}.

\subsection{Chirality}
\label{sec:chirality}

Finally, let us examine how to probe modifications in the propagation of the different polarizations $h_{+,\times}$ due to chiral GW oscillations. For that, one needs to be sensitive to each polarization. However, in general, the polarizations are degenerate with the location in the sky and the inclination angle. Having a network of ground-based detectors across the surface of the Earth can break this degeneracy. Moreover, if the source is located with an EM counterpart, the capability to probe different polarizations increases. Since the two LIGO detectors are aligned to maximized the joint sensitivity, the role of Virgo has been crucial to start performing tests of the types of polarizations. For instance, with the three-detector detection GW170814 \cite{Abbott:2017oio} it was possible to contrast the hypothesis of the signal being purely tensor against it being purely vector or scalar. A much stronger result favoring purely tensor polarizations was obtained with GW170817 \cite{Abbott:2018lct,LIGOScientific:2019fpa} since the sky position was determined with high accuracy thanks to the EM counterparts.

In order to test chiral GW oscillations it would be needed to distinguish each polarization and measure their amplitude as a function of frequency $h_{+,\times}(f)$. Alternatively, one could measure the luminosity distance as a function of redshift for each polarization. In principle, with a network of detectors these effects could be probed. However, one should remember that this is going to be a small effect since it is suppressed by the wavenumber. In fact, in the large-$k$ (or shortwave) limit, we have seen that the chirality $\chi_h$ vanishes (see discussion in section \ref{sec:chiral_mixing}). We leave the study of particular scenarios in which the chiral mixing is enhanced for future work.

\section{Conclusions}\label{sec:conclu}

Gravitational wave astronomy has opened a new window to explore the cosmos, its evolution and its different components. In this work we have studied how GWs are sensitive to cosmological fields that behave as tensor modes. These tensor perturbations could be linked to the fundamental nature of the additional cosmological fields or to the re-arragement of their internal symmetries with the background isometries. In any of these cases, these new modes could mix over cosmological scales with the tensor perturbations of the metric coupled to matter, leading to \emph{GW oscillations}.

Additional cosmological tensor perturbations could arise in theories with multiple dynamical metrics but also in scenarios with gauge fields. We survey a landscape of theories including  bigravity, multi-Proca and gaugid models. We show that GW oscillations is a quite generic phenomenon that can be realized in different interaction terms. For this reason, we adopt an agnostic and phenomenological perspective and study the phenomenon of GW oscillations in full generality, including all possible mixings. In order to solve the coupled evolution of the tensor modes, we develop two approximation schemes based on a WKB approximation and a large-$k$ (or short-wave) expansion. To gain further insight on the effect of each possible mixing, through the friction, velocity, chiral and mass matrices, we analyze them individually. We provide analytical solutions for the amplitude and phase of the GW.

We find a rich phenomenology associated to GW oscillations. If the mixing depends on the frequency, as it is the case of a mass mixing, the GW wave-form will be modulated according to the period of oscillation. More generically, since the emitted GW will transform into the second tensor and back, the amplitude detected will vary with the distance. This leads the GW luminosity distance to oscillate in redshift. As a cumulative effect, this will be best constrained with LISA \cite{Belgacem:2019pkk}. In addition, we find that if the extra mode does not propagate at the speed of light and there is a mixing, there will be a net anomalous propagation speed for the GWs. Lastly, for chiral mixings, the amplitude of each polarization will evolve differently, becoming a target for a network of ground-based interferometers.  

Looking to the future, our results and framework could be applied and extended in several manners. For example, the formalism developed in this work could be used to study concrete cosmological scenarios enabling to set new constraints on their additional cosmological fields. In particular, we have shown that GW oscillations are a common phenomenon in theories with multiple cosmological vector fields. A rich realm basically unexplored except for \cite{Caldwell:2016sut,Caldwell:2018feo}. Moreover, the phenomenology that we have described could be empowered by devoting specific analysis for different GW detectors. Another possible direction to extend our framework is to give up on some of our assumptions. For instance, abandoning the isotropy of the background\footnote{In the late time universe, violations of isotropy are only permitted at the $10^{-3}$ level by the CMB dipole. If the dipole is assumed to be entirely due to our motion with respect to the CMB, then isotropy can only be violated at the $10^{-5}$ level.} allows for a much richer phenomenology including the mixing of GWs with other helicity modes mediated by the background-isotropy-breaking operators. Along these lines, a particularly interesting case where our methodology could be applied is in the study of the GW propagation over spatially dependent backgrounds such as the ones generated by models featuring screening mechanisms.

Finally, although we have focused on the effect of GW oscillations on GWs from resolvable compact binaries, this work could be extended to stochastic GW backgrounds. Other extensions such as the inclusion of modifications in the emission, new couplings to matter or the propagation over modified cosmological backgrounds are left for future work.

\section*{Acknowledgments}

We are grateful to Max Isi for his comments on the draft. 
JBJ acknowledges support from the {\textit{ Atracci\'on del Talento Cient\'ifico en Salamanca}} programme and the projects PGC2018-096038-B-I00, FIS2014-52837-P and FIS2016-78859-P (AEI/FEDER). JME is supported by NASA through the NASA Hubble Fellowship grant HST-HF2-51435.001-A awarded by the Space Telescope Science Institute, which is operated by the Association of Universities for Research in Astronomy, Inc., for NASA, under contract NAS5-26555. He is also supported by the Kavli Institute for Cosmological Physics through an endowment from the Kavli Foundation and its founder Fred Kavli. 
LH is supported by funding from the European Research Council (ERC) under the European Unions Horizon 2020 research and innovation programme grant agreement No 801781 and by the Swiss National Science Foundation grant 179740. 
This article is based upon work from COST Action CA15117, supported by COST (European Cooperation in Science and Technology). 
This research has made use of data, software and/or web tools obtained from the LIGO Open Science Center (https://losc.ligo.org), a service of LIGO Laboratory, the LIGO Scientific Collaboration and the Virgo Collaboration. 
In particular, we have used pyCBC \cite{alex_nitz_2019_3546372} to generate the General Relativity templates displayed in Fig \ref{fig:modified_strain}. 
LIGO is funded by the U.S. National Science Foundation. Virgo is funded by the French Centre National de Recherche Scientifique (CNRS), the Italian Istituto Nazionale della Fisica Nucleare (INFN) and the Dutch Nikhef, with contributions by Polish and Hungarian institutes. 

\appendix

\section{Review on the analogue one dimensional problem}
\label{app:1D}

As a warm-up exercise for the formalism developed in section \ref{sec:genFrame}, we present in this appendix the analogue one dimensional problem. This corresponds to solving the evolution of a second order differential equation with time-varying parameters 
\be
\phi''+\nu(\eta)\phi'+(c(\eta)^2k^2+\pi(\eta) k + m(\eta)^2)\phi=0\,,
\ee
where we can define $\omega(\eta)^2=c(\eta)^2k^2+\pi(\eta) k + m(\eta)^2$. 
We apply to this problem the two different approximations schemes that we use in sections \ref{sec:WKB} and \ref{sec:largek}, a WKB approximation and a large-$k$ expansion respectively. The main simplification will certainly be that we do not have to deal with matrices and worry about their commutation.

\subsection{WKB expansion for one variable}

Within the WKB approximation we state that the time variation of the parameters is much smaller than the frequency of the wave. We can formalize this introducing a dimensionless, small parameter $\epsilon$ suppressing the time derivatives
\be
\epsilon^2\phi''+\epsilon\nu(\eta)\phi'+(c(\eta)^2k^2+\pi(\eta) k + m(\eta)^2)\phi=0\,.
\ee
We can solve this problem using a plane-wave ansatz
\be
\phi=e^{\frac{i}{\eps}\int\theta d\eta}\lp\phi_0 +\epsilon\phi_1 + \cdots\rp\,,
\ee
where we decompose the amplitude in serie of $\epsilon$. Solving in increasing powers of $\epsilon$, we obtain
\begin{align}
&\eps^0:  & &\lp \omega^2 + i \nu\theta-\theta^2\rp \phi_0=0\,, \\
&\eps^1:  & &\lp2\theta-i\nu\rp\phi'_0+\theta'\phi_0=0\,, \\
&\eps^2: & &\phi_0'' +i\lp2\theta-i\nu\rp\phi'_1+i\theta'\phi_1=0\,.
\end{align} 
The first equation determines the phase of the wave $\theta$ which reads
\be
\theta_{1,2}=\frac{i\nu\pm\sqrt{4\omega^2-\nu^2}}{2}\,.
\ee
The second equation fixes the leading term in the amplitude
\be
\phi_{0}=c_0\,e^{-\int\frac{\theta'}{2\theta-i\nu}d\eta}\,,
\ee
where $c_0$ is a constant. In the absence of friction, the scaling is just $\phi_0\sim 1/\sqrt{\theta}$. 
The last equation gives the first correction
\be
\phi_1=e^{-\int\frac{\theta'}{2\theta-i\nu}d\eta}\lb\pm\int\frac{ie^{-\int\frac{\theta'}{2\theta-i\nu}d\eta}}{2\theta-i\nu}\phi''_0 d\eta+c_1\rb\,.
\ee

\subsection{Large-$k$ expansion for one variable}

Alternatively, we solve our problem by assuming that the wavenumber is large compared to the rest of the parameters. Therefore, we make an expansion in $k\rightarrow k/\epsilon$ where $\epsilon$ is again a small, dimensionless parameter. Before that, we can first absorb the friction term
\be
\phi=e^{-\frac{1}{2}\int\nu dt}\tilde{\phi}
\ee
so that
\be
\tilde{\phi}''+(w^2-\frac{1}{2}\nu'-\frac{1}{4}\nu^2)\tilde{\phi}=0\,.
\ee
Taking that $w^2=c^2k^2+m^2$ and that $\nu=\mathcal{O}(k^0)$, we make an expansion for large $k$ applying the ansatz
\be
\tilde{\phi}=(\phi_0+\epsilon\phi_1+\cdots)e^{(i/\epsilon)\int \theta dt}\,.
\ee
Solving order by order one obtains
\begin{align}
&\epsilon^{-2}:~~~~~~~\lp c^2k^2-\theta^2\rp\phi_0=0\,, \\
&\epsilon^{-1}:~~~~~~~~2i\theta\phi'_0+i\theta'\phi_0+(c^2k^2-\theta^2)\phi_1=0\,, \\
&\epsilon^{0}:~~~~~~~~~\phi_0''+(m^2-\frac{1}{2}\nu'-\frac{1}{4}\nu^2)\phi_0+2i\theta\phi'_1+i\theta'\phi_1=0\,. 
\end{align} 
The first equation solves the phase 
\be
\theta=\pm c\cdot k\,.
\ee
The second equation fixes the leading term in the amplitude
\be
\phi_{0}=\frac{c_0}{\sqrt{\theta}}\,,
\ee
where $c_0$ is a constant. 
The last equation gives the first correction
\be
\phi_1=\frac{1}{\sqrt{\theta}}\lb\pm\int\frac{i}{2\sqrt{\theta}}\lp\phi''_0+(m^2-\frac{1}{2}\nu'-\frac{1}{4}\nu^2)\phi_0\rp dt+c_1\rb\,.
\ee
All together, the large $k$-solution stands as
\be
\phi=\frac{e^{-\frac{1}{2}\int\nu dt}}{\sqrt{\theta}}\lb c_0\pm\int\frac{i}{2\sqrt{\theta}}\lp\phi''_0+(m^2-\frac{1}{2}\nu'-\frac{1}{4}\nu^2)\phi_0\rp dt+\cdots\rb e^{\pm i\int\theta dt}\,,
\ee
where a general solution can be constructed by summing two independent particular solutions. One can see that in this limit, the mass term does not affect the phase and only enters at second order as a correction to the amplitude. This is contrasts with the WKB where the mass enters already at leading order in the phase.

\bibliographystyle{JHEP}
\bibliography{GWsOscillations_refs}
\end{document}